\newcommand{\beq}{\begin{equation}}
\newcommand{\eeq}{\end{equation}}
\newcommand{\eqna}{\begin{eqnarray}}
\newcommand{\eqne}{\end{eqnarray}}
\newcommand{\mb}{\mathbf}
\newcommand{\vecr}{\mathbf{r}}
\documentclass[preprint,preprintnumbers,amsmath,amssymb,prd,tightenlines]{revtex4}
\usepackage{graphicx}
\usepackage{color}
\begin{document}
\setlength{\voffset}{.5cm}
\title{Feshbach resonances and their interaction in light scattering off photonic crystal slabs
}
\author{I. Evenor  $^{{\rm a}}$}  \author{E. Grinvald $^{{\rm a}}$}    \email{eran.grinvald@weizmann.ac.il}
\author{F. Lenz $^{{\rm b}}$}    \email{flenz@theorie3.physik.uni-erlangen.de}
\author{S. Levit$^{{\rm a}}$
}    \email{shimon.levit@weizmann.ac.il}
\affiliation{$^{{\rm a}}$ Department of Condensed Matter \\ The Weizmann Institute of Science\\ Rehovot Israel   \\ \\
$^{{\rm b}}$ Institute for Theoretical Physics III \\
University of Erlangen-N\"urnberg \\
Staudstrasse 7, 91058 Erlangen, Germany\\ }
\date{July 20, 2011}
\begin{abstract}

The concept of Feshbach resonances developed for  quantum mechanical scattering is applied in the analysis of
classical light scattering off photonic crystal slabs. It is shown that this concept can be realized almost
perfectly in these systems.  As an application guided-mode resonances
in the grating waveguide structure (GWS) are  studied in detail.
Using simple resonance dominance approximation the characteristic properties of isolated  Feshbach resonances in light scattering  are  exhibited. Formation and  interaction  of overlapping resonances are investigated.   The  relevant parameters of the GWS  are identified which control the shape of the reflectivity of interacting resonances as well as the enhancement of the electromagnetic field. The differences in the properties of TE and TM resonances  is emphasized for both isolated and interacting resonances.

\end{abstract}
\maketitle
\section{Introduction}
A  general property of scattering of waves, e.g., of  light or matter waves is the appearance of resonances. Resonances are observed in acoustic waves as  well as in scattering of waves associated with elementary particles at wavelengths ranging from meters to $10^{-15}\,$m.   Crudely speaking there are two classes of resonances, the shape (or potential) resonances and Feshbach resonances. Shape resonances occur in a variety of classical and quantum mechanical systems in which  wavelength and the size of the resonator or target are of the same order of magnitude.

The concept of Feshbach resonances \cite{FESH58,FESH62} on the other hand  applies to  quantum mechanical  scattering on many-body systems such as atomic nuclei \cite{FESH92} and has found recently important applications in atomic and molecular physics \cite{TTHK99,CGJT09}. A Feshbach resonance occurs if the kinetic energy  of the incident particle is close to an almost stable intermediate ``molecular'' state. In the field of cold atoms,  it has been instrumental  that the  condition for appearance of Feshbach resonances can be manipulated by tuning  the strength of  an external magnetic field to which the magnetic moments of the atoms are coupled.

In this work we will  show that Feshbach resonances  occur in the  very different context of  scattering of  light  off photonic crystal slabs. The fundamental mechanism for the formation of resonances is the process of turning a bound state (guided mode) into a resonance ``state'' (barely radiating mode).

We will focus our studies on a particular photonic crystal slab  - the grating waveguide structure (GWS). Guided mode resonances in GWS have been studied both theoretically and experimentally since 1985, Refs.\,\cite{GSST85,MP85}, and have already been incorporated into a wide variety of applications, cf.\,Refs.\,\cite{WM93,MCWEC08,BFCBBDKTS10,FLPFB10,MSJ10,KLGY10}. Theoretically they have been investigated using several different approaches including a scattering matrix approach \cite{FMS02}, temporal coupled mode theory \cite{FSJ03,FAJO02}, Wigner-Weisskopf formalism \cite{K03} as well as coupled mode theory \cite{RSA97}. In the majority of these theoretical approaches  parameter fitting is required to make comparison with exact numerical calculations \cite{SSS82,MBPK95}. Our approach is similar to the coupled mode approach of Rosenblatt et al. \cite{RSA97} which was in turn based on the work of Kazarinov et al. in distributed feedback lasers \cite{KH85}. Yet the origin of our treatment is different being based on the formalism of Feshbach resonances. In this way  are able to derive accurate analytical expressions for essentially all resonance properties.  Miroshnichenko \cite{M09,MFK10} and others \cite{KCLG11} have pointed out that various photonic resonances can be treated as Fano-Feshbach resonances, yet since no  analytic expressions for the coupling exist a parameter fitting procedure had to be be used. In addition most of the theories concentrated on the  properties of the TE resonances. To the best of our knowledge non of the fully analytic theories of guided mode resonances treated the TM polarized resonances and their differences from the TE resonances.

As in the case of atomic physics, the light scattering  off GWS will be shown to be tunable   in a  variety of ways generating isolated as well as overlapping  Feshbach resonances. We will apply  a simple resonance dominance approximation and will obtain essentially analytic expressions for all relevant observables.  In this way we will provide the tools to  produce resonances with  properties  desired in applications by adjusting the  the GWS  parameters. The resonance dominance approximation will  also allow us to study systematically the interaction of two  or more Feshbach resonances.


\section{Resonance scattering of light - formal development}

\begin{figure}[ht] \centering
\vspace{.2cm}\hspace{-8cm}\includegraphics[width=.52\linewidth]{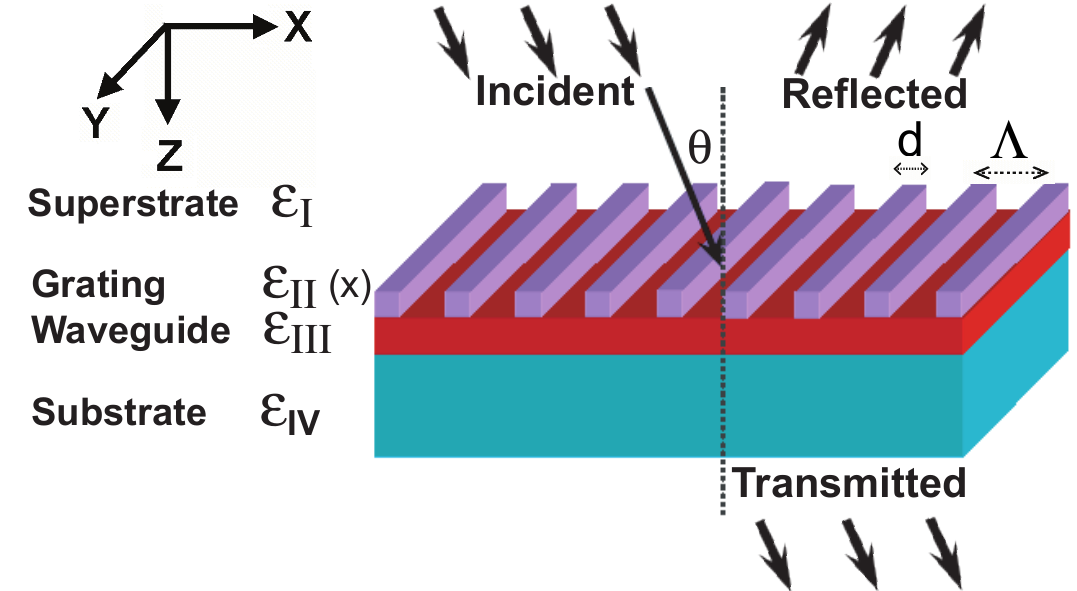}\vskip-5.05cm\hspace{8cm}\includegraphics[width=.4\linewidth]{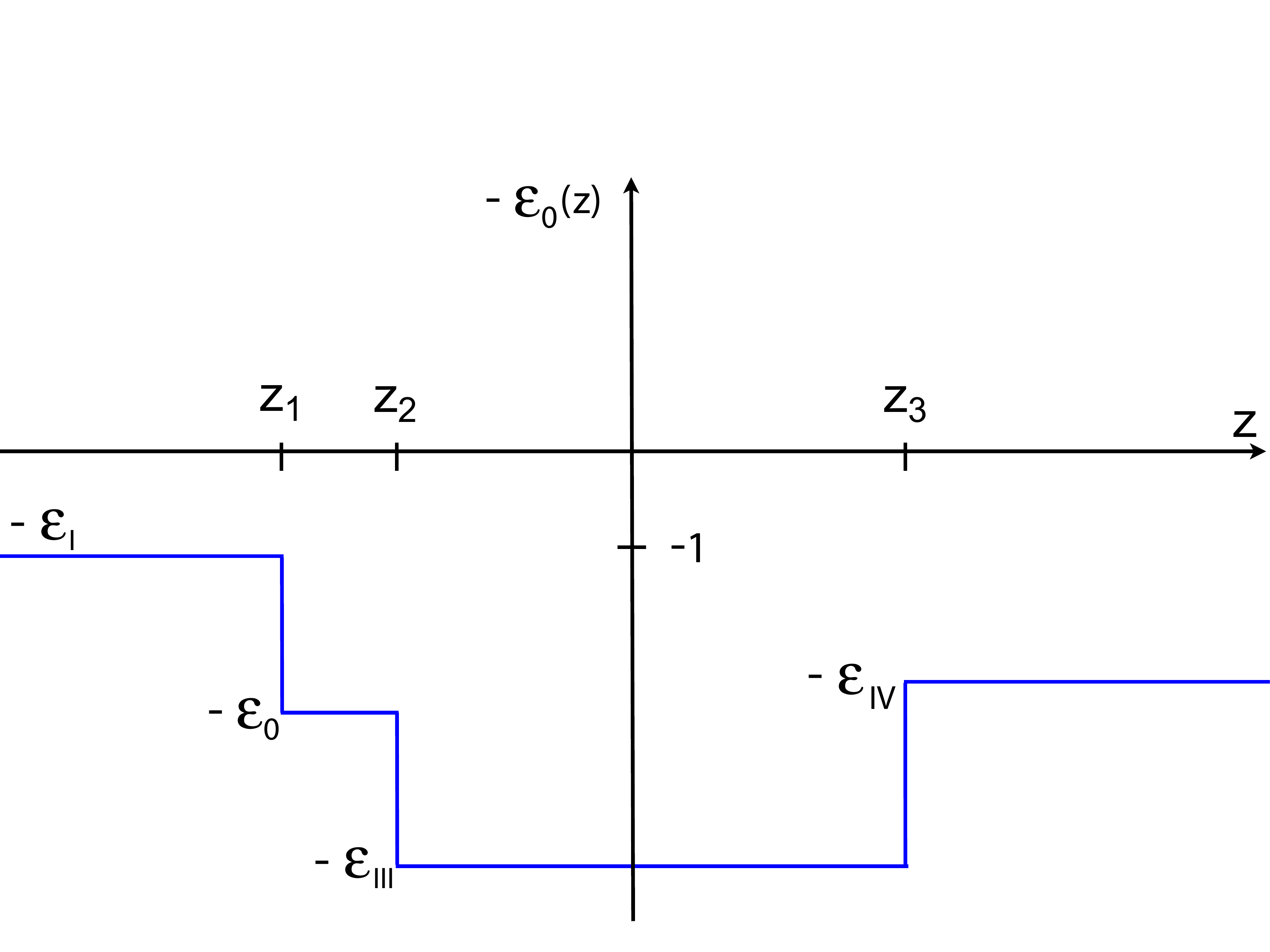}
\caption{Left:\,Light incident on a   grating waveguide structure (GWS) composed of piecewise $z-$independent  layers with dielectric constants $\epsilon_I ...\epsilon_{IV}$.The geometry of the  grating layer (II) is characterized by the grating period $\Lambda$ and the duty cycle $d/\Lambda$. Right:\, Polarization dependent effective dielectric constant  (cf.\,Eqs.\,(\ref{epFocode}) and (\ref{eq:defofgamman}))  as a function of $z$ for the GWS on the left. }
\vskip -.1cm
\label{dispn}
\end{figure}

In this section we will develop a formalism  which will make explicit the role of Feshbach resonances in light scattering  off photonic crystals. For this purpose and as an example we will consider  the so called   grating  waveguide structure \cite{RSA97}  (GWS) which consists of a plane waveguide with  one dimensional grating  on the top of it.  The left part of Fig.\,\ref{dispn} shows an  example of a GWS with piecewise constant grating layer where the waveguide,  the grating, the substrate and the superstrate  layers are shown  as well as the incident, reflected and transmitted light. The right part of the figure shows the corresponding  dielectric  function  where we replaced the grating layer $\epsilon_{II} \left( x \right)$ with a homogeneous layer with an effective dielectric constant $\epsilon_0$. The exact  definition of  $\epsilon_0$ depends on the polarization of the incident light, cf.\, Eq.\,(\ref{epFocode})  and Eq.\,(\ref{eq:defofgamman})  below.  In our formal studies we will treat on equal footing both piecewise constant as well as continuous  transitions of $\epsilon$ between the dielectric layers.

In the present work we investigate only the case of classical incidence where the incident light wave vector is perpendicular to the grating grooves, i.e. $k_y=0$ .  We start  with the TE polarization.  The modifications needed for the TM case  will be considered in Section \ref{TMwv}.

\subsection{TE waves}
 In the TE polarization the electric field  of the incident light  is parallel to the grating grooves which  in our geometry (cf.\,Fig.\ref{dispn}) implies
\begin{equation}
\boldsymbol{E}(\boldsymbol{r})= E(x,z)\,\boldsymbol{e}_y\,.
\label{vecE}
\end{equation}
The  stationary  Maxwell equation for  a time harmonic electric field of frequency $\omega$ (with the speed of light set to unity)
\begin{equation}
\boldsymbol{\nabla}\times\boldsymbol{\nabla}\times \boldsymbol{E}(\boldsymbol{r}) = \epsilon(\boldsymbol{r})\omega^2 \boldsymbol{E}(\boldsymbol{r})\,,
\label{TED}
\end{equation}
then simplifies to
\begin{equation}
 \Big(-\frac{\partial^2}{\partial z^2} -\frac{\partial^2}{\partial x^2} -\epsilon(x,z) \omega^2 \Big)E(x,z)= 0\,.
\label{swva1}
\end{equation}
 To describe light incident on GWS from $z\to-\infty$  with the wave vector in the $x$-$z\,$ plane  ${\bf k}=(k_x,0,k_z)$ we impose  the boundary conditions
that for $z\rightarrow\infty$ we have only an outgoing wave. With this choice the asymptotic behavior of $E(x,z)$ is
\beq
\lim_{z\rightarrow-\infty} E(x,z)=e^{ik_x x}(e^{ik^{-}_z z} + r(\omega,k_x)e^{-ik^{-}_z z}) \;\; , \;\;  \lim_{z\rightarrow \infty} E(x,z)=t(\omega,k_x)e^{ik_x x} e^{ik^{+}_z z}\,.
\label{eq:boundcondTEE}
\eeq
The choice of the normalization is free and, as is common, we have normalized the amplitude of  the incident wave to be 1. As a consequence, the reflected and transmitted fields are given in units of the incident field. The quantities $r(\omega,k_x)$  and $t(\omega,k_x)$ are respectively the reflection and transmission amplitudes. One of our goals is to devise a simple method to calculate these amplitudes and to show  that  their resonant behavior is typical of Feshbach resonances.  Note that $k_x^2+(k_z^{-})^2=\epsilon_{\text{I}}\omega^2$ while $k_x^2+(k_z^{+})^2=\epsilon_{\text{IV}}\omega^2$

The x-dependence of the  dielectric constant is limited to the  grating layer ($z_1 < z < z_2$). It is periodic with the grating period $\Lambda$
\begin{equation}
\epsilon(x+\Lambda,z)=\epsilon(x,z)\,.
\label{peri}
\end{equation}
and can be represented  in terms of its Fourier-components
 \begin{equation}
\epsilon_n(z)=\frac{1}{\Lambda}\int_{-\Lambda/2}^{\Lambda/2} dx \,\epsilon(x,z) e^{-in K_g x}\,.
\label{epFocode}
\end{equation}
Here we consider  real valued $\epsilon(x,z)$
\begin{equation}
\epsilon_{-n}(z)=\epsilon_n^{\,\star}(z)\,.
\label{cocu}
\end{equation}
If $\epsilon(x,z)$ is symmetric or antisymmetric around an appropriately chosen  center of the elementary interval, the Fourier-components $\epsilon_n(z)$ are real or imaginary respectively. In the  coordinate system  shown in Fig.\ref{dispn}, the periodicity of $\epsilon(x,z)$ implies that the x dependence of $E(x,z)$  contains only Fourier components with the discrete  x-components of the wave vectors of the form
\begin{equation}
 k_x+nK_g\,,\quad K_g= \frac{2\pi}{\Lambda}.
\label{disv}
\end{equation}
Inserting
\begin{equation}
E(x,z) = \sum_{n=-\infty}^{\infty} E_n(z) e^{-i(k_x+n K_g)x}\,,
\label{fexp}
\end{equation}
 into Eq.\,(\ref{swva1}) we rewrite the latter  as a system of ordinary differential equations (here and in the following we will suppress the summation limits $\pm\infty$)
\begin{equation}
\big(-\frac{d^2}{dz^2} +(k_x+n K_g)^2 -\epsilon_0(z)\omega^2\big) E_n(z) = \omega^2\sum_{m\neq n}\epsilon_{n-m}(z) E_m(z)\,.
\label{syseq}
\end{equation}
To see  what the asymptotic conditions (\ref{eq:boundcondTEE}) imply for this system we note that for $m\neq n$ $\epsilon_{n-m}(z)=0$ when $z\rightarrow\pm\infty$ so that the equations decouple at large $|z|$. We will further assume  that we deal with  the so called subwavelength zero order grating, i.e. such that
$$ \omega^2-(k_x+nK_g)^2\geq 0\quad \text{only \;\; for}\;\; n=0.$$
Therefore all the components $E_n(z)$ with $n\ne 0$ are required to decay  exponentially with   $z\rightarrow\pm\infty$.  The non vanishing asymptotic conditions (\ref{eq:boundcondTEE})  apply only to  $E_0(z)$
\beq
\lim_{z\rightarrow-\infty} E_0(z)=e^{ik^-_z z} + r(\omega,k_x)e^{-ik^-_z z} \;\; , \;\;  \lim_{z\rightarrow \infty} E_0(z)=t(\omega,k_x) e^{ik^+_z z}\,.
\label{bcte2}
\eeq
We convert  (\ref{syseq})  into a system of integral equations and  introduce  to this end  the Green's functions satisfying
 \begin{equation}
 \Big(-\frac{d^2}{d z^2} + (k_x+nK_g)^2 -\epsilon_0(z) \omega^2 \Big)g_n(z,z^\prime)= -\delta(z-z^\prime)\,,
\label{gzp}
\end{equation}
and obtain
\begin{equation}
E_n(z)= E_0^{(+)}(z)\delta_{n0}-\omega^2\int dz^\prime g_n(z,z^\prime) \sum_{m\neq n} \epsilon_{n-m}(z^\prime)E_m(z^\prime)\,.
\label{E0+}
\end{equation}
Here we use the fact that light is incident only in the $n=0$ channel. The term $E_0^{(+)}(z)\delta_{n0}$
satisfies (\ref{syseq}) with vanishing right hand side so that  $E_0^{(+)}(z)$ is a solution of
 \begin{equation}
\Big(-\frac{d^2}{d z^2} + k_x^2 -\epsilon_0(z) \omega^2 \Big)E_0^{(+)}(z)=0\,.
\label{hom}
\end{equation}
and we impose the same boundary condition as in Eq.\,(\ref{eq:boundcondTEE})  with reflection and transmission amplitudes which we will denote by $r_0(\omega,k_x)$ and $t_0(\omega,k_x)$ respectively. It is the z-dependent part of the electric field $E_0^{(+)}(z)e^{ik_x x}$ which propagates in and is scattered off the effective dielectric structure defined by $\epsilon_0(z)$  shown in the right part of Fig.\,\ref{dispn}. We will refer to this scattering as   ``background scattering'' upon which  Feshbach resonances are formed by coupling to guided modes. In   Appendix \ref{sec:bagr} the appropriate  ``background''  Green's function is explicitly constructed.

Systems of equations of coupled channels such as Eqs.\,(\ref{syseq},\,\ref{E0+}) are underlying the description of
Feshbach resonances  in atomic and nuclear physics.  A Feshbach resonance occurs if, by neglecting  certain  couplings, a bound state of the entire system exists. Accounting for the couplings converts this state into a resonance.In this spirit let us  consider the above system (\ref{syseq}) in absence of the coupling $\epsilon_n(z) =0,\; n\neq 0$ of the Fourier components $E_n(z)$. In this approximation, the photonic crystal slab is described by the effective dielectric constant  in the right part of Fig.\,\ref{dispn} and the resulting guided modes are the
 progenitors of Feshbach resonances.

As the wave functions of bound states in quantum mechanics,  guided modes in  dielectric  slabs are localized in the z-direction
\begin{equation}
 z\to \pm \infty\,, \quad E(x,z) \to 0\,,
\label{loc}
\end{equation}
and  are determined by the eigenvalue equation
\begin{equation}
 \Big(-\frac{d^2}{d z^2} -\epsilon_0(z) \omega^2 \Big)\mathcal{E}_\eta(z)= \eta\, \mathcal{E}_\eta(z)\,,
\label{swvafc}
\end{equation}
with the frequency dependent eigenvalues $\eta =\eta(\omega)$. The electric  field  is given by
\begin{equation}
E(x,z)= e^{i\beta  x}  \mathcal{E}_\eta(z)\,.
\label{efi}
\end{equation}
Inserting this ansatz into the wave equation (cf.\,Eq.\,(\ref{swva1}))
\begin{equation}
 \Big(-\frac{\partial^2}{\partial z^2} -\frac{\partial^2}{\partial x^2} -\epsilon_0(z) \omega^2 \Big)E(x,z)= 0\,,
\label{swva0}
\end{equation}
the dispersion relation between frequency $\omega$  and the x-component  $\beta$ of the wave vector is obtained
\begin{equation}
\beta^{2} = -\eta(\omega)\,.
\label{disp2}
\end{equation}

\begin{figure}[ht] \centering
\includegraphics[width=.5\linewidth]{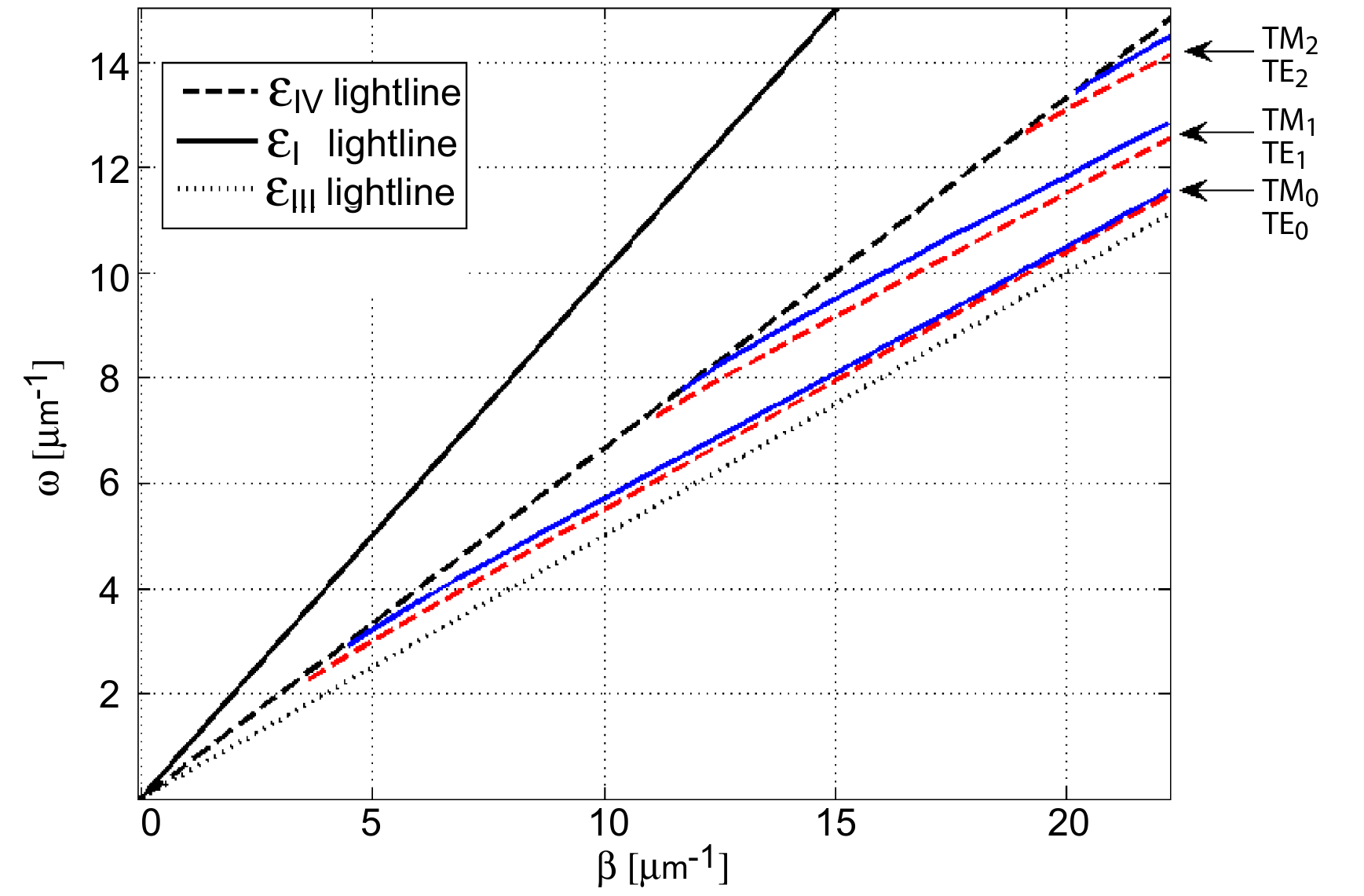}
\caption{Guided mode frequencies $\omega_i(\beta)$ as a function of the propagation constant $\beta$ for a medium with the z-dependent dielectric constant $\epsilon_0(z)$ of Fig.\,(\ref{dispn}).  The numerical values of the dielectric constants are given in Eqs.\,(\ref{size}) and (\ref{gratin}) and $i$ is the mode order. Dashed curves are for TE guided modes (red color online), solid curves TM (blue color online). Also shown are the lightlines $\omega=\beta/\sqrt{\epsilon}$ for $\epsilon_I$, $\epsilon_{III}$ and $\epsilon_{IV}$.}
\label{lcone}
\end{figure}

In Fig. (\ref{lcone}) we show as an  example the dispersion curves of the first three TE and TM guided modes for a medium with the z-dependent dielectric constant $\epsilon_0(z)$ of Fig.\,(\ref{dispn}) and numerical values of the dielectric constant given in Eqs.\,(\ref{size}) and (\ref{gratin})

The subset of equations  (\ref{E0+}) with $n\neq 0$ can be  rewritten as
\begin{equation}
 E_n(z) = -\int dz^\prime \left\{  \sum_i \frac{\mathcal{E}_{\eta_i}(z) \mathcal{E}_{\eta_i}(z^\prime) }{-\eta_i-(k_x+nK_g)^2}+ G_c(z,z^\prime)\right\}\omega^2\sum_{m\neq n}\epsilon_{n-m}(z^\prime) E_m(z^\prime)\,,
\label{inteq}
\end{equation}
where  the first term in the curly brackets  is the contribution of the guided modes  $\mathcal{E}_{\eta_i}(z)$ to the  the Green's function $g_n(z,z^\prime)$ (cf.\,Eq.\,(\ref{gzp})) and $ G_c(z,z^\prime)$ the contribution of the continuum (radiating) modes. The guided modes are described by   normalizable functions and we  require
\begin{equation}
\int_{-\infty}^\infty
 dz  \mathcal{E}_{\eta_i}(z)\mathcal{E}^{\star}_{\eta_i}(z) =1\,.
\label{nrgu}
\end{equation}
With this exact reformulation of the original wave equation  we are in the position to formulate the criterion for appearance and dominance of Feshbach resonances. Theoretically, an isolated Feshbach resonance is expected to occur for sufficiently small coupling  ($\epsilon_n(z)\;n\neq 0$) and if the kinematics ($k_x$ and $\omega$) is chosen such that (cf.\,Eq.\,(\ref{inteq}))
\begin{equation}
\eta_i+(k_x+nK_g)^2 \approx 0\,.
\label{cfsh}
\end{equation}
In this case the scattering of light will be dominated by one of the   terms  in the discrete part of the spectrum, i.e., by one of the guided modes localized in $z$. The coupling to the extended mode delocalizes these modes and  turns them  thereby into resonances, i.e.,  guided modes become ``slightly radiating''.
 Narrow subwavelength resonances in this system were observed  (cf. Refs. \cite{RSA97,BFCBBDKTS10,LTSYM98,DTZYJF00,BLTFS09}) in a wide range of the parameters,\, i.e.  wavelength, angle of incidence and the dielectric structure parameters. This supports the assumption of weak coupling of the guided to  the extended modes.


The  criterion (\ref{cfsh}) is essentially identical to the quantum mechanical criterion for Feshbach resonances which requires that the incident energy coincides approximately with that of a ``molecular'' state which in the limit of vanishing coupling becomes a true bound state. As in scattering of ultracold atoms, the properties of the Feshbach resonances can be manipulated in light scattering as well. They can be tailored towards a particular application  by variation of the parameters of the system or externally by using electro-optical materials\cite{KYFMHZ05}.

For the analysis of the resonances observed in  scattering of light off photonic crystal slabs  the  reformulation of the wave equation (\ref{swva1}) by the system of equations ((\ref{gzp}),\,(\ref{inteq})) suggests to neglect  for $n\neq 0$ the continuum contributions  to  the Greens functions $g_n(z,z^\prime)$, i.\,e., to approximate Eq.\,(\ref{inteq}) by
\begin{equation}
 E_n(z) \approx \omega^2\int dz^\prime  \sum_{i} \frac{\mathcal{E}_{\eta_i}(z) \mathcal{E}_{\eta_i}(z^\prime) }{\eta_i+(k_x+nK_g)^2}\sum_{m\neq n}\epsilon_{n-m}(z^\prime) E_m(z^\prime)\,, \quad n\neq 0.
\label{redo}
\end{equation}
We will refer to this approximation as ``resonance dominance''. In general the condition (\ref{cfsh}) is satisfied for one guided mode only and the contributions from the other guided modes can be neglected. With a a judicious choice of the parameters, a simultaneous excitation of two or more  Feshbach resonances can be achieved and the interaction of these ``overlapping'' resonances can be studied.  A particular class of two interacting  Feshbach resonances  will be discussed later.

Notwithstanding  the formal connection between Feshbach resonances in quantum mechanical scattering of particles and in classical scattering of TE polarized light   the  content of the corresponding wave equations is very different. We mention in particular the role of the frequency. The Maxwell equation (\ref{TED}) implies that the frequency plays a twofold role. On the one hand, both  in quantum mechanics and in electrodynamics,  the frequency   determines the asymptotic properties  of the incident particle or  light. Simultaneously  in electrodynamics, the frequency also determines  the strength of the interaction of light with the dielectric medium.  Peculiar consequences are the frequency dependence of the eigenvalues $\eta$ and the  electric fields of the guided modes $\mathcal{E}_\eta(z)$  (cf.\,Eq.(\ref{swvafc})) as well as the independence of these quantities  on the x-component of the wave vector. As we will see below, these properties not only distinguishes TE modes from quantum mechanical waves but also from the TM modes.

\subsection{TM waves}\label{TMwv}
In  many aspects the treatment of the TM polarization is identical to the TE case so we will present it briefly.
In the  TM polarized waves the magnetic field is parallel to the grating grooves which in our geometry means  that
\beq
\mb{H}(\vecr)=H(x,z)\mb{e}_y\,.
\eeq
Inserting this  into the Maxwell equation  (we set c=1)
\beq
\nabla\times\frac{1}{\epsilon(\vecr)}\nabla\times \mb{H}=\omega^2\mb{H}
\eeq
we obtain
\beq
\label{eq:eq3forH}
\left[-\partial_x\frac{1}{\epsilon(x,z)}\partial_x-\partial_z\frac{1}{\epsilon(x,z)}\partial_z\right]H(x,z)=\omega^2 H(x,z)\,.
\eeq
As discussed in  \cite{JJWM08}  this equation is analogous to a Schrodinger equation in two dimensions with no potential but coordinate dependent mass. Accordingly we write it in the form
\beq
\label{eq:eq4forH}
\Theta H=\omega^2 H
\eeq
where
\beq
\Theta=-\partial_x\frac{1}{\epsilon(x,z)}\partial_x-\partial_z
\frac{1}{\epsilon(x,z)}\partial_z\,.
\eeq
As in the TE case we  need to solve the  above equation under the condition that  only outgoing wave is present at $z\rightarrow\infty$
\beq
\lim_{z\rightarrow-\infty} H(x,z)=e^{ik_x x}(e^{ik^{-}_z z} + r(\omega,k_x)e^{-ik^{-}_z z}) \;\; , \;\;  \lim_{z\rightarrow \infty} H(x,z)=t(\omega,k_x)e^{ik_x x} e^{ik^{+}_z z}
\label{eq:boundcondTE}
\eeq
where as before  $r(\omega,k_x)$  and $t(\omega,k_x)$ are respectively the reflection and transmission amplitudes  for the present TM scattering.
Using the  analog of the expansion  in Eq. (\ref{fexp})
\beq \label{eq:expanofH}
H(x,z)=\sum_{n=-\infty}^{\infty} H_n(z) e^{-i(k_x+n K_g)x}\,,
\eeq
and inserting in (\ref{eq:eq4forH}) we obtain a set of coupled equations similar to (\ref{syseq})
\beq
\label{eq:eq2forxin}
[\omega^2-\Theta_{nn}]H_n(z)=
\sum_{m\ne n}
\Theta_{nm}H_m(z)\,,
\eeq
where we introduced
\beq  \label{eq:defofThetanm}
\Theta_{nm}=-\partial_z \gamma_{n-m}(z) \partial_z
 + (k_x+nK_g)(k_x+mK_g)\gamma_{n-m}(z)\,,
\eeq
and  defined
\beq \label{eq:defofgamman}
\gamma_n(z)\equiv\frac{1}{\Lambda}\int_{-\Lambda/2}^{\Lambda/2} dx \frac{1}{\epsilon(x,z)}e^{-i n K_g x}\;.
\eeq
The effective dielectric constant $\epsilon_{0}(z)$ shown in Fig.\,\ref{dispn}  is (for the TM polarization) just  $1/\gamma_0(z)$.

We note that there exist an ambiguity in our definition of  $\gamma_{n-m}(z)$ in (\ref{eq:defofThetanm}). If we view it as a matrix $\gamma_{nm}(z)$ we can either use the above definition $\gamma_{nm}(z)=\gamma_{n-m}(z)$  as  given by
Eq. (\ref{eq:defofgamman}) or we could also have defined it as the  inverse of the matrix $\epsilon_{n-m}$ with $\epsilon_{n}$ given by Eq. (\ref{epFocode}). We will comment upon and test this ambiguity below in our numerical examples.

As in the TE case the subwavelength grating condition imply that all the components $H_n(z)$ with $n\ne 0$ should have  exponential decay as the boundary conditions at $z\rightarrow\pm\infty$. The zeroth component $H_0(z)$ should  asymptotically behave  as
\beq
\lim_{z\rightarrow-\infty} H_0(z)=e^{ik^-_z z} + r(\omega,k_x)e^{-ik^-_z z} \;\; , \;\;  \lim_{z\rightarrow \infty} H_0(z)=t(\omega,k_x) e^{ik^+_z z}\,.
\label{eq:boundcondTH1}
\eeq
Let us now convert the set (\ref{eq:eq2forxin}) into integral equations  similar to what we did in
 Eq. (\ref{E0+})
 \beq \label{eq:inteqforH}
 H_n(z)=H_0^{(+)}(z)\delta_{n0}+\int dz' g_n(z,z')\sum_{m\ne n} \Theta_{nm}H_m(z')\,,
 \eeq
where in analogy to Eq.\,(\ref{hom}) the component $H_0^{(+)}(z)$  and the Green's functions  $g_n(z,z')$  solve respectively  the differential equations
\begin{equation}
[\omega^2-\Theta_{00}] H_0^{(+)}(z)=0\,,
\label{bgh}
\end{equation}
and
\beq
[\omega^2-\Theta_{nn}]g_n(z,z') = \delta(z-z')\,.
\label{eq:TM-green}
\eeq
Here $H_0^{(+)}(z)$ describes  the TM scattering off the effective  structure defined by $\epsilon(z)=\gamma_0^{-1}(z)$.
As in the TE case we choose it to  satisfy the  boundary conditions similar to (\ref{eq:boundcondTH1})   denoting the corresponding reflection and transmission amplitudes by  $r_0(\omega,k_x)$ and $t_0(\omega,k_x)$  respectively. Accordingly we chose the boundary conditions  for  the Green's function $g_0(z,z')$
such that  the full  solution $H_0(z)$ satisfies   Eq. (\ref{eq:boundcondTH1}).

It  is useful to consider the eigenfunctions of the operators $\Theta_{nn}$ with different $n$'s
\beq
\label{eq:eqforpsi}
\Theta_{nn}\mathcal{H}_{n,\nu}(z) = \eta_{n,\nu} \mathcal{H}_{n,\nu}(z)\,.
\eeq
Note that for each $n$ there is a complete set of eigenfunctions $\mathcal{H}_{n,\nu}(z)$ with the corresponding eigenvalues $\eta_{n,\nu}$ distinguished  by the index $\nu$.

From the comparison with the Maxwell equation (\ref{eq:eq3forH}) one can easily see that the eigenfunctions $\mathcal{H}_{n,\nu}(z)$  determine the z dependent part of the TM photonic modes  with propagation constant $k_x+nK_g$ and frequency $\omega_{n,\nu}^2=\eta_{n,\nu}$,
 \beq
 H_{(k_x+nK_g),\nu}(x,z)=e^{i(k_x+nK_g)x}\mathcal{H}_{n,\nu}(z)
\label{eq:eqforpsi2}
 \eeq
 of the  effective  structure defined by   $\epsilon(z)=\gamma_0^{-1}(z)$.  We will call such a structure  "unperturbed".  We will base the treatment of the effects of the x-dependence of the grating upon solutions for this structure which we assume known.




We now concentrate on the equations (\ref{eq:inteqforH}) with $n \ne 0$  and use the spectral representation of  $g_n(z,z')$ as in
Eq. (\ref{inteq})
\begin{equation}
 H_n(z) = \int_{I_g} dz^\prime \left\{  \sum_i \frac{\mathcal{H}_{n,\nu_i}(z) \mathcal{H}_{n,\nu_i}(z^\prime) }{\omega^2-\eta_{n,\nu_i}}+G_c(z,z')\right\}
\sum_{m\ne n} \Theta_{nm}H_m(z')
\label{eq:inteqforHnwithnonzeron}
\end{equation}
where  the first term in    the curly brackets  is the contribution of the guided modes  (denoted by discrete index $\nu_i$) to the  the Green's function $g_n(z,z^\prime)$ while
 $ G_c(z,z^\prime)$ is the contribution of the continuum.

We have assumed the normalization of the guided modes
\begin{equation}
\int_{-\infty}^\infty
 dz  \mathcal{H}_{n,\nu_i}(z)\mathcal{H}^{\star}_{n,\nu_i}(z) =1\,.
\label{eq:nrforH}
\end{equation}
Note that since $\Theta_{nm}$ with $n \ne m$ is vanishing outside the grating interval  $I_g$ , the integrals in (\ref{eq:inteqforHnwithnonzeron}) are over $I_g$

As in the TE case  our main approximation will consist in neglecting  the continuum contributions  to  the Greens functions $g_n(z,z^\prime)$ for $n\neq 0$.  Thus we approximate Eq. (\ref{eq:inteqforHnwithnonzeron}) as
\beq
H_n(z) = \int_{I_g} dz^\prime \left\{ \sum_i\frac{\mathcal{H}_{n,\nu_i}(z) \mathcal{H}_{n,\nu_i}(z^\prime)} {\omega^2-\eta_{n,\nu_i}}\right\}
\sum_{m\ne n} \Theta_{nm}H_m(z')\,.
\label{redoTM}
\eeq
 This is the ``resonance dominance'' approximation in the TM case. An isolated TM Feshbach resonance occurs when the physical parameters are such that
 \beq  \label{eq:TMresoncond}
\eta_{n,\nu_i}\approx \omega^2
\eeq
and the contributions from the other guided modes can be neglected.  Of special interest to us will also be cases of
a simultaneous excitation  and interaction of two or more  Feshbach resonances.

\section{Isolated resonances}
\subsection{TE resonances}
In this section we will discuss in detail properties of an  isolated Feshbach resonance. By comparison with the results of exact numerical evaluations, we will determine the degree of validity of resonance dominance in light scattering.
To be concrete, we will carry out this study  in the   grating waveguide structure (GWS) as shown on the left in Fig.\,\ref{dispn}.
The grating,  periodic in $x$  (cf.\,Eq.\,(\ref{peri})),  is restricted in the z-direction to  the region II  which we denote as the grating interval $I_g$.
The effective dielectric constant  is given by (cf.\,Eq.\,(\ref{epFocode}) and the right part of Fig.\,\ref{dispn})
\begin{equation}
\epsilon_{0}(z) =  \epsilon_{\text{I}}\theta(z_1-z) + {\epsilon_0}\theta(z-z_1)\theta(z_2-z) + \epsilon_{\text{III}}\theta(z-z_2)\theta(z_3-z)+\,\epsilon_{\text{IV}}\theta(z-z_3)
\,.
\label{teps}
\end{equation}
while
\begin{equation}
\epsilon_n(z) = \epsilon_n\,\theta(z-z_1)\theta(z_2-z)\;\;\;\; {\rm for} \;\;\; n \ne 0 \,.
\label{epti}
\end{equation}
In these expressions $\epsilon_n$ denote constants given by Eq. (\ref{epFocode}) for the  layered structure   when $z$ is restricted to the grating interval.

Let us assume that these parameters of the GWS  as well as the angle of illumination  and the frequency are such  that the condition (\ref{cfsh})  is satisfied for a single value of    $n=\nu$ and a particular
eigenvalue $\eta_0$   of the guided mode equation (\ref{swvafc}). Let us denote by $\mathcal{E}_{\eta_0}$  the corresponding eigenfunction.   Beyond the  resonance dominance (cf.\,Eq.\,(\ref{redo})) we truncate the system of equations ((\ref{E0+}),\, (\ref{redo})) further and take into account only  two modes, the extended mode of the incident light  (n=0) and the guided mode ($\eta_0,\;n=\nu$ ).  We thus reduce  the system of equations (\ref{E0+})  to
\begin{eqnarray}
\label{car1}
&&E_0(z)= E_0^{(+)}(z) - \epsilon_{-\nu}\,\omega^2\int_{I_g} dz^\prime g_0(z,z^\prime) E_{\nu}(z^\prime)\,,\\
\label{car2}
&& E_{\nu}(z)=   \frac{ \epsilon_{\nu}\,\omega^2}{\eta_0+(k_x+\nu K_g)^2}\,\mathcal{E}_{\eta_0}(z) \int_{I_g} dz^\prime\mathcal{E}_{\eta_0}(z^\prime)  E_{0}(z^\prime)\,,
\end{eqnarray}
and for simplicity will refer to this combined approximation to ``resonance dominance''.
This system can be solved analytically.  As  Eq. (\ref{car2}) shows the resonance component of the electric field  $ E_{\nu}$  is proportional to the resonating guided mode
\begin{equation}
 E_{\nu}(z) = \sigma_{\nu} \mathcal{E}_{\eta_0}(z)
\label{sigma}
\end{equation}
with  the proportionality coefficient $\sigma_\nu$   measuring the degree of the excitation of the guided mode.
Inserting this expression for $E_{\nu}(z)$ into Eq.\,(\ref{car1}), multiplying the  resulting  equation with  $\mathcal{E}_{\eta_0}(z)$  and integrating  we  obtain
the following expression for the field enhancement coefficient
\begin{equation}
\sigma_{\nu}= \epsilon_{\nu}\,\omega^2\frac{\mathcal{C}^{(+)}}{\rho}\,,
\label{sign}
\end{equation}
with $\mathcal{C}^{(\pm)}$ denoting the  coupling  to  the guided mode $\mathcal{E}_{\eta_0}$ of the  electric fields $E_0^{(\pm)}(z)$ incident  from either  $-\infty$
or  $\infty$ (cf.\,Eq.\,(\ref{asym}) for  their precise definitions) ,
\begin{equation}
\mathcal{C}^{(\pm)}=\int_{I_g} dz \mathcal{E}_{\eta_0}(z) E_0^{(\pm)}(z)\,.
\label{EI}
\end{equation}
The integral   is carried out over the grating interval $I_g$.
The   denominator in (\ref{sign})
\begin{equation}
\rho=\eta_0+ (k_x+\nu K_g)^2+ |\epsilon_{\nu}|^2 \omega^4\Sigma\,,
\label{rho}
\end{equation}
contains the resonance condition (cf.\,Eq.\,(\ref{cfsh})) modified by the ``self-coupling''  $\Sigma$ of the guided mode
\begin{equation}
\Sigma = \int_{I_g} dz \int_{I_g} dz^\prime  \mathcal{E}_{\eta_0}(z)
g_0(z,z^\prime)   \mathcal{E}_{\eta_0}(z^\prime)\,.
\label{shga}
\end{equation}
$\Sigma$ is generated by transitions from the  guided to
the extended mode, the propagation in the extended mode   and then the back transition to the guided mode. The strength of this contribution to $\rho$  is determined  by the corresponding Fourier coefficient $\epsilon_\nu$ of the dielectric function (cf.\,Eqs.\,(\ref{epFocode}),\,(\ref{cocu}),\,(\ref{epti})).
$\Sigma$ is complex with the  imaginary part  accounting for the loss of intensity from the guided to the extended mode. It gives rise to a shift of the resonance position and to a width. Using the identity (\ref{Imag}) derived in Appendix B,  the following expression for the width of the resonance is obtained
\begin{equation}
\frac{\tilde{\Gamma}}{2} = -|\epsilon_\nu|^2\omega^4\,\text{Im}\Sigma= \frac{\omega^4}{2 k_z^{-}}\big|\epsilon_\nu\, \mathcal{C}^{(+)}\big|^2\,,
\label{res}
\end{equation}
with the z-component of the wave vectors  $k^{\pm}_z$  defined in\,Eq.\,(\ref{asmo}). The  strength $\sigma_{\nu}$ of the guided mode excitation also determines the resonance behavior of reflection and transmission amplitudes which
are found  from   asymptotics of $E_0(z)$  in Eq.\,(\ref{car1}). Using  the expression for $g_0(z,z^\prime)$ and its asymptotics
derived in the Appendix B (Eq.\,(\ref{asg0})) as well Eq.\,(\ref{sigma}) we find
\begin{equation}
 r(\omega,k_x) - r_0(\omega,k_x)= \frac{i}{2 k_z^{-}} \sigma_{\nu}\, \epsilon_{-\nu}\,\omega^2\, \mathcal{C}^{(+)}=  -\frac{i}{2 k_z^{-}} \frac{\big|\epsilon_\nu\big|^2\omega^4\mathcal{C}^{(+)\,2}}{-\eta_0- (k_x+\nu K_g)^2- |\epsilon_{\nu}|^2 \omega^4\Sigma}\,.
\label{tore}
\end{equation}
Similarly the corresponding expression for the transmission amplitude is obtained (cf.\,Eq.\,(\ref{asyp}))
\begin{equation}
t(\omega,k_x) - t_0(\omega,k_x) =  \frac{i}{2 k_z^{+}} \sigma_{\nu}\, \epsilon_{-\nu}\,\omega^2\, \mathcal{C}^{(-)} = -\frac{i}{2 k_z^{+}}  \frac{\big|\epsilon_\nu\big|^2\omega^4\mathcal{C}^{(+)}\mathcal{C}^{(-)}}{-\eta_0- (k_x+\nu K_g)^2- |\epsilon_{\nu}|^2 \omega^4\Sigma}\,.
\label{totr}
\end{equation}
Reflection and transmission amplitudes are related to each other.   The resonance dominance approximation shares with the exact system of equations the property of power conservation. It is straightforward to derive the following identity
\begin{equation}
\big|r(\omega,k_x)\big|^2+\frac{k_z^+}{k_z^-}\big|t(\omega,k_x)\big|^2=1\,,
\label{unitt}
\end{equation}
which is valid  irrespective of the number of guided modes.  It thus applies to the case of isolated as well as overlapping resonances  which we study below and  is easily generalized to the case when more than one extended mode is present.

The background scattering is not only present in the  first term in Eq. (\ref{car1}) and  in the reflection and transmission amplitudes ($r_0,\,t_0$).   It  also affects  the resonance  terms  via $\mathcal{C}^{(+)}$ and  $\Sigma_{\nu}$, Eqs. (\ref{EI}) and (\ref{shga}).  Only in this way it is possible to have  the absolute squares of the total reflection and transmission amplitudes $r(\omega,k_x),\,t(\omega,k_x)$  satisfying  the identity (\ref{unitt}) (cf. also Eq.\,(\ref{asmo})). The interference between background and resonance contributions will be one  source of the asymmetry of the shape of the resonance curve akin to the phenomenon of Fano resonances in quantum mechanical systems \cite{FANO35,FANO61}.

Fan et al. \cite{FAJO02} have previously shown that the reflectivity curves for guided mode resonance can be described analytically as Fano resonances where the spectral bandwidth of the resonance is an external fitting parameter from exact numerical simulations. Our investigations support their claim and in addition produce the full analytic form of the resonant width (cf. Eqs. \ref{sign}-\ref{totr}), including the weak spectral dependence of $\Sigma$.

In general, the background scattering also exhibits resonances. These shape resonances  owe their existence to the properties of  the ``potential'' $\omega^2\epsilon_0(z)$ (Eq.\,(\ref{teps})) and do not involve transformation of a localized mode into a (delocalized) resonance.

We have restricted our numerical studies to  GWS with piecewise constant dielectric structures. Minor changes only are required to  account for a general $z$-dependence. We have to replace
$$ \epsilon_{\nu} \mathcal{C}^{(\pm)}\to  \int  dz\,\mathcal{E}_{\eta_0}(z)\epsilon_{\nu} (z)\, E_0^{(\pm)}(z)\,,\quad \big|\epsilon_{\nu}\big|^2\Sigma \to  \int dz \int dz^\prime  \mathcal{E}_{\eta_0}(z)\epsilon_{\nu}
(z)\,g_0(z,z^\prime)\, \epsilon_{-\nu}(z)  \mathcal{E}_{\eta_0}(z^\prime)\,. $$

In all our   numerical studies we kept  fixed the thickness of the layers and the  dielectric constant of the superstrate. Using the notation of Eqs. (\ref{teps}) and (\ref{epti})

\begin{equation}
\ell_g= z_2-z_1= 0.1\,\mu m,\quad \ell=z_3-z_2= 0.4\,\mu m\,,\quad \epsilon_{\text{I}}=1.
\label{size}
\end{equation}
In the first application we choose  the following profile of the grating interval, the dielectric constants, the grating period $\Lambda$, the duty cycle $d/\Lambda$  and  the  grating   contrast  $\epsilon_g-1$
\begin{eqnarray}
&&\epsilon_{\text{II}}(x) = \Big[1+ (\epsilon_g-1)\theta\big(d^2-x^2\big)\Big]\theta(x-x_0)\theta(\Lambda+x_0-x)   \,,\nonumber\\
&&\epsilon_{\text{III}}=4,\quad \epsilon_{\text{IV}}= 2.25\,,\quad \Lambda=0.81 \,\mu m,\quad  d=\frac{\Lambda}{2}\,,\quad  \epsilon_g=4.
\label{parep}
\label{graep}
\label{gratin}
\end{eqnarray}
These parameters have been chosen such that, for  the wavelength $\lambda=1.5\,\mu m$ and the angle  $\theta=5^\circ$ of the incident light, the extended mode resonates with the guided mode with $\nu=-1$ (cf.\,Eqs.\,(\ref{car1}), (\ref{car2})).

The numerical evaluation of the  expression for the reflection amplitude (\ref{tore}) proceeds in two steps. In the first step,
  the extended  $E_0^{\pm}(z)$ and guided $\mathcal{E}_{\eta_0}(z)$ modes together with the guided mode  eigenvalue $\eta_0$ are calculated for the effective medium (\ref{teps}) where the dielectric constant in the grating layer is replaced by its  averaged value $\epsilon_0= 2.5$ for the choice of the grating profile (\ref{graep}) and the parameters (\ref{gratin}).  The functions $E^\pm(z)$ and $\mathcal{E}_{\eta_0}$ have an analytic form and only  the eigenvalue $\eta_0$  requires numerical solution  of a transcendental equation. Given these building blocks it is straightforward to calculate in the second step observables such as the reflectivity, the electric field, etc.\,.

\begin{figure}[ht] \centering
\includegraphics[width=.458\linewidth]{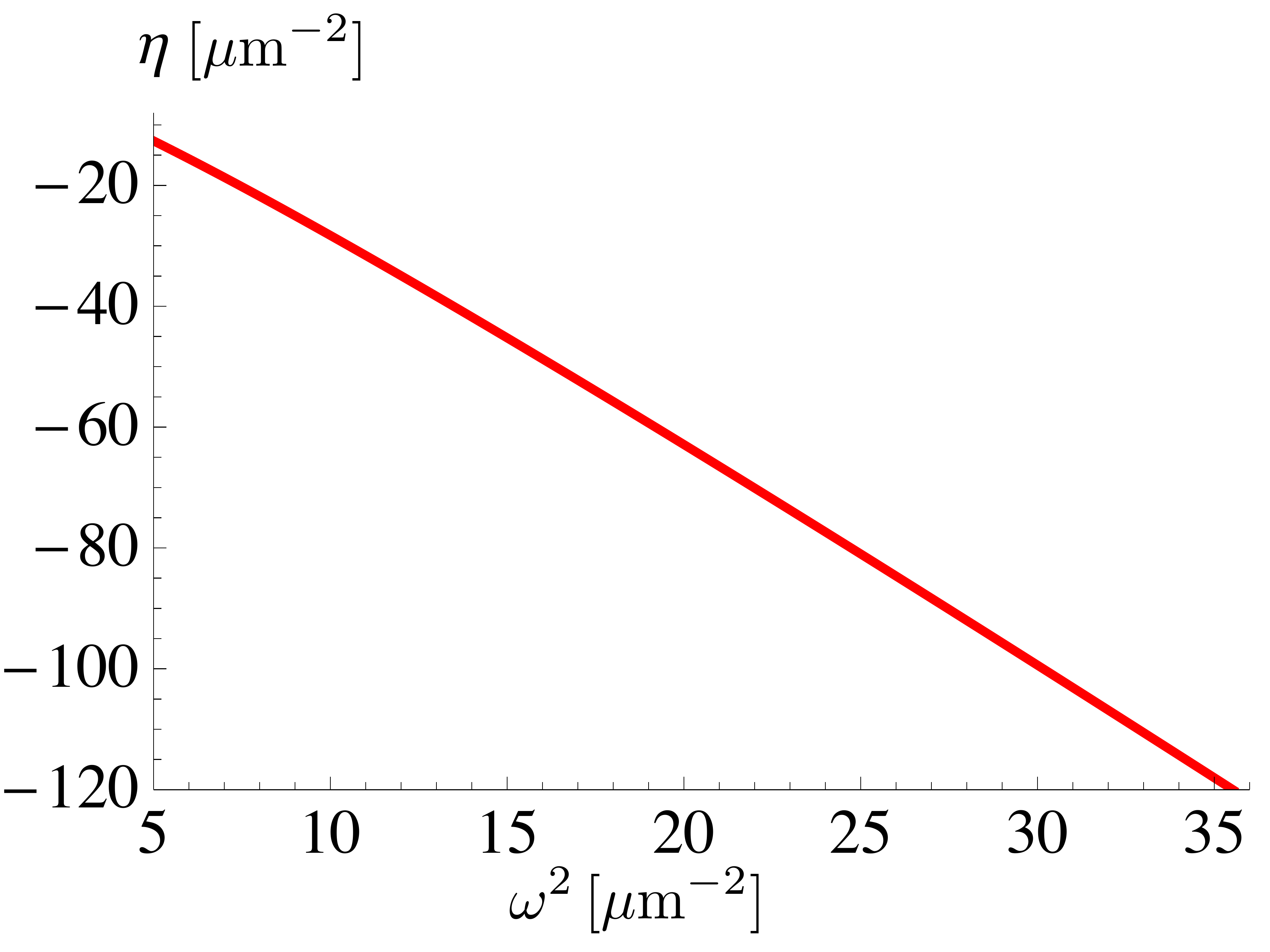}\hspace{.5cm}\includegraphics[width=.48\linewidth]{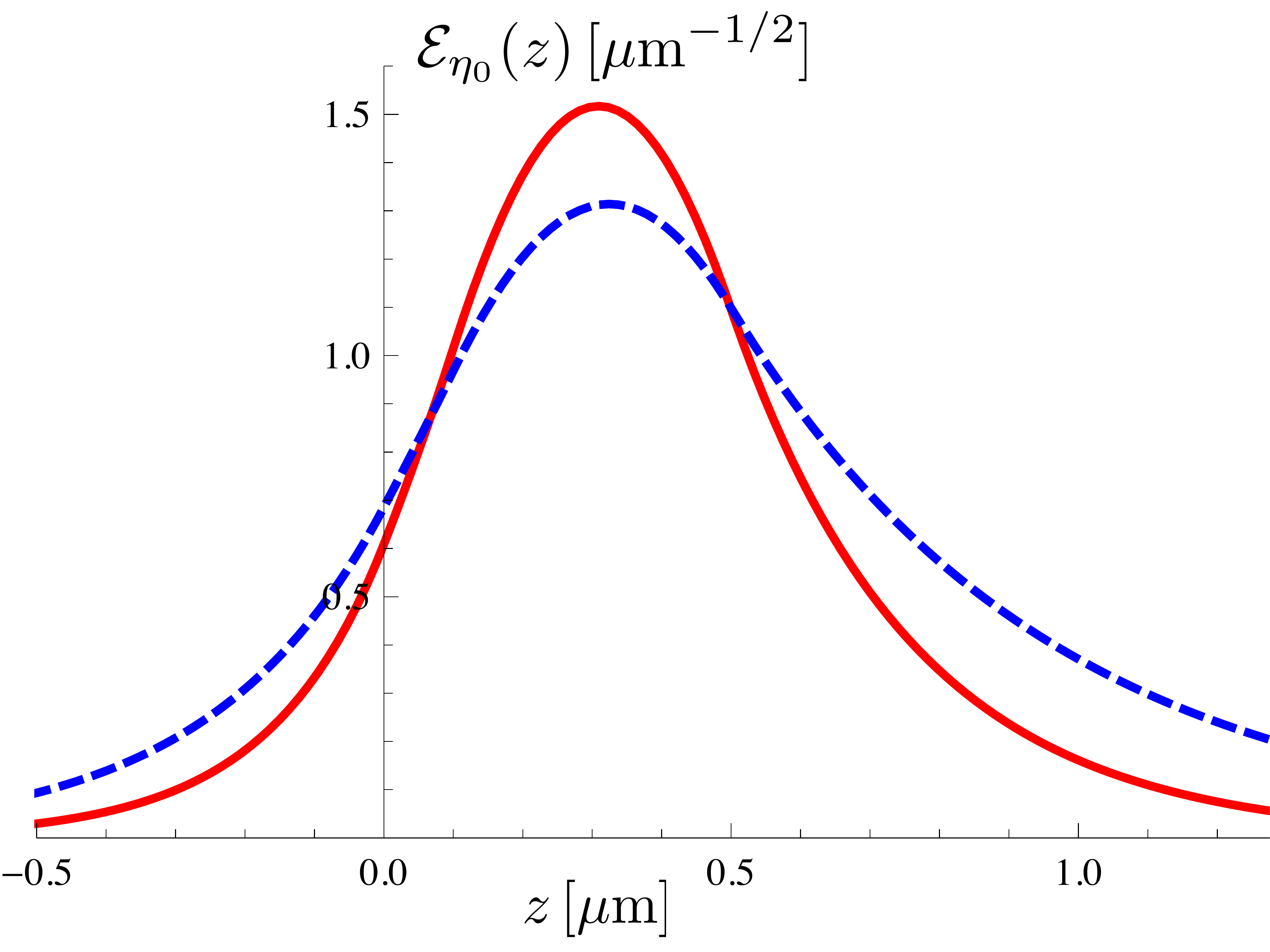}
\caption{Results for GWS with parameters  specified in Eqs.(\ref{size})\,-\,(\ref{gratin}). Left:\,the eigenvalue $\eta$ (cf.\,Eq.\,(\ref{swvafc})) as a function of $\omega^2$.
Right: normalized electric fields of guided modes for $\lambda=1.5\,\mu$m  (solid) and $2.1\,\mu$m (dashed line).
}
\vskip -.1cm
\label{guidmo}
\end{figure}

\begin{figure}[ht] \centering
\hspace{-8.5cm}\includegraphics[width=.42\linewidth]{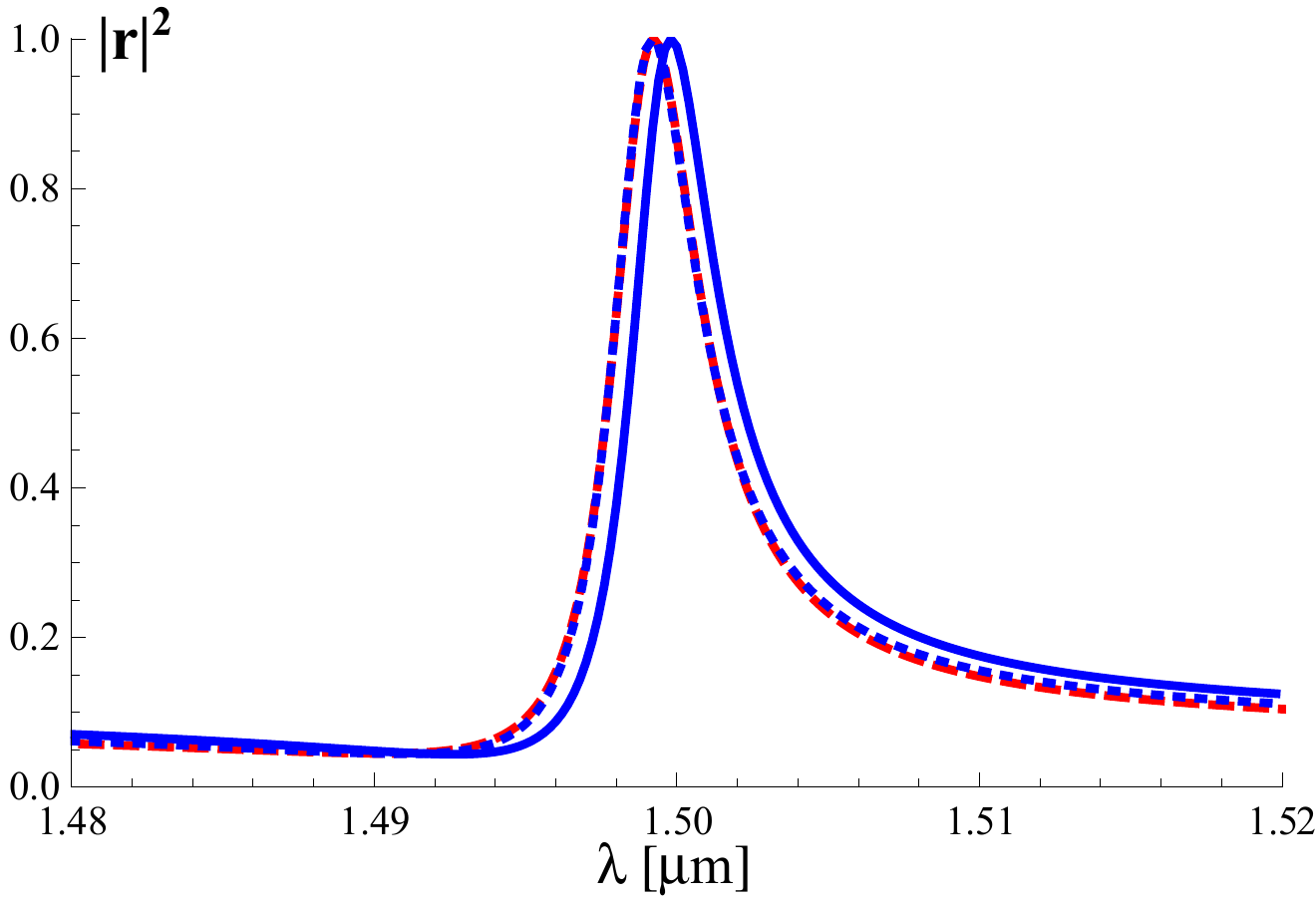}\vskip-4.9cm\hspace{7cm}\includegraphics[width=.42\linewidth]{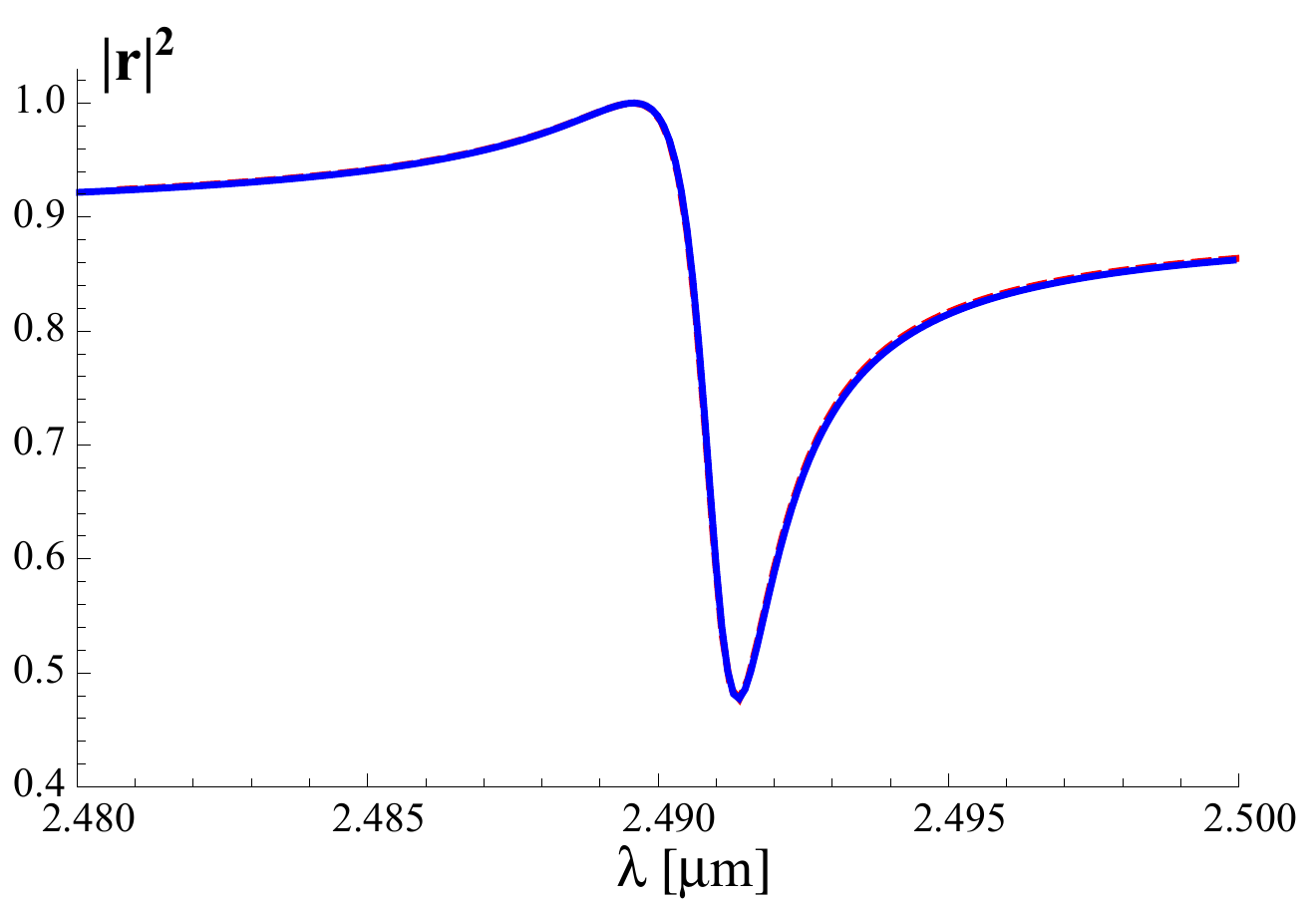}
\caption{Reflectivity  (cf\, Eq.\,(\ref{tore})) as a function of the wavelength.
Left: angle of incidence $5^\circ$ and parameters of the GWS specified in Eqs.\,(\ref{size})-(\ref{gratin}). Solid lines: exact values,  dashed lines: values obtained in the  resonance dominance approximation. Dotted line (in the left figure and coinciding with  the dashed line): values obtained by truncation of Eq.\,(\ref{inteq}) to $n,m=0,-1$, see the text.
Right: angle of incidence $75^\circ$ with changes in the parameters (cf.\,Eq.\,(\ref{gratin2})). Solid line: exact values,  dashed lines: values obtained in the  resonance dominance approximation and shifted by  0.75\,nm   towards larger wavelengths to facilitate the comparison.}
\label{refc1}
\end{figure}

The eigenvalue $\eta_0$ as a function of $\omega^2$ and the electric fields  of the guided modes
are displayed in Fig.\,\ref{guidmo}.
The numerical results of the evaluation of Eq.\,(\ref{tore}) are presented on the left of  Fig.\,\ref{refc1}.

In order to assess the accuracy of our theory we have compared it with  numerical results obtained by applying transfer matrix techniques (cf.\,\cite{MASO08}) to the system of equations (\ref{syseq}) truncated to a finite number of channels. Convergence was typically achieved    with 5\,-\,10 channels.  We shall refer to these converged  results  as "exact".
Such exact  values of the resonance position $\lambda_{\text{res}}$ and the full width at half maximum $\Gamma$  are
\beq
\lambda_{\text{res}}= 1.5000\,\mu m,\quad  \Gamma= 3.91168\,\text{nm}\,.
\eeq
The deviations from these values obtained when truncating  the coupled equations (\ref{syseq}) to just two (i.e. $n=-1$ and $n=0$) and  in the resonance approximation are respectively
\beq
\delta \lambda_{\text{res}}= -0.6\,\text{nm}, \quad \delta \Gamma= -0.0087\,\text{nm}\,,\quad
\delta \lambda_{\text{res}}= -0.6\,\text{nm}, \quad \delta \Gamma= -0.0129\,\text{nm}\,.
\label{resval}
\eeq
The exact resonance position is shifted by $\delta \lambda_{\text{res}}=17.4\,$nm  relative to the eigenvalue of  the guided mode. The resonance dominance approximation reproduces  this  shift as well as the resonance width  with an accuracy of 3.3 \%.


The coupling of the  $\nu=-1$ guided mode  to other guided modes neglected in this simplest version of the resonance dominance approximation is the most likely mechanism to account for the   few \% deviation of the resonance position from the exact result. It is not difficult to show that, independent of any details, the resulting shift due to such additional couplings  is always towards longer wavelengths.  In addition, the essentially identical results obtained in  resonance dominance and  at the lowest level truncation  rule out a significant contribution from the  $\nu=-1$ continuum contributions (cf.\,Eq.\,(\ref{inteq})).

Similar results and the same level of agreement have been found for other values of the angle of the incident light.  In order to enhance the effect of the background scattering  and to  study  in detail the changing shapes generated by  the interference between resonance and background contributions (Fano resonances) we have changed to the following parameters  of the dielectric medium (cf.\,Eq.\,(\ref{parep})),
\begin{equation}
\epsilon_{\text{IV}}=1\,,\quad \Lambda=1\,\mu m.
\label{gratin2}
\end{equation}
With this choice  one finds a strong  background  reflection with $|r_0(\omega,,k_x)|=0.9 $. As shown in the right part of  Fig.\,\ref{refc1}, as a result,  a drastic change in the shape of the resonance curve is obtained due the interference between background and resonance amplitudes.   To facilitate the comparison we have shifted the resonance dominance curve by $0.75$\,nm. An almost perfect agreement is observed indicating that the interference between background and resonance amplitudes is treated correctly.

A peculiar feature of the TE case in sharp contrast to quantum mechanical scattering is the dependence of the  strength of the interaction of  light with the  dielectric   on the  frequency  $\sim (\epsilon(x,z)-1) \omega^2$  (cf.\,Eq.\,(\ref{swva1})). This implies that the eigenvalues and eigenfunctions of the guided mode equation (\ref{swvafc}) depend on the frequency  (cf.\,Fig.\,\ref{guidmo}). At the same time, as in quantum mechanics, the frequency determines asymptotically the behavior of the incident or  scattered  light. As a consequence of this twofold role of the frequency, the observables like   width and position of the resonance have a rather intricate dependence on the frequency. To illustrate the consequences we consider the width  and expand the resonance denominator $\rho$ (\ref{rho})  around $\omega_r$,  the zero of its real part
\begin{equation}
\text{Re}\,\rho(\omega_r)=0\,,\quad \rho(\omega) \approx \gamma \big( \omega-\omega_r -i \gamma^{-1} |\epsilon_{\nu}|^2 \omega_r^4  \text{Im}\,\Sigma(\omega_r))\,,
\label{zere}
\end{equation}
with
\begin{equation}
\gamma= (K_g-\omega_r \sin\theta) \sin\theta +\Big[ 2\omega\frac{d\eta_0}{d\omega^2} - |\epsilon_{\nu}|^2 \frac{d}{d\omega}(\omega^4\text{Re}\,\Sigma)\Big]_{\omega=\omega_r}\,.
\label{dvt}
\end{equation}
In the range of frequencies of interest to us, the dominant contribution to $\gamma$ arises from the second term. Without a frequency dependent interaction, $\gamma \approx -2\omega_r $. Instead, as can be read off from Fig.\,\ref{guidmo} , we find   $\gamma \approx -7\omega_r $.
Thus the frequency dependence of the interaction leads to a significant narrowing of the resonance structure.


\subsection{TM resonances}\label{sec:TMresonances}
We consider here the simplest case when only one resonating  term, i.e. a term with small denominator  is present in the sums
 over  the modes in Eq. (\ref{eq:inteqforHnwithnonzeron}).   We stress that our assumption means that such a term occurs only
 for one particular value of n, i.e. that the resonant condition (\ref{eq:TMresoncond}) is fulfilled for one particular pair $(n,\nu_i)$. We denote it by $(k,\kappa)$.  We approximate by truncating the system (\ref{eq:inteqforH}) to only two equations - that of $n=0$ (in which the incoming light appears) and the resonating $n=k$. This is the truncation approximation. It leads to two coupled equations
 \eqna
\label{eq:truncforTM}
 H_0(z)&=&H_0^{(+)}(z)+\int_{I_g} dz' g_0(z,z') \Theta_{0k}H_k(z')\,,    \\
 H_k(z) &=& \int_{I_g} dz^\prime \left\{  \sum_\nu \frac{\mathcal{H}_{k,\nu}(z) \mathcal{H}_{k,\nu}(z^\prime) }{\omega^2-\eta_{k,\nu}}\right\}\Theta_{k0}H_0(z')\,.
\label{hkz}
\eqne
 Furthermore in the equation for $H_k$  we retain only the resonating mode $k,\kappa$. This is the approximation of the resonance dominance.  We thus obtain
\eqna \label{eq:eqforH0}
 H_0(z)&=&H_0^{(+)}(z)+\int_{I_g} dz' g_0(z,z') \Theta_{0k}H_k(z')\,,    \\
 \label{eq:eqforHk}
 H_k(z) &=& \frac{\mathcal{H}_{k,\kappa}(z) }{\omega^2-\eta_{k,\kappa}} \int_{I_g} dz^\prime  \mathcal{H}_{k,\kappa}(z^\prime)\,
\Theta_{k0}H_0(z')\,.
\label{hkz2}
\eqne
Exactly as in the TE case  this system of equations can be solved analytically, cf., Eqs. (\ref{car1},\ref{car2}). Equation (\ref {hkz2})  shows that the $ H_k(z)$ is just proportional to the guiding mode $\mathcal{H}_{k,\kappa}(z)$
\beq \label{eq:relofHhtoguidmode}
H_k(z)=\sigma \mathcal{H}_{k,\kappa}(z)\;\;\;\; ,\;\;\;\; \sigma=\frac{1 }{\omega^2-\eta_{k,\kappa}} \int_{I_g} dz^\prime  \mathcal{H}_{k,\kappa}(z^\prime)\Theta_{k0}H_0(z')\,.
\eeq
Inserting $H_k(z)=\sigma \mathcal{H}_{k,\kappa}(z)$ in (\ref{eq:eqforH0}) we obtain
\beq \label{eq:expresforH0}
H_0(z)=H_0^{(+)}(z)+\sigma\int dz' g_0(z,z') \Theta_{0k}\mathcal{H}_{k,\kappa}(z')\,.
\eeq
We then multiply both sides of this equality  by  $\mathcal{H}_{k,\kappa}(z)\Theta_{k0}$,  integrate over $z$ and solve for the field enhancement coefficient $\sigma$. We obtain
\beq \label{eq:TMexpforsigma}
\sigma=\frac{\mathcal{C}^{(+)}}{\omega^2-\eta_{k,\kappa}-\Sigma}\,,
\eeq
where we have introduced
\eqna
\label{eq:TMexpforCplus}
\mathcal{C}^{(\pm)}&=&\int_{I_g} dz \mathcal{H}_{k,\kappa}(z)\Theta_{k0}H_0^{(\pm)}(z)\,, \\
\Sigma&=&\int_{I_g} dz dz' \mathcal{H}_{k,\kappa}(z)\Theta_{k0}g_0(z,z')\Theta_{0k}\mathcal{H}_{k,\kappa}(z')\,.
\label{eq:TMexpforSigma}
\eqne
Here $H_0^{(-)}$  is the matching  pair of the solution $H_0^{(+)}$
as defined in  the TM part of Appendix \ref{sec:bagr}.

Equations (\ref{eq:relofHhtoguidmode}) through  (\ref{eq:TMexpforSigma})  provide a complete solution of the problem.
It is expressed through the solutions $H_0^{(+)}(z)$ and $\mathcal{H}_{k,\kappa}$ of the unperturbed structure.
From (\ref{eq:expresforH0}) one can easily extract the reflection and transmission amplitudes  from the asymptotic behavior of $H_0(z)$  and $g_0(z,z')$.   Using Appendix \ref{sec:bagr} we obtain
\eqna  \label{eq:asymtofH0forTM}
  \mathop {\lim }\limits_{z \to -\infty }H_{0}(z)&=&e^{ik^{-}_{z} z} + r\; e^{-ik_{z}^{-}z},\;\; r=r_{0} -
\frac{i \epsilon_{I}}{2k_z^{-}}\frac{|\gamma_k|^2\mathcal{\tilde{C}}^{(+)\,2}}{\omega^2-\eta_{k,\kappa}-\Sigma} , \\
\mathop {\lim }\limits_{z \to \infty }H_{0}(z)&=& \quad\quad\quad \;\, t\; e^{ik^+_z z},\;\;\;\; t=t_0-\frac{i \epsilon_{IV}}{2k_z^{+}} \frac{\;\;|\gamma_k|^2\mathcal{\tilde{C}}^{(+)}\mathcal{\tilde{C}}^{(-)}\;\;}{\omega^2-\eta_{k,\kappa}-\Sigma}\,.
\nonumber
\eqne
Here $r_0$ and $t_0$ are the "background" values which appear in $H_0^{(+)}(z)$.  Note also that we have used
  the replacement $ \Theta_{nm}\rightarrow\gamma_{n-m}\tilde{\Theta}_{nm}$ explained in Appendix \ref{app:singular}
  where
 \beq
 \tilde{\Theta}_{nm}=\overleftarrow{\partial_z}\;\overrightarrow{\partial_z}+(k_x+nK_g)(k_x+mK_g)\,.
 \eeq
to define $\mathcal{C}^{(\pm)}=\gamma_k \mathcal{\tilde{C}}^{(\pm)}$  with
\beq
\mathcal{\tilde{C}}^{(\pm)}=\int_{I_g} dz\left[ \partial_z\mathcal{H}_{k,\kappa}(z)\partial_z H_0^{(\pm)}(z)  + (k_x+kK_g)k_x\mathcal{H}_{k,\kappa}(z)H_0^{(\pm)}(z)\right]\,,
\eeq
As in the TE case the  self coupling  $\Sigma$ which appears in the denominator of  (\ref{eq:TMexpforsigma})  and (\ref{eq:asymtofH0forTM})  plays a crucial role. Its real part     shifts the resonance frequency away from the eigenvalue $\eta_{k\kappa}$ of the resonating guided mode while the resonance width is given by the imaginary part of $\Sigma$.

 Although formally the expressions for $r$ and $t$ are  very similar to those for the TE case one should note the important and crucial differences   in the
expressions (\ref{eq:TMexpforCplus},\ref{eq:TMexpforSigma}) for  $\mathcal{C}^{(\pm)}$  and  $\Sigma$ which contain the coupling operators $\Theta_{mn}$.

As already mentioned the TM formalism has an ambiguity of using $\gamma_{m-n}(z)$ as given by
 Eq. (\ref{eq:defofgamman}). We could   alternatively regard $\epsilon_{m-n}(z)$ as a matrix  $\boldsymbol{\epsilon}_{mn}\equiv\epsilon_{m-n}(z)$ and inverting it to find $\gamma_{m-n}=(\boldsymbol{\epsilon}^{-1})_{mn}$. We have found that the second alternative leads to improved comparison with exact results and adopted it in our examples below.  We do not have a good argument to justify  this improvement.
In the literature, cf., Refs \cite{Li96,GRGU96,LAMO96,PONE00,LYPT07}, this ambiguity  in the truncation has been discussed and resolved in favor of the ``inverse'' method by considering the asymptotic convergence properties  of the Fourier components of $H$ (cf.\,Eq.\,(\ref{eq:expanofH})) which is an important issue in high precision numerical studies.

In our examples below we have used the same parameters of the structure (\ref{size}) and (\ref{parep}) as  in the TE case. In the first step,
the extended  $H_0^{\pm}(z)$ and guided $\mathcal{H}_{k,\kappa}(z)$ modes together with the guided mode  eigenvalue $\eta_{k,\kappa}$ were calculated for the effective medium (\ref{teps}). They were computed analytically apart from finding  eigenvalue $\eta_{k,\kappa}$ which required solving numerically a transcendental equation.

The main difference at this step as compared with the TE case is the peculiar boundary conditions for the derivatives of the solutions at the boundaries. The combinations  $\epsilon_0(z)\partial_z H_0^{\pm}(z)$  and $\epsilon_0(z)\partial_z \mathcal{H}_{k,\kappa}(z)$ rather than the derivatives themselves must be continuous. This leads to a subtlety when the integrals (\ref{eq:TMexpforCplus}) and (\ref{eq:TMexpforSigma}) for $\Sigma$ and $\mathcal{C}^{(\pm)}$ are computed. Since the derivative terms in the operators  $\Theta_{k0}$ and $\Theta_{0k}$ contain
$\partial_z\epsilon_{k}(z)\partial_z$ rather than $\partial_z\epsilon_{0}(z)\partial_z$ combinations they produce $\delta$-function like singularities when acting on  $H_0^{\pm}(z)$, $\mathcal{H}_{k,\kappa}$ and $g_0(z,z')$.   We refer to Appendix \ref{app:singular} in which we explain how we dealt with such singularities.  Adopting this approach, the quantities   $\Sigma$ and $\mathcal{C}^{(\pm)}$  essentially  can be evaluated analytically with  the results used to calculate   reflectivity and other observables.

\begin{figure}[ht] \centering
\vskip -.2cm
\includegraphics[width=.45\linewidth]{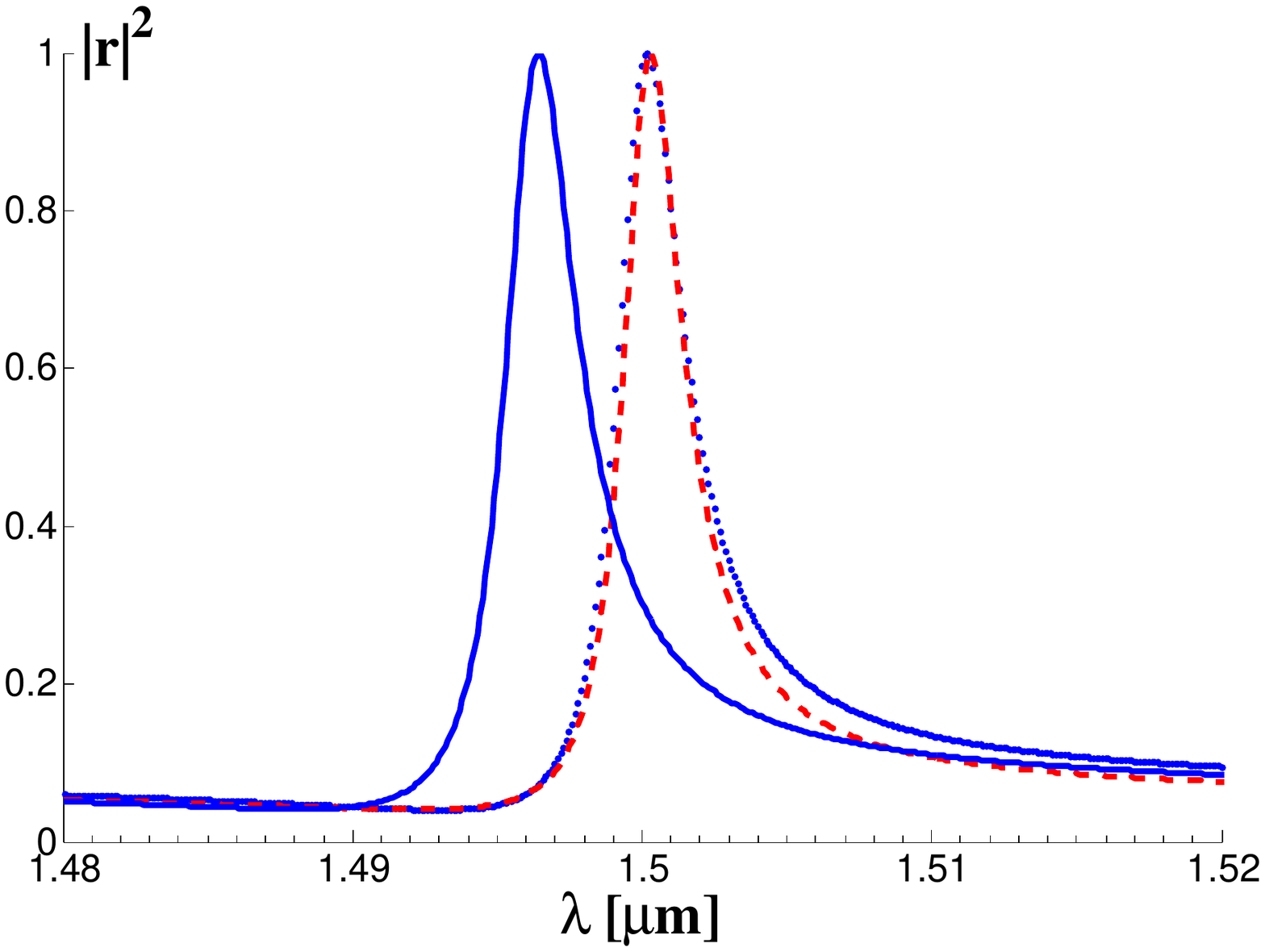}\hspace{1cm}\includegraphics[width=.45\linewidth]{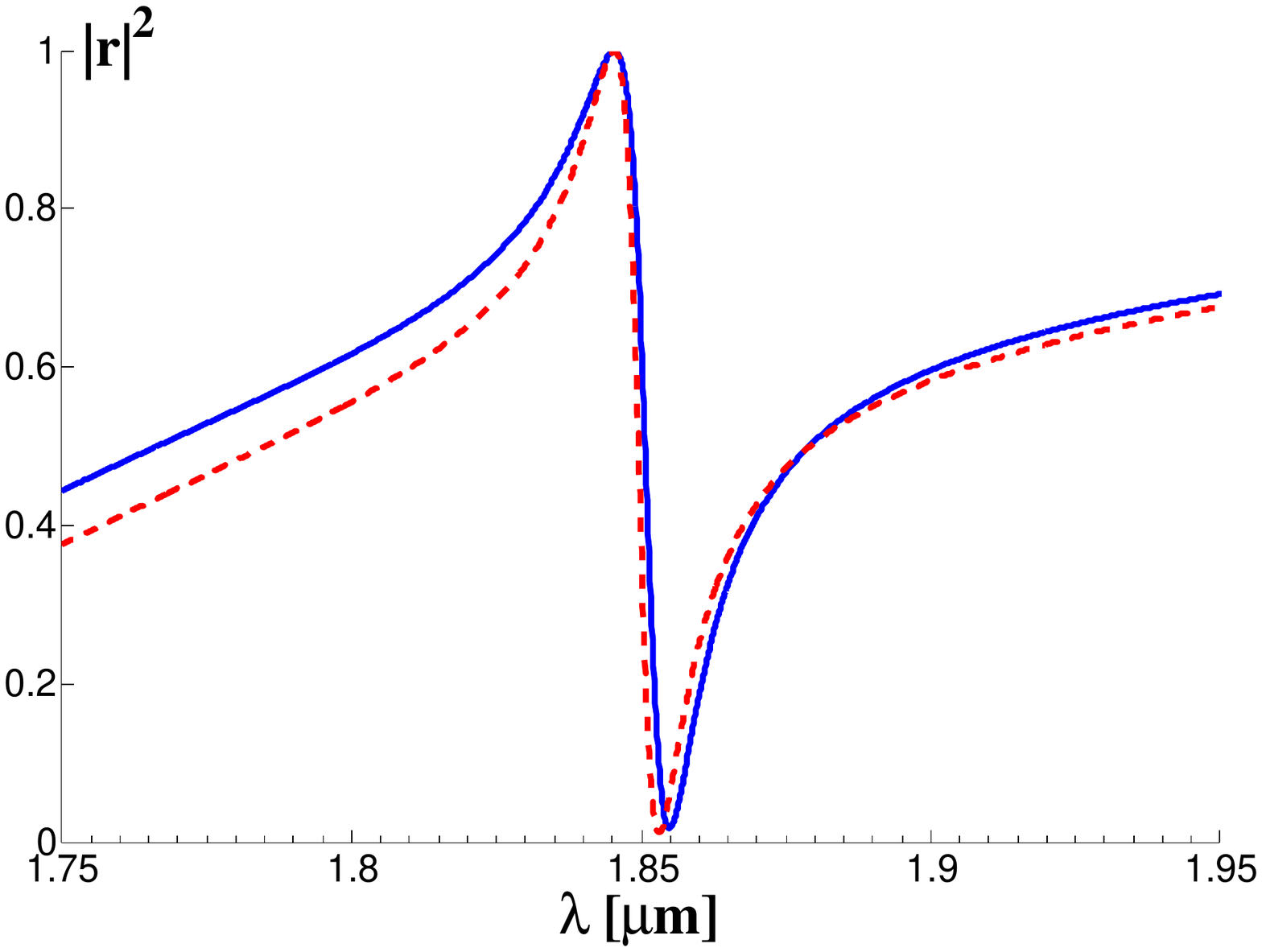}
\caption{Reflectivity  $|r(\omega,k_x)|^2$, cf. Eq. (\ref{eq:asymtofH0forTM}),  as a function of the wavelength. Left: angle of incidence $5^\circ$.  The other  parameters  are as in  Eqs.\,(\ref{size})-(\ref{gratin}) apart of the grating period which is  0.87\,$\mu m$. Solid lines: exact values,  dashed lines: values obtained in the  resonance dominance approximation. Dotted line (in the left figure and coinciding with  the dashed line): values obtained by truncation of Eq.\,(\ref{eq:eq2forxin}) to $n,m=0,-1$, see the main text.
Right: angle of incidence $86^\circ$ with changes in the parameters to $\epsilon_{IV}=1$  and  $\Lambda=0.8\,\mu m$. Solid line: exact values,  dashed lines: values obtained in the  resonance dominance approximation and shifted by 6.4\,nm respectively towards smaller  wavelengths.}
\vskip -.1cm
\label{refTM1}
\end{figure}

   The numerical results for the reflectivity  $|r|^2$ (cf. Eq.\,(\ref{eq:asymtofH0forTM})) are presented in  Fig.\,\ref{refTM1}.

With the indicated parameters of the grating the resonating  guided mode   $\mathcal{H}_{k,\kappa}$ has $k=-1$ and $\kappa=0$.  Our results are compared with  numerical results obtained by applying transfer matrix techniques after truncating the system of equations (\ref{eq:eq2forxin}).  This is similar to what is used in the rigorous coupled wave analysis (RCWA), cf. Ref. \cite{MBPK95}. Convergence was typically achieved with  20\,-\,30 channels - slower than in the TE case.  In the left part of Fig. \ref{refTM1} we also compare with the results of truncation of  the system of Eqs. (\ref{eq:eq2forxin}) to just two channels, $n,m=0,-1$.

 As in the TE case we have also examined the structure parameters for which the Fano interference between the resonance and the background is particularly pronounced.  This is shown in the right part of Fig.\,\ref{refTM1}.
We observe  a good qualitative agreement in both graphs.  The same quality of  agreement has been found for other values of the angle of the incident light.

 At the same time it is seen  that on the quantitative level the comparison with the exact results for TM modes is less satisfactory than for the TE modes.
For example in  the particular resonances of Fig.\,\ref{refTM1}  the discrepancy in the  resonance position is $-3.9$ $nm$ (left) and $-6.4$ $nm$ (right) as compared to $0.6$ $nm$ and $0.7$ $nm$ in the TE case.  The discrepancy in the width in the left figure is $0.639$ $nm$ between the exact results and resonance
 dominance as compared to just $-0.013$ $nm$ in the TE case.
 What is particularly notable is the discrepancy with the two channel truncated results where  an almost perfect agreement for the TE case has been obtained.

 We tend to attribute these  discrepancies to the peculiar TM boundary condition at the borders of the grating layer.  These  conditions for the guided mode channel are violated in the resonance  dominance approximation which we use. We have verified  that the discrepancies decrease with decreasing
  contrast of the grating or decreasing size of the grating interval.
We have also noticed that resonance dominance begins to fail in the TM case for regular binary gratings with very small or very large duty cycle for which  $|\gamma_3|\sim|\gamma_2|\sim|\gamma_1|$.

A more systematic analysis is required to clarify the above issues.  In this respect let us note the  following.
The general framework  for the resonance dominance approximation  is provided by the Feshbach projection operators formalism as it is outlined in Appendix \ref{app:feshbach}. There the exact resonating state is  one of the  eigenstates  in the Q subspace. In our approach we have approximated this state by neglecting the coupling between the Fourier components $H_n$  with $n\ne k$, c.f.,\,Eq.\,(\ref{eq:truncforTM}).  Using an expansion of the modes in terms of Bloch waves rather than plane waves appears as a promising tool to improve the treatment of the TM modes.


\section{Interaction of Feshbach resonances} \label{ss2res}
\subsection{Overlapping TE resonances}
\subsubsection{Formal development}
In this section we will extend our discussion and address the issue of overlapping resonances and their interaction. For  photonic crystal slabs of the structure shown in Fig.\,\ref{dispn}   overlapping resonances  always exist at illumination close to normal incidence. This has its origin in the time reversal symmetry. It is straightforward to verify that with $E_n(k_x,k_z^\pm,z)$ also  $E_{-n}^*(-k_x,-k_z^{\pm},z)$ solves the system of equations (\ref{syseq}) and satisfies the boundary condition (\ref{bcte2}).
This implies that resonances occurring  for  incident light with $k_x\to 0$  always involve two guided modes,  i.e. a resonance with $n=\nu$   is always  accompanied by the resonance for $n=-\nu$. In the resonance dominance approximation this conclusion follows  readily from the resonance condition (\ref{cfsh}). If it is satisfied for a guided mode $\mathcal{E}_{\eta_0}(z)$ and  $n=\nu$ at $k_x\to 0$ it will also be satisfied for the same guided mode and $n=-\nu$. The physics of this condition is that the incoming light with $k_x\to 0$ is coupled by the grating  to both right ($\beta=k_x+\nu K_g$) and left ($\beta=k_x-\nu K_g$) propagating guided modes having the same $z$-profile $\mathcal{E}_{\eta_0}(z)$.

We consider in detail this class of overlapping resonances, i.\,e.\, we include together with $E_{\nu}$ also  $E_{-\nu}$ resulting in a system of equations with 3 components (cf.\,Eqs.\,(\ref{car1}),\,(\ref{car2}))
\begin{eqnarray}
\label{car21}
\hspace{-.5cm}E_0(z)&=& E_0^{(+)}(z)-\omega^2\int_{I_g} dz^\prime g_0(z,z^\prime)  \big(\epsilon_{-\nu}E_{\nu}(z^\prime)+\epsilon_{\nu}E_{-\nu}(z^\prime)\big)\,,\\
\label{car22}
\hspace{-.5cm} E_{\pm \nu}(z) &=&   \frac{\omega^2}{\eta_0+(k_x\pm \nu K_g)^2}\,\mathcal{E}_{\eta_0}(z) \int_{I_g} dz^\prime\mathcal{E}_{\eta_0}(z^\prime) \big( \epsilon_{\pm \nu} E_{0}(z^\prime)+\epsilon_{\pm 2 \nu} E_{\mp \nu}(z^\prime)\big)\,.
\end{eqnarray}
Here we have made use of the peculiar property of the TE modes that the value of $\nu$  gives rise to shifts of the  the eigenvalue  $\eta_0$) but does  not affect the eigenmode $\mathcal{E}_{\eta_0}$ (cf.\,Eq.\,(\ref{swvafc})).
According to Eq.\,(\ref{car22}) the resonance waves $E_{\pm \nu}(z)$  differ only in their (resonance enhanced)  strengths $\sigma_{\pm \nu}$ (cf.\,Eq.\,(\ref{sigma})). Up to these unknown normalizations they are given by the guided mode $ \mathcal{E}_{\eta_0}$
\begin{equation}
 E_{\pm\nu}(z) = \sigma_{\pm\nu}\, \mathcal{E}_{\eta_0}(z)\,.
\label{sigma2}
\end{equation}
As in the case of an isolated resonance (cf.\,Eq.\,(\ref{tore2})), the reflection amplitude is determined by the asymptotics of $E_0(z)$ in (\ref{car21}). Using Eqs.\,(\ref{E0+}) and (\ref{sigma2}) it is easily seen that Eq.\,(\ref{car21})  is obtained from Eq.\,(\ref{car1}) by replacing  $\sigma_{\nu}\,\epsilon_{-\nu}$ by $\sigma_{\nu}\,\epsilon_{-\nu}+\sigma_{-\nu}\,\epsilon_{\nu}$. Accordingly the expression (\ref{tore}) for the reflection amplitude of an isolated resonance is replaced by
\begin{equation}
r(\omega,k_x) = r_0(\omega,k_x)+ \frac{i}{2k_z^-} \big(\sigma_{\nu}\,\epsilon_{-\nu}+\sigma_{-\nu}\,\epsilon_{\nu}\big) \omega^2 \mathcal{C}^{(+)}\,.
\label{tore2}
\end{equation}

It remains to determine  the unknown  strength parameters $\sigma_{\pm \nu}$. To this end one inserts Eq.\,(\ref{car21}) into the two equations (\ref{car22}) and obtains after using (\ref{sigma2}) a $2\times 2$ linear system system of equations for the  unknown variables $\sigma_{\pm \nu}$,
\begin{equation}
W\left(
   \begin{array}{ccc}
   \sigma_{+\nu} \\
    \sigma_{-\nu} \\
   \end{array}\right)= \omega^2\, \mathcal{C}^{(+)} \left(
   \begin{array}{ccc}
   \epsilon_{\nu} \\
   \epsilon_{-\nu}  \\
   \end{array}\right)\,,
\label{sigsy}
\end{equation}
with the coupling matrix $W$
\begin{equation}
W=
    \left(\begin{array}{ccc}
     \eta_0+(k_x+\nu K_g)^2+|\epsilon_{\nu}|^2 \omega^4\,\Sigma & \epsilon_{\nu}^{2}\omega^4\,\Sigma -\epsilon_{2\nu}\omega^2\,V \\
 \epsilon_{-\nu}^{2}\omega^4\,\Sigma -\epsilon_{-2\nu}\omega^2\,V & \eta_0+(k_x
-\nu K_g)^2+|\epsilon_{\nu}|^2\omega^4\,\Sigma \\
   \end{array}
\right)\,.
\label{Msig}
\end{equation}
The diagonal elements of the matrix $W$
contain the complex  valued self interaction $\Sigma$ defined in Eq.\, (\ref{shga}). The off-diagonal elements contain  a direct coupling via $\epsilon_{\pm 2 \nu}$  between the guided modes
with
\begin{equation}
V=\int_{I_g} dz  \mathcal{E}_{\eta_0}(z)   \mathcal{E}_{\eta_0}(z)\,,
\label{v2}
\end{equation}
as well as the indirect coupling terms $\epsilon_{\nu}^{2}\Sigma$ and  $\epsilon_{-\nu}^{2}\Sigma$  via the extended mode.
In terms of the matrix elements  $W_{ij}$, the inverse of $W$ is given by
\begin{equation}
W^{-1}=
    \frac{1}{W_+ W_-}\left(\begin{array}{ccc}
     W_{22} &- W_{12} \\
 -W_{21} & W_{11}\\
   \end{array}
\right)\,,
\label{Minv}
\end{equation}
where the eigenvalues of $W$ are given by
\begin{eqnarray}
W_{\pm}=&& \eta_0 +k_x^2+\nu^2K_g^2 +\big|\epsilon_{\nu}\big|^2\omega^4\, \Sigma
\nonumber\\
\pm&&
\sqrt{(2k_x\nu K_g)^2 + \omega^4\,\Big(\big(|\epsilon_{\nu}|^2 \omega^2\,\Sigma -|\epsilon_{2\nu}| \cos \varphi_\epsilon  V\big)^2+|\epsilon_{2\nu}|^2 \sin^2 \varphi_\epsilon \, V^2\Big)}\,.
\label{solev}
\end{eqnarray}
{We have introduced the relative phase between $\epsilon_{2\nu}$  and  $\epsilon_{\nu}^2$
\begin{equation}
e^{i\varphi_\epsilon}=\frac{\epsilon_{2\nu}\epsilon_\nu^{\star\,2}}{|\epsilon_{2\nu}\epsilon_\nu^2|}\,,
\label{tau}
\end{equation}
which will be seen to distinguish
different types of interacting resonances.  The importance of the relative phase between the different Fourier coefficients has been previously  emphasized  by Barnes et al. \cite{BPKS96} in the similar phenomenon of surface plasmon resonance off metallic gratings. Here it is not the relative phase ($\sim \epsilon_2/\epsilon_1$) which matters. Rather the phase $\varphi_\epsilon$ results from the interference of the 2-step process via the extended mode and the one step process connecting directly the 2 guided modes.
The combination of the strengths $\sigma_{\pm \nu}$ which determines the reflection amplitude is easily calculated
\begin{equation}
\sigma_{\nu}\epsilon_{-\nu}+\sigma_{-\nu}\epsilon_{\nu} =  2\omega^2|\epsilon_{ \nu}|^2\mathcal{C}^{(+)}\frac{W_0}{W_+ W_-}\,,\quad W_0=
\eta_0+k_x^2+\nu^2K_g^2+  |\epsilon_{2\nu}|\cos  \varphi_\epsilon \,\omega^2 V \,.
\label{sieps}
\end{equation}
This is the central result of this section. We will use it in the next section to analyze
possible patterns of overlapping resonances.

\subsubsection{Reflectivity of overlapping resonances}
The equations  (\ref{tore2}) and (\ref{sieps}) show that the zeros of the real parts of $ W_+$  and $W_-$ define the positions of the two resonances while their widths are given respectively by  Im\,$W_+$ and Im\,$W_-$. According to Eq.\,(\ref{solev})
 the sum of the  resonance widths  and  the midpoint of their  positions
are not affected by the interaction of the resonances. It follows that broadening of one of the resonances is accompanied by narrowing of the other. This is reminiscent of the phenomenon of motional narrowing \cite{BlPP48} as well as the phenomenon of superradiance \cite{D54} and subradiance \cite{PCPCL85} for coupled emitters. The distance $\big|\text{Re}( W_+-W_-)\big|$  between the two resonances can either shrink or expand  depending on the  structure of the GWS. If
\begin{equation}
\text{Re}\Big(\big(|\epsilon_{\nu}|^2 \omega^2\,\Sigma -|\epsilon_{2\nu}| \cos \varphi_\epsilon  V\big)^2+|\epsilon_{2\nu}|^2 \sin^2 \varphi_\epsilon \, V^2\Big) < 0,
\label{condat}
\end{equation}
and if the distance  $|4k_x\nu K_g|$ between the  non-interacting resonances is sufficiently large this distance decreases due to the interaction.  The  condition (\ref{condat})  is actually satisfied for the choice of the parameters (\ref{gratin}) where $|$Im\,$\Sigma|\approx 2.5\, |$Re\,$\Sigma|$ and $\epsilon_{2\nu}=0$.  Obviously, for sufficiently large values of the term proportional to $V$ repulsion of the resonances results. Below we will discuss such a case. For $k_x=0$ the distance trivially has to increase or to remain 0 as a consequence of the interaction. In the regime in between these limits the change in distance depends on the details of the various quantities in (\ref{condat}).  It is remarkable that the system of overlapping resonances can be tuned to exhibit either ``level'' repulsion or attraction by variation of an external parameter $(k_x)$.
A peculiarity of the interacting resonances is the appearance of a zero in the resonance amplitude,\,i.e. $W_0(\omega,\theta)=0$ in Eq. (\ref{sieps}).  The presence of this zero may distort significantly the shape of the reflectivity or obscure the presence of two resonances  if  the distance between  the zero of $W_0$ and  the resonance position of $W_-$  is of the same size as or smaller than the corresponding width.

Finally, the ratio of the   kinematical  term $(2k_x\nu K_g)^2$ and the interaction induced 2nd term in the square root of  Eq.\,(\ref{solev}) controls  the transition from overlapping to isolated resonances. For sufficiently large $k_x$ the interaction induced term,  i.e. the off-diagonal  matrix elements of $W$ in  (\ref{Msig}) can be neglected and two isolated resonances are described by $W$.


We now will give a brief overview of the possible structures  of the resonances generated by Eq.\,(\ref{sieps}) and we will discuss their  dependence on the properties of the GWS. In these qualitative studies, but  not in the numerical results,   we disregard the contribution  $r_0(\omega,k_x)$ to the reflection amplitude  $r(\omega,k_x)$ (cf.\,Eq.\,(\ref{tore2})). We first consider the case when  $|\epsilon_{2\nu}|=0$ for which $\varphi_{\epsilon_0}$ is ill defined.  This is  realized for a grating with 50\% duty cycle.   According to Eq. (\ref{solev}),  for  $k_x\rightarrow 0$ the transition from two separated  resonances to one resonance takes place. At $k_x=0$ only one resonance is present  with a  width twice as large as that of an isolated resonance ($W_-$ is exactly canceled by $W_0$), while at arbitrarily small non zero angles  two peaks will appear due to the small imaginary part of $W_-$.


 This is illustrated in the  left part of Fig.\,\ref{thresh2} for the case where the  resonance conditions are satisfied for modes with $\nu=\pm 1$.  In the resonance region  $|\epsilon_1|^2\omega^4\Sigma\approx  (0.07+0.16\,i)\,\mu\text{m}^{-2}$.  The distance between the resonances is thus smaller than the wide resonance width and indeed  no clear separation of the resonances  is observed until the incidence  angle
 $|\theta|$  is increased to values of the order of or larger than
$$ |\theta|  \ge -|\epsilon_1| ^2 \omega^3 \text{Im}\,\Sigma/K_g \approx 0.3^\circ. $$

One can also see the effect caused by the zero of $W_0$ in Eq. (\ref{sieps}). It produces the dip in the combined peak in the left part of Fig.\,\ref{thresh2}. The appearance of the zero can also be interpreted as arising  from a cancelation between the contribution from the two eigenstates of $W$ (\ref{Msig}) associated with the eigenvalues $W_\pm$. This destructive interference of the contributions from the two eigenstates causes a transparency window within the resonance and is closely related to the ``EIT'' phenomenon (electromagnetically induced transparency), cf., Ref. \cite{H97}. Thus  for an appropriate choice of parameters  guided mode resonances provide  yet another classical analog of EIT, cf.,\,\cite{AMN02,YSWF04}.

\begin{figure}[ht] \centering
\vskip -.2cm
\includegraphics[width=.45\linewidth]{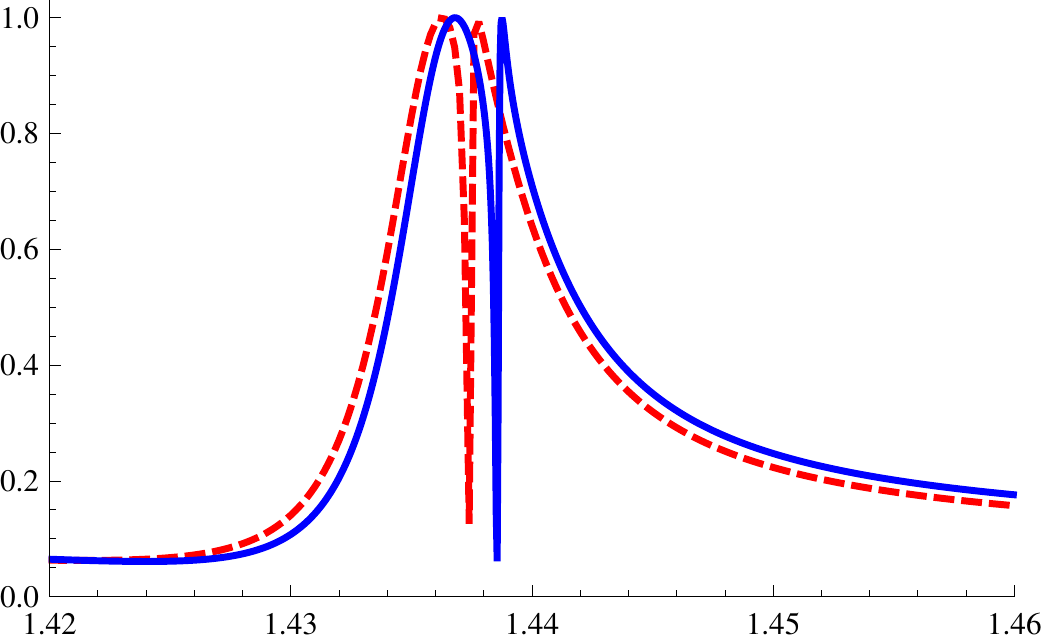}\hspace{.5cm}\includegraphics[width=.45\linewidth]{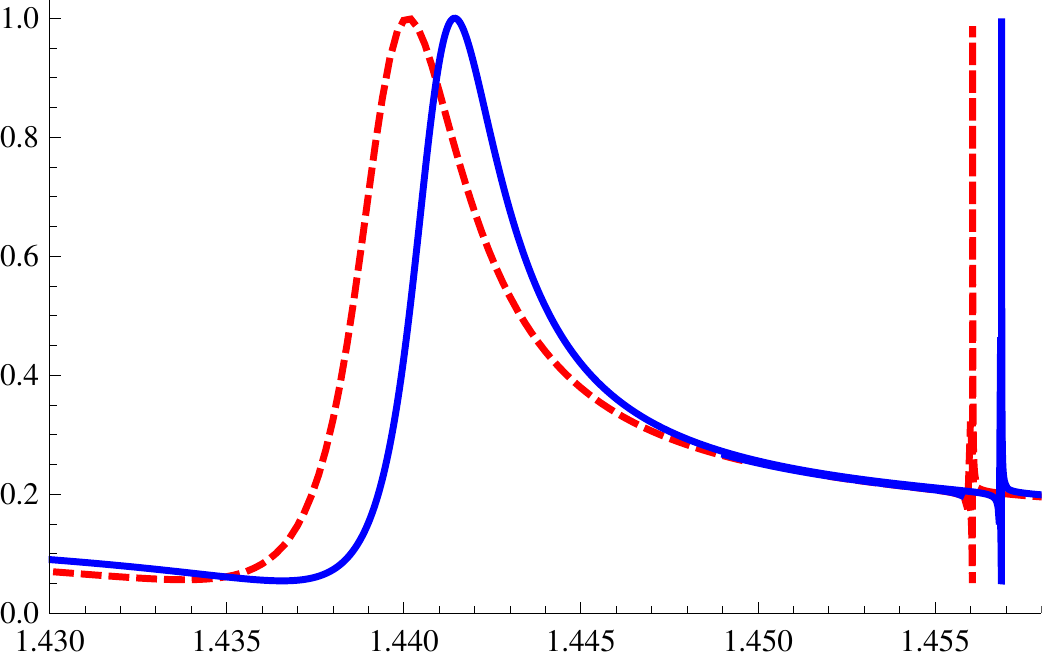}
\caption{Reflectivity as a function of the wavelength for the  angle of incidence  $\theta=  0.05^\circ$,   structure and  parameters as specified in Eqs.\,(\ref{size}-\ref{gratin}). Solid lines: exact values,  dashed lines: values obtained in the  resonance dominance approximation.  Left: duty cycle $d/\Lambda=0.5$.  Right: duty cycle $d/\Lambda=0.75$. }
\label{thresh2}
\end{figure}
We now consider the case where $\epsilon_{2\nu}$ and $\epsilon_\nu^2$ are, up to a sign,  in phase, i.\,e., for
\begin{equation}
\varphi_\epsilon=0,\,\pi\,,
\label{iph}
\end{equation}
 Then Eq.\,(\ref{solev}) simplifies  for small $k_x$,
\begin{equation}
W_{\pm}\approx
\eta_0 +k_x^2+\nu^2K_g^2 +\big|\epsilon_{\nu}\big|^2\omega^4\, \Sigma \pm\Big[\big|\epsilon_{\nu}\big|^2\omega^4\, \Sigma  -  |\epsilon_{2\nu}|  \omega^2 V +
\frac{2k_x^2\nu^2 K_g^2}{\omega^2\,\big(|\epsilon_{\nu}|^2 \omega^2\,\Sigma
-|\epsilon_{2\nu}| \cos \varphi_\epsilon
V\big)}\Big]\,.
\label{iplim}
\end{equation}

When the  direct interaction $V$ is strong  in comparison to the self coupling $\Sigma $  one will have a clear separation of the two resonances. Since $V$ is real one of the resonances will still have a vanishing width at $k_x=0$ while another will have the full width.  This is illustrated in the right part of Fig.\,\ref{thresh2} for the duty cycle of $75$\%. The shape of the reflectivity is drastically changed as compared to the left part of this figure  showing two well separated resonances  with widths which  differ  by a factor of 350.

It should be noted that any simple binary grating of arbitrary duty cycle (cf. left part of Fig (\ref{dispn})) are symmetric under x-mirror reflection over a plane defined by $x=x_0$. Any grating with such symmetry will either have $\epsilon_{2\nu}=0$ or will fulfill the condition (\ref{iph}). Since  the incident field is symmetric under x-mirror reflection symmetry only for $k_x=0$ we may interpret the vanishing width at normal incidence for such symmetric gratings as a symmetry selection rule. Thereby it is only possible to excite the combination of guided modes which is symmetric under the x-mirror reflection.
Interacting resonances with similar widths   at small or vanishing angle of incidence  can be generated only if the relative phase $\varphi_\epsilon$ (\ref{tau}) deviates significantly from 0 or $\pi$.

We consider the case
\begin{equation}
\varphi_\epsilon=\pm\frac{\pi}{2}\,.
\label{pi2}
\end{equation}
and realize this value by choosing the profile of the grating shown in Fig.\,\ref{epII} with  parameters
\begin{equation}
\Lambda=0.81\,\mu m,\quad \epsilon_{\text{min}}=1,\quad  \epsilon_{\text{max}}=4,\quad d/\Lambda= 0.2\,.
\label{par2}
\end{equation}

\begin{figure}[ht] \centering
\vskip -.3cm
\includegraphics[width=.3\linewidth]{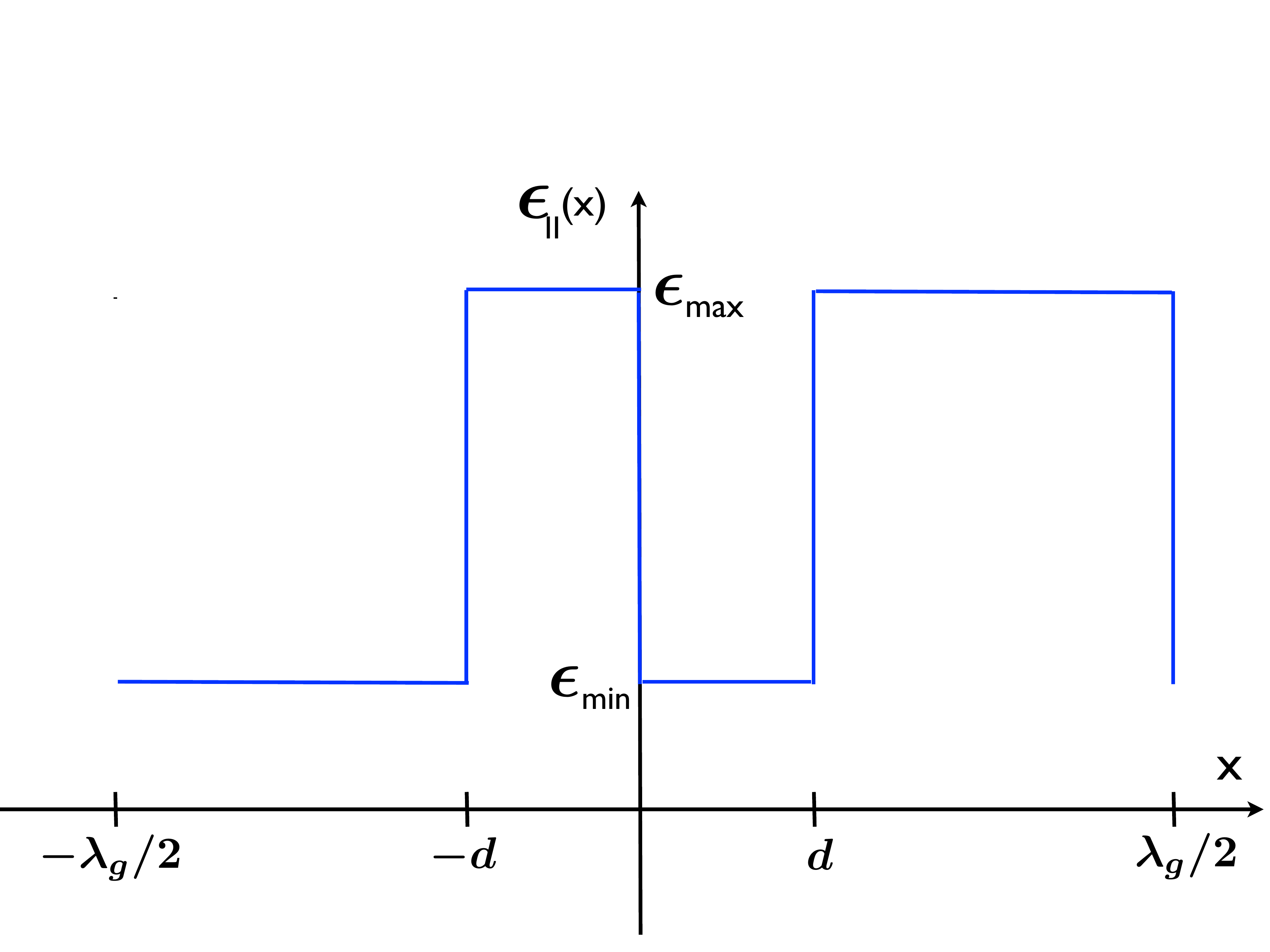}
\caption{Profile of the dielectric constant in an elementary interval of the grating layer with grating period $\Lambda$.  }
\vskip-.1cm
\label{epII}
\end{figure}

The  structure of the resonance curves  depend on the  strength $\alpha$ of the direct interaction of the 2 resonances in comparison  to  the indirect interaction  via  the  extended  mode
\begin{equation}
\alpha = \frac{|\epsilon_{2\nu}| V}{|\epsilon_{\nu}^2| \omega^2|\Sigma|}\,.
\label{rs}
\end{equation}
With the above choice of the parameters (\ref{par2}) the direct coupling dominates, i.e.\,$\alpha\gg 1\,,$ and we find for small $k_x$
\begin{equation}
W_{\pm}=
\eta_0 +k_x^2+\nu^2K_g^2 +\big|\epsilon_{\nu}\big|^2\omega^4\, \Sigma \pm\Big[\omega^2 |\epsilon_{2\nu}|V + \frac{|\epsilon_{\nu}|^4\omega^8\Sigma^2+4k_x^2\nu^2K_g^2}{2\omega^2|\epsilon_{2\nu}|  V}\Big]\,.
\label{iplim3}
\end{equation}
The direct interaction $V$ generates a repulsion of the two resonances; to leading order in $\alpha$,  their widths are equal  that of a non-interacting resonances. Also in this limit the numerical results confirm our analytical analysis as is seen in Fig.\,\ref{thresh2b}.

\begin{figure}[ht] \centering
\vskip -.2cm
\includegraphics[width=.45\linewidth]{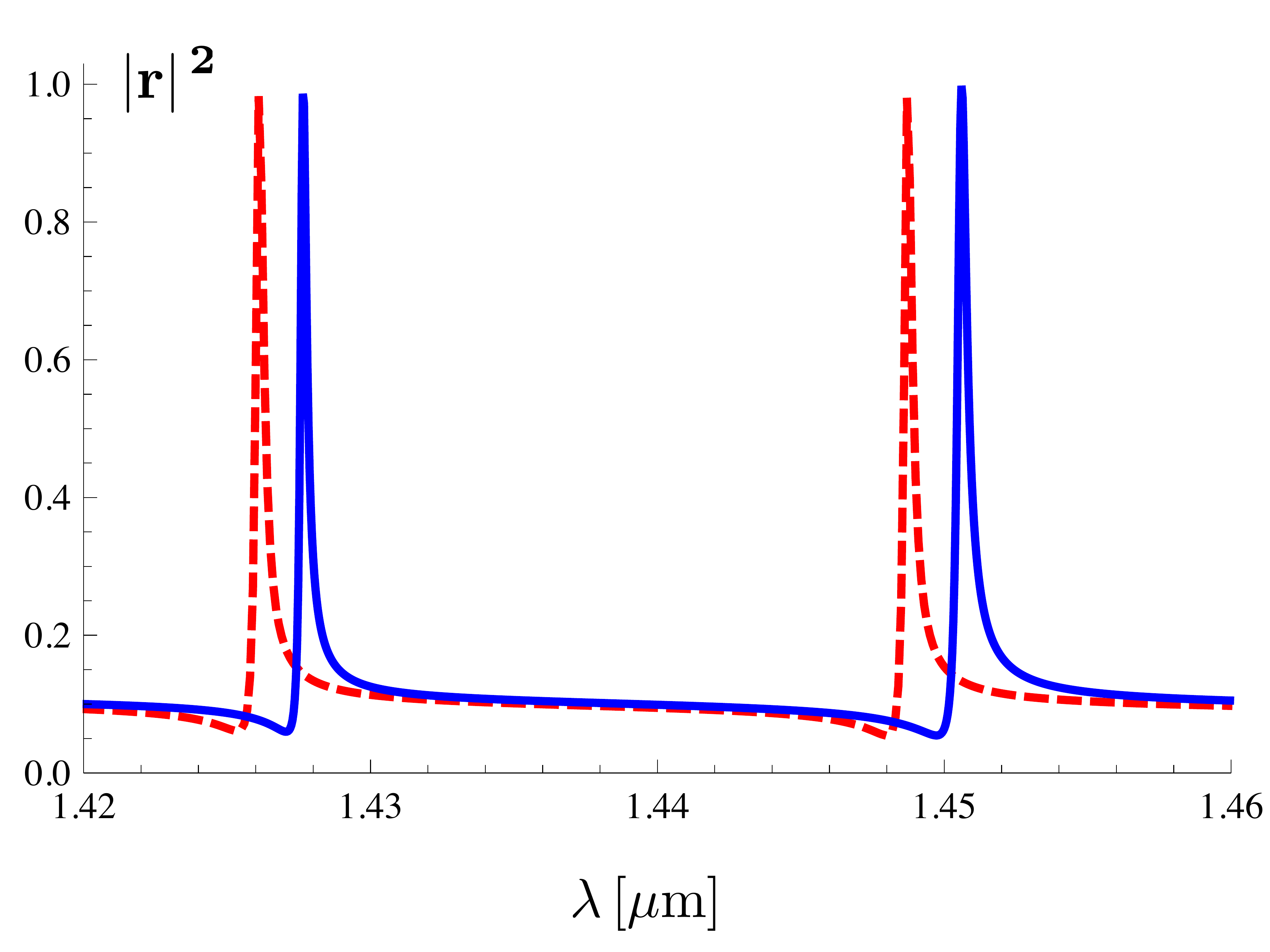}
\caption{Reflectivity as a function of the wavelength for the angle of incidence $\theta=0.005^\circ$, grating structure shown in \,Fig.\,\ref{epII} and  parameters (\ref{size})\,,(\ref{parep}) and (\ref{par2}). Solid lines: exact values,  dashed lines: values obtained in the  resonance dominance approximation. }
\label{thresh2b}
\end{figure}

\subsubsection{Contour plots, band gaps and curvatures}
Contour plots of the reflectivity in the $\theta$\,-\,$\lambda$ plane offer a   convenient way to survey  the scattering in different kinematical regimes including the separated and strongly interacting resonances.  In Figs.\,(\ref{th3}) and (\ref{th3b}) such contour plots are shown which correspond to structures and parameters of Figs.\,(\ref{thresh2}) and  (\ref{thresh2b}).
Quantities which are important for applications and which can be extracted from the contour plots are the band gap, i.e. the difference in wavelength between the peaks $|r|^2=1$ in the limit $\theta\to 0$ and the curvature of the $|r|^2=1$ curves at these  points. We will calculate these quantities and establish their dependence on the parameters of the GWS.

An analytic understanding is possible only under the assumption that the coupling of extended and guided mode is weak, i.e., $\big|\epsilon_\nu \mathcal{C}^{(+)}\big|^2\omega \ll 1$. This condition is satisfied  in all of the examples to be discussed. It implies that  values of $|r(\omega,k_x)|^2$ of the order of 1 can be realized only if     $\omega^2 |W_0/W_+W_-|\gg 1$, i.e. if  Re\,$W_+$ or Re\,$W_- \approx 0.$ In the following analytical studies we therefore replace the contour lines of the reflectivity by the contour lines of the equations
\begin{equation}
\text{Re}\,W_\pm (\omega,\theta) \approx 0\,.
\label{contla}
\end{equation}
Fig.\,\ref{th3b}  displays   the  level repulsion arising from the direct coupling term in Eq.\,(\ref{iplim3}). We know from  Fig.\,\ref{thresh2}  that    the strength of the level repulsion is correctly reproduced in resonance dominance.
To calculate band gap and curvature we drop the sub leading terms proportional to $\Sigma$ in Eq.\,(\ref{iplim3}), expand around $\omega_0=2\pi/\lambda_0$, the zero of $\eta(\omega_0)+\nu^2K_g^2$ and obtain the following expressions for  the band gap $\delta \lambda$ and curvature $\kappa$
\begin{equation}
\delta \lambda \approx \frac{\lambda_0}{-\eta_0^\prime}|\epsilon_{2\nu}| V\,,\quad \kappa= \frac{d\lambda}{d\theta^2} \approx \frac{\lambda_0}{-\eta_0^\prime}\Big(1\pm \frac{\nu^2 \lambda_0^2}{|\epsilon_{2\nu}|V\Lambda^2}\Big)  \,,\quad-\eta_0^\prime = \frac{d\eta_0(\omega^2)}{d\omega^2}\Big|_{\omega^2=\omega_0^2}\,.
\label{bacu0}
\end{equation}

\begin{figure}[ht] \centering
\includegraphics[width=.47\linewidth]{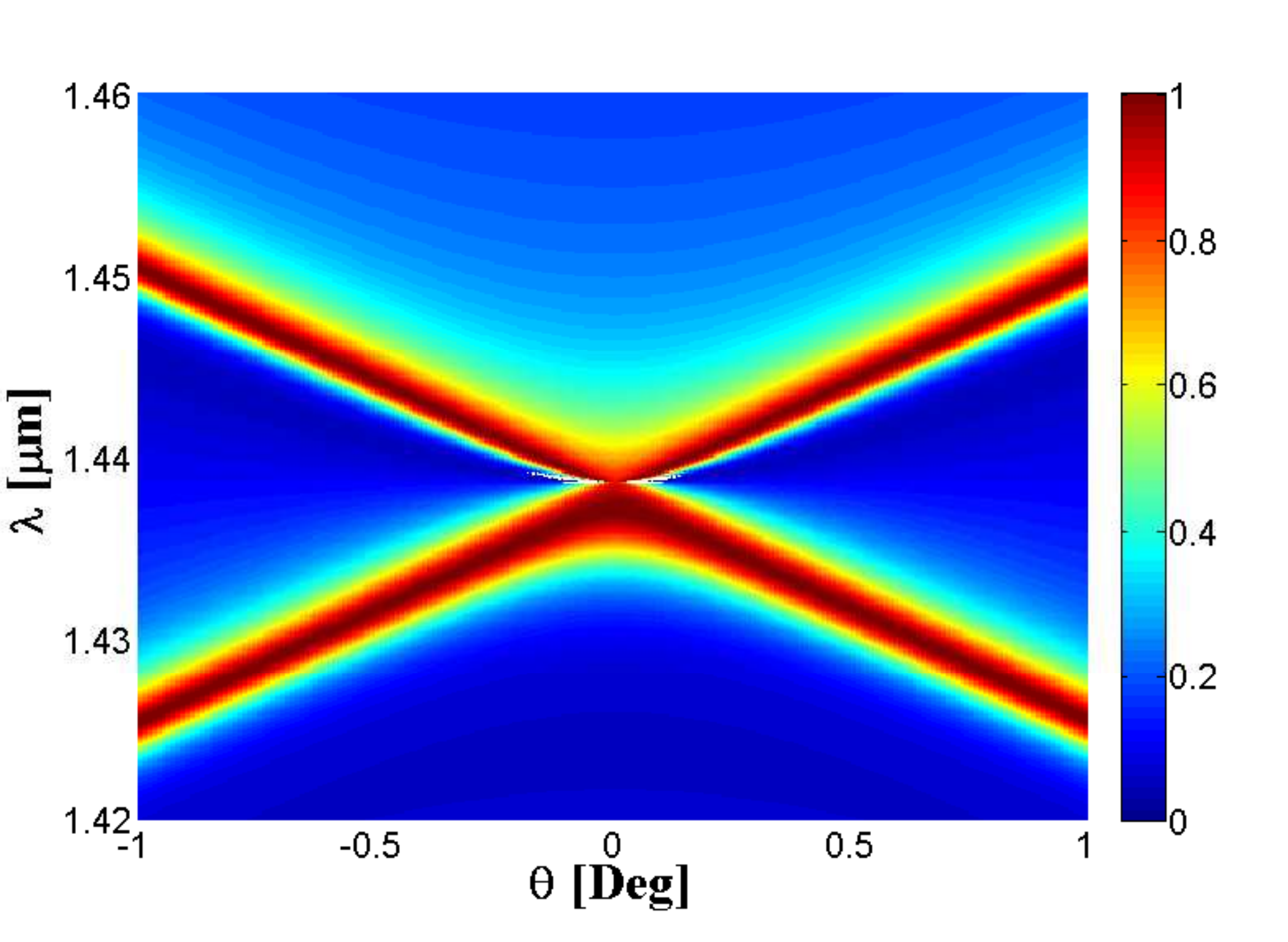}
\includegraphics[width=.47\linewidth]{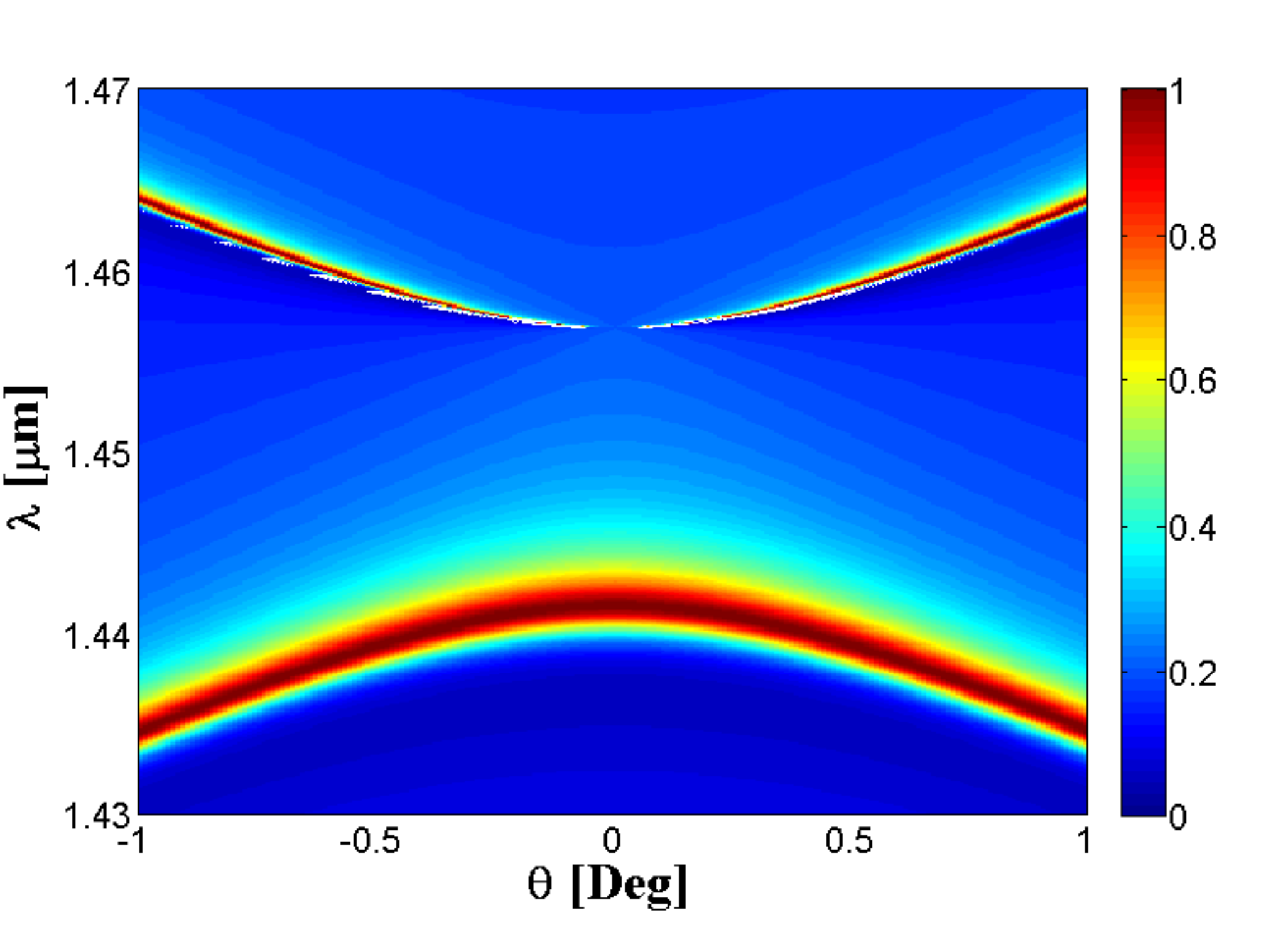}
\caption{Contour lines of the reflectivity in the $\theta\,$-$\,\lambda$ plane. Structure and  parameters  are specified in Eqs.\,(\ref{size}-\ref{gratin}). In the right part of the figure the duty cycle has been changed from 50 \% to 75\% }
\label{th3}
\end{figure}

\begin{figure}[ht] \centering
\includegraphics[width=.50\linewidth]{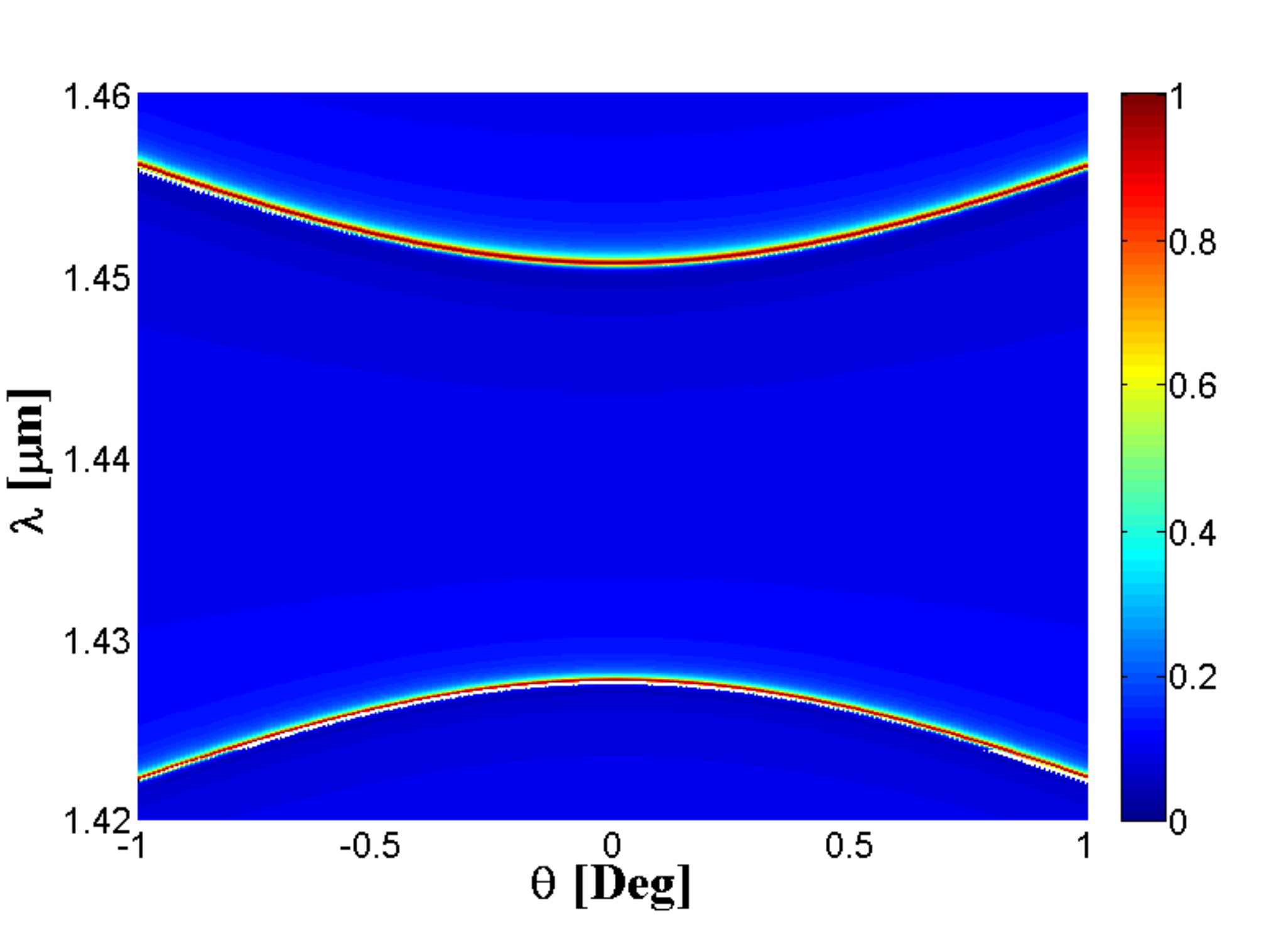}
\caption{Contour lines of the reflectivity in the $\theta\,$-$\,\lambda$ plane. Structure and  parameters  are specified in Eqs.\,(\ref{size})\,,(\ref{parep}) and (\ref{par2}).}
\label{th3b}
\end{figure}
These estimates yield for the structure of Fig.\,\ref{th3b}  $\delta\lambda\approx 22\,\text{nm},$ and   for the curvature of the 2 contour lines $\kappa\approx 0.3 \pm 6.8\,\text {nm} \,,$ and agree with the numerical results of Fig.\,\ref{th3b} .

Analytical results for the contour plot  on the left hand side of Fig.\,\ref{th3} are harder to obtain due to the presence of the zero (cf.\,Fig.\,\ref{thresh2}) of the resonance contribution to the reflection amplitude (cf.\,Eq.\,(\ref{sieps}))
for the parameters (\ref{gratin}) of the   GWS. We note, that  $\epsilon_2=0$ for  the choice of the duty cycle $d/\Lambda= 1/2$. As discussed above, the transition from isolated to interacting resonances takes place if
\begin{equation}
  \Big|\frac{\epsilon_1^2\omega^4 \Sigma}{2\omega \theta K_g}\Big| \approx 1\,.
\label{thresh}
\end{equation}
With  $|\epsilon_1|^2\omega^4 \Sigma \approx (0.06+i 0.16) \mu\text{m}^{-2}$ this condition is satisfied for  $ \theta = 0.15^\circ$\,.
For larger values of $\theta$ we expect a linear dependence of the wavelength on the angle of the incident light in agreement with the numerical results.
For significantly smaller values of $\theta$ we consider  the resonance associated with $W_+$ which is not directly affected  by the presence of the zero of $W_0$. We identify the lower branch of the contour lines of the reflectivity with the contour lines  Re$\,W_+=0$.  Expanding as above $W_+$ around $\omega_0$  we obtain
\begin{equation}
\kappa\approx -\frac{\lambda_0}{\eta_0^\prime}\Big(1-\text{Re}\frac{2K_g^2}{|\epsilon_1|^2\omega^4_0 \Sigma}\Big) \approx 110 \,\text{nm}\,,
\label{kap2}
\end{equation}
i.e., these  estimates  account for the order of magnitude difference of the curvatures in Fig.\,\ref{th3b} and the left part of Fig.\,\ref{th3}.

\subsubsection{Electric fields of overlapping resonances}
So far, we have concentrated our discussion on one observable, the reflectivity. An independent observable is the resonating electric field with components $E_{\pm\nu}$ (cf.\,Eq.(\ref{sigma2})). With only  minor modifications the above analysis of the reflectivity can be applied.   In terms of $\sigma_\pm$,  the guided mode part of the electric field is given by (cf.\,\,Eq.\,)
\begin{equation}
E_\nu(x,z)+E_{-\nu}(x,z)= -i\Big( e_+\cos \nu K_g x
+ e_- \sin \nu K_g x \Big) e^{ik_x x} \mathcal{E}_{\eta_0}(z)\,,
\label{efie1}
\end{equation}
with
\begin{equation}
e_+=i(\sigma_\nu+\sigma_{-\nu}),\quad e_-=\sigma_\nu-\sigma_{-\nu}\,.
\label{epmf}
\end{equation}
The calculation of the two components  via  Eqs.\,(\ref{sigsy})\,and\,(\ref{Minv}) is simplified if we choose $\epsilon_\nu$ to be real (which  always can be achieved with an appropriate choice of the origin of $x$ in the integral (\ref{epFocode})). It follows immediately from Eqs.\,(\ref{sieps}) and (\ref{tore2})  that $e_+$ is proportional to the resonating part of the  reflection amplitude
\begin{equation}
e_+= \frac{ 2i \epsilon_\nu\omega^2 \mathcal{C}^{(+)} W_0}{W_+ W_-}= - \frac{2  k_z^-\big(r(\omega,k_x) - r_0(\omega,k_x)\big)}{\epsilon_\nu \omega^2\mathcal{C}^{(+)}}\,,
\label{epl}
\end{equation}
This expression is valid for any value of $\theta$. It applies  also to the the case of isolated resonances.   Up to the normalization, the coefficient $e_+$ is  given by the (resonating part of the)  reflection amplitude. As seen in Fig.\,\ref{e+e-}, $e_+$  vanishes at $\omega_0$ and exhibits the same asymmetry around the corresponding wavelength as the reflectivity in Fig.\,\ref{thresh2}. The absolute value of $e_+$ is related to the half width of the isolated resonances (cf.\, Eq.\,\ref{res}) by
\begin{equation}
|e_+|= 2 \sqrt{k_z^-/\tilde{\Gamma}}\,\,|r(\omega,k_x) - r_0(\omega,k_x)|\,,
\label{e+nor}
\end{equation}
i.e. the strength of this component of the electric field increases with decreasing half width $\tilde{\Gamma}$.
In  region III of the  GWS (Fig.\,\ref{dispn}), the strength of the electric field proportional to $e_+$ can be larger than the incident field by  by up to a  factor of 10 (cf.\,Fig.\,\ref{guidmo}).
The $e_-$-component
\begin{equation}
\label{efie2}
 e_-= \frac{ 2 \epsilon_\nu\omega^2 \mathcal{C}^{(+)}}{W_+ W_-} \big(-2k_x\nu K_g +i |\epsilon_{2\nu}|\omega^2 V \sin \varphi_\epsilon \big)\,
\end{equation}
of the resonating electric field   differs from $e_+$ by the factor $W_0^{-1}$. For
$$\varphi=0,\pi$$
 the appearance of $W_0$ in   Eq.\,(\ref{sieps}) is necessary to keep the reflection amplitude finite by canceling   the zero of $W_-$  (cf.\,Eq.\,(\ref{iplim})). In turn this implies that $e_-$ is singular for $\theta\to 0$ and
$\omega=\omega_0$ where $\omega_0$ denotes zero of $W_0$
\begin{equation}
W_0(\omega_0)=\eta_0(\omega_0^2)+\nu^2K_g^2+\omega_0^2|\epsilon_{2\nu}| V(\omega_0^2)=0.
\label{om0}
\end{equation}
\begin{figure}[ht] \centering
\includegraphics[width=.4\linewidth]{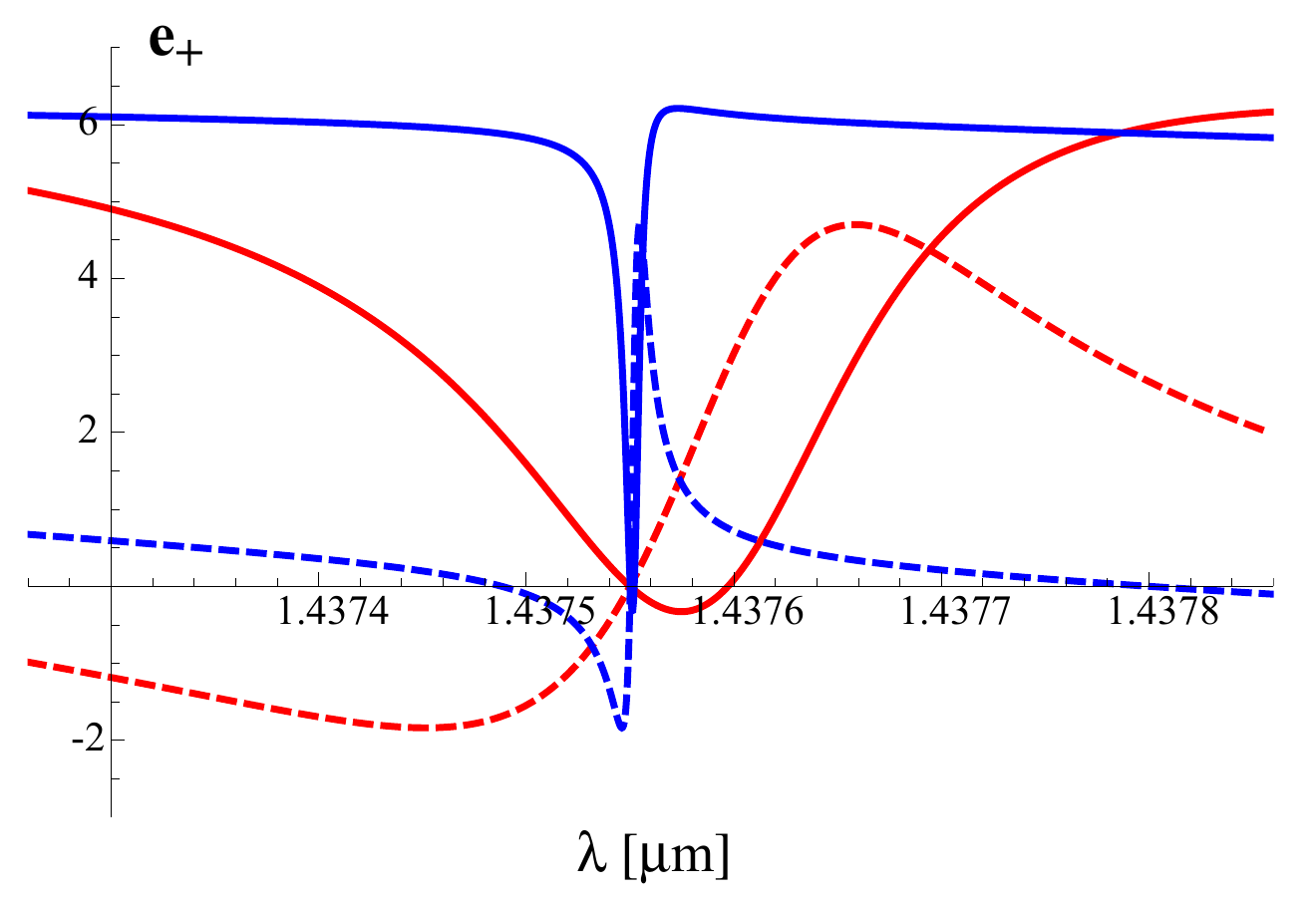}\hspace{1cm} \includegraphics[width=.40\linewidth]{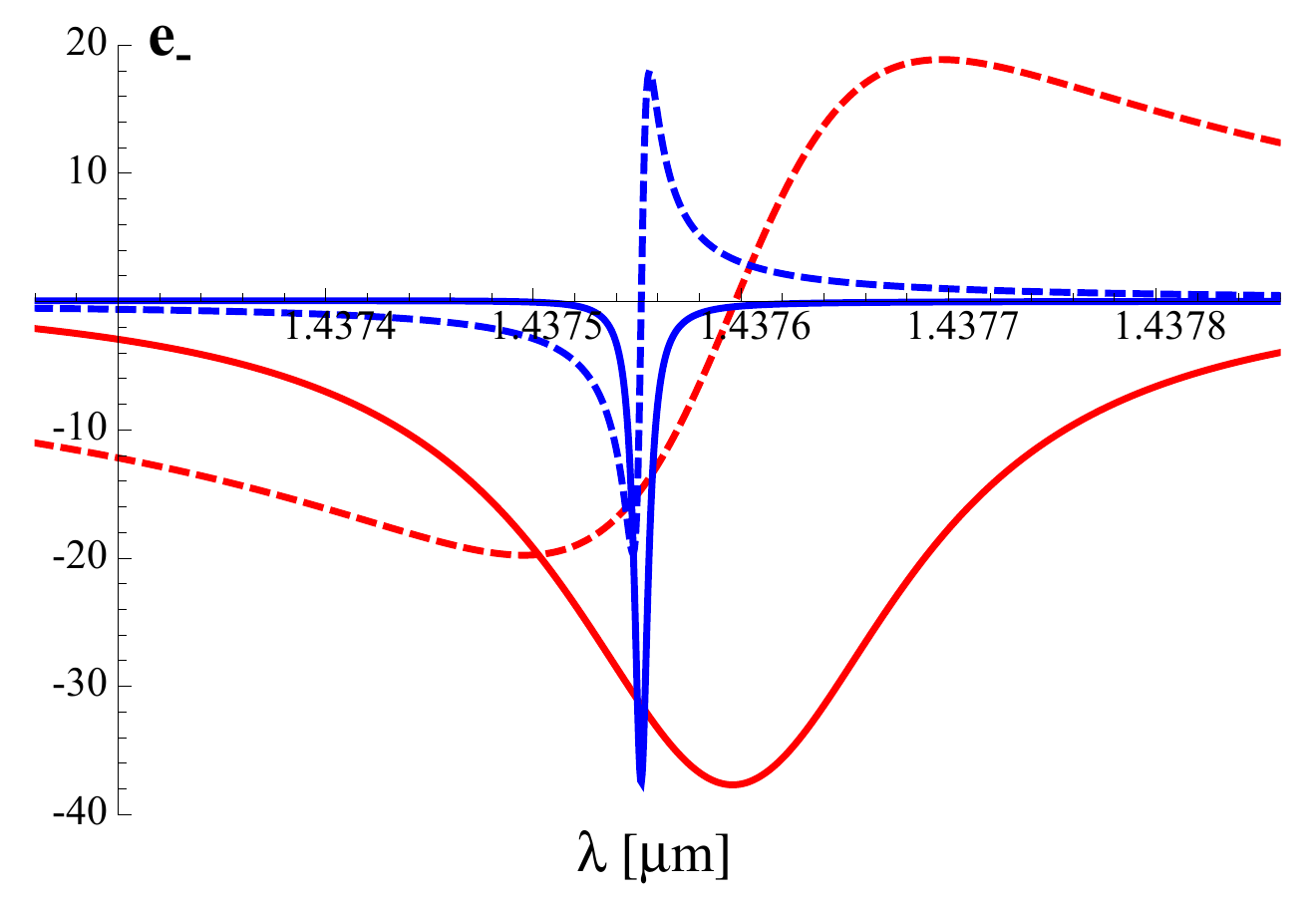}
\caption{Electric fields  $e_+$ (left) and $e_-$  (right)  (cf.\,Eqs.\,(\ref{epl}) and (\ref{efie2})) in units of the amplitude of the incident field as a function of the wavelength for the interacting resonances of the left hand side of Fig.\,\ref{thresh2}. Solid lines: real part, dashed lines: imaginary part, red (color online) $\theta=0.05^\circ$, blue  (color online) $\theta=0.01^\circ$. The values for $e_-$ at  $\theta=0.01^\circ$ have been divided by a factor of 5. }
\label{e+e-}
\end{figure}
For values  of ($\theta$, $\omega^2$) sufficiently close  to the singular point ($0$, $\omega_0^2$) in the $\theta,\omega^2$ plane,     $e_-$ is given by
\begin{equation}
 e_-= \frac{4\omega_0 \epsilon_\nu\mathcal{C}^{(+)}}{\nu K_g}\, \frac{\theta\, \Sigma_-}{\omega - \omega_0 +\theta^2\,\Sigma_-}\,,\quad \Sigma_-= \frac{2\nu^2 K_g^2}{W_0^\prime(\omega_0) \Big(-|\epsilon_{\nu}\big|^2\omega_0^2\, \Sigma(\omega_0)+  |\epsilon_{2\nu}|  V(\omega^2_0)\Big)}\,.
\label{sing0}
\end{equation}
Considered as a function of $\omega$, for fixed $\theta$, the component $e_-$ has a Lorenzian (Breit Wigner) shape.
It reaches its maximal value   at $\omega=\omega_0-\theta^2\text{Re} \Sigma_-$
\begin{equation}
 |e_-|_{\text{max}}= \frac{4\sqrt{\tilde{\Gamma}k_z^-}}{|\nu|\omega_0 K_g}\frac{|\Sigma_-|}{\theta |\,\text{Im} \Sigma_-|}\,.
\label{maxva}
\end{equation}
As Fig. \ref{e+e-} demonstrates the increase in the electric field strength with decreasing angle $\theta$ of the incident light is accompanied by a decrease of the half width $\theta^2\, \text{Im}\,\Sigma_-$ of the Breit-Wigner function (\ref{sing0}). In the homogeneous layer III (cf., Figs.\,\ref{dispn},\,\ref{guidmo}) the resonating part of the electric field reaches values which are up to a factor $3/\theta[^\circ]$ larger than the field of the incident light.

\subsection{Overlapping TM resonances}\label{ss2resTM}
\subsubsection{Formal development}
We now consider interacting resonances in the TM case.  We follow the developments  in Section \ref{ss2res} and consider the case when there are two eigenvalues $\eta_1\equiv\eta_{n_1,\nu_1}$ and $\eta_2\equiv\eta_{n_2,\nu_2}$ with values  close  to $\omega^2$. As the comparison of Eqs. (\ref{swvafc}) and  (\ref{eq:eqforpsi}) shows, unlike in the TE case, the guided modes $\mathcal{H}_{n,\nu_i}$    depend on the index $n$ which makes the following analysis more involved. The same complications will occur in the TE case if overlapping resonances associated with different eigenvalues $\eta_{n_{1,2}}$ are considered.

We retain both  resonating terms in the system of equations (\ref{eq:inteqforHnwithnonzeron}) and write coupled equations for the resonating   and the zero components. We will  then use the resonance dominance approximation retaining only a single term with the guided  mode $\mathcal{H}_{n_i,\nu_i}(z)$ in the sums over $i$   in Eq.(\ref{redoTM})
\begin{eqnarray}
H_0(z) & = & H_0^{(+)}(z) + \int_{I_g} dz' g(z,z') \left( \Theta_{0,n_1} H_{n_1}(z') + \Theta_{0,n_2} H_{n_2}(z') \right)\,, \label{eq:TM3_H0} \\
H_{n_1}(z) & = & \frac{1}{\omega^2 - \eta_{1}} \mathcal{H}_{n_1,\nu_1}(z) \int_{I_g} dz' \mathcal{H}_{n_1,\nu_1}(z') \left( \Theta_{n_1,0} H_0(z') + \Theta_{n_1,n_2} H_{n_2}(z') \right) \label{eq:TM3_Hn1}\,, \\
H_{n_2}(z) & = & \frac{1}{\omega^2 - \eta_{2}} \mathcal{H}_{n_2,\nu_2}(z) \int_{I_g} dz' \mathcal{H}_{n_2,\nu_2}(z') \left( \Theta_{n_2,0} H_0(z') + \Theta_{n_2,n_1} H_{n_1}(z') \right)\,. \label{eq:TM3_Hn2}
\end{eqnarray}
As in the case of the single resonance the components $H_{n_1}$  and $H_{n_2}$ are proportional to the guided modes
\beq  \label{eq:relHtotwomodes}
H_{n_1}(z)=\sigma_{n_1}\mathcal{H}_{n_1,\nu_1}(z)\,, \;\;\;H_{n_2}(z)=\sigma_{n_2}\mathcal{H}_{n_2,\nu_2}(z)\,,
\eeq
but the equations determining $\sigma_{n_i}$'s show mixing of the two resonating modes.  Inserting
(\ref{eq:relHtotwomodes}) into (\ref{eq:TM3_Hn1}) and (\ref{eq:TM3_Hn2}) together with (\ref{eq:TM3_H0}) we obtain closed equations for   the field enhancement coefficients
\begin{equation} \label{eq:coupledTMeqsforsigmas}
W \left( \begin{array}{c} \sigma_{n_1} \\ \sigma_{n_2} \end{array} \right) =
\left( \begin{array}{c}\gamma_{n_1} \mathcal{C}_{n_1}^{(+)} \\ \gamma_{n_2}\mathcal{C}_{n_2}^{(+)} \end{array} \right)
\end{equation}
where, in analogy with the TE case (Eqs.\,(\ref{sigsy}),\,(\ref{Msig}), we have introduced
the coupling matrix
\begin{equation}
W = \left[ \begin{array}{cc}
\omega^2 - \mu_{1}&
-\gamma_{n_1-n_2}V -\gamma_{n_1}\gamma_{-n_2} \Sigma_{n_1,n_2} \\
-\gamma_{n_2-n_1}V - \gamma_{-n_1}\gamma_{n_2}\Sigma_{n_2,n_1} &
\omega^2 - \mu_{2}
\end{array} \right]\,,
\label{eq:TM3_Wmatrix}
\end{equation}
and
\beq
\mu_i=\eta_{i} + |\gamma_{n_i}|^2\Sigma_{n_i,n_i}\,.
\eeq
We have used  the replacement explained in Appendix \ref{app:singular}
$ \Theta_{nm}\rightarrow\gamma_{n-m}\tilde{\Theta}_{nm}$  with the notation
 \beq
 \tilde{\Theta}_{nm}=\overleftarrow{\partial_z}\;\overrightarrow{\partial_z}+(k_x+nK_g)(k_x+mK_g)\,.
 \eeq
The quantities
 \begin{eqnarray}
\mathcal{C}_{n_i}^{(+)}& = & \int_{I_g} dz' \mathcal{H}_{n_i,\nu_i}(z') \tilde{\Theta}_{n_1,0} H_0^{(+)}(z')\,,
\label{eq:TM3_C} \\
\Sigma_{n_i,n_j} & = & \int_{I_g} dz' \int_{I_g} dz'' \mathcal{H}_{n_i,\nu_i}(z') \tilde{\Theta}_{n_i,0} g_0(z',z'') \tilde{\Theta}_{0,n_j} \mathcal{H}_{n_j,\nu_j}(z'') \;,\;\;i,j=1,2 .\label{eq:TM3_Sigmas}
\end{eqnarray}
are generalizations of the definitions (\ref{eq:TMexpforCplus}) and (\ref{eq:TMexpforSigma}) for isolated resonances. Here a new quantity appears
\begin{equation}
V  =  \int_{I_g} dz'  \mathcal{H}_{n_1,\nu_1}(z') \tilde{\Theta}_{n_1,n_2} \mathcal{H}_{n_2,\nu_2}(z')
\label{eq:TM3_V}
\end{equation}
which
 controls the direct coupling between the two guided modes.  An indirect coupling between the guided modes via the interaction with the extended mode  is generated by $\Sigma_{n_1,n_2} =\Sigma_{n_2,n_1}$.
The quantities $\Sigma_{n_i,n_i}$ are the self-interactions of each guided mode via the extended mode. Note that for the dielectric structure in which $\gamma_{n_1}=\gamma_{n_2}=0$ but with $\gamma_{n_1-n_2}\ne0$ we obtain the standard avoided level crossing problem, i.e. the non radiating photonic band gap case.

The eigenvalues of $W$  and the field enhancement coefficients are given by
\eqna \label{eq:TM3_Wpm}
W_{\pm} &=& \omega^2 - (\mu_1 + \mu_2)/2     \\
 && \pm \sqrt{ (\mu_1 - \mu_2)^2/4+( \gamma_{n_1-n_2}V +\gamma_{n_1}\gamma_{-n_2} \Sigma_{n_1,n_2} )( \gamma_{n_2-n_1}V + \gamma_{-n_1}\gamma_{n_2}\Sigma_{n_2,n_1} ) }\,, \nonumber
\eqne

\begin{eqnarray}
\sigma_{n_1} & = & \frac{ \gamma_{n_1}( \omega^2 - \mu_2) \mathcal{C}_{n_1}^{(+)} + \gamma_{n_2}(\gamma_{n_1-n_2}V +\gamma_{n_1}\gamma_{-n_2} \Sigma_{n_1,n_2}) \mathcal{C}_{n_2}^{(+)} }
{W_+W_-}\,, \label{eq:TM3_sigman1} \\
\sigma_{n_2} & = & \frac{ \gamma_{n_1}( \gamma_{n_2-n_1}V + \gamma_{-n_1}\gamma_{n_2}\Sigma_{n_2,n_1}) \mathcal{C}_{n_1}^{(+)} +  \gamma_{n_2}(\omega^2 - \mu_1) \mathcal{C}_{n_2}^{(+)}  }{W_+W_-}\,. \label{eq:TM3_sigman2}
\end{eqnarray}

Using Eq. (\ref{eq:TM3_H0}) and the asymptotics of the Green's function (cf.\,Appendix \ref{sec:bagr})  the reflection amplitude is given by
\begin{equation}
r(\omega,k_x) = r_0(\omega,k_x) - \frac{i \epsilon_{I}}{2k_z^-} \left( \sigma_{n_1}\gamma_{-n_1} \mathcal{C}_{n_1}^{(+)} + \sigma_{n_2}\gamma_{-n_2} \mathcal{C}_{n_2}^{(+)} \right)
\label{eq:TM3_refcoeff1}
\end{equation}


%

\subsubsection{Overlapping resonances close to normal incidence.} To proceed with the analysis we now specify to the special case of the vicinity of $k_x=0$.  We note that a solution of Eq. (\ref{eq:eqforpsi2}) with a given $k_x$ and $n$ is also a solution with   $k_x\rightarrow-k_x$ and $n\rightarrow-n$. This means that at   $k_x=0$ the resonance condition  (\ref{eq:TMresoncond}) will always be satisfied by a pair of guided modes. Let us consider such a resonating pair with the eigenvalues $\eta_{n,\nu}=\eta_{-n,\nu}$.   For such a pair
$n_1=n=-n_2$
 and at $k_x=0$ all the components of $ \mathcal{C}^{(+)}$ and of $\Sigma$ are identical and given by
\begin{eqnarray} \label{eq:TMexpressions}
\mathcal{C}^{(+)}&=& \mathcal{C}_{n_i}^{(+)}= \int_{I_g} dz' [\partial_{z'}\mathcal{H}_{n,\nu}(z')] [\partial_{z'} H_0^{(+)}(z')] \;\;, \nonumber \\
 \Sigma&=&\Sigma_{n_i,n_j} =  \int_{I_g} dz' dz'' [\partial_{z'}\mathcal{H}_{n,\nu}(z')] [\partial_{z'}\partial_{z''} g_0(z',z'') ] [\partial_{z''}\mathcal{H}_{n,\nu}(z'')]\,, \\
V & = &\int_{I_g} dz'  \left\{ [\partial_{z'} \mathcal{H}_{n,\nu}(z')]^2
 - (nK_g)^2 [\mathcal{H}_{n,\nu}(z')]^2\right\}\;. \nonumber
\end{eqnarray}
For  small deviations from   $k_x=0$  we find
\beq
\Delta \eta_{\pm n,\nu}=\pm2\zeta nK_g k_x, \;\;\; \zeta=\int_{-\infty}^{\infty} dz \gamma_0(z)\mathcal{H}^2_{n,\nu}(z)\,,
\eeq
where we have used Eq. (\ref{eq:eqforpsi})  and
\beq
\frac{\partial \eta_{n\nu}}{\partial k_x}=\int_{-\infty}^{\infty} dz \;
\mathcal{H}_{n,\nu}\frac{\partial \Theta_{nn}}{\partial k_x}\mathcal{H}_{n,\nu}\,.
\eeq
Identifying for simplicity $\Sigma$ and $V$ with their values at $k_x=0$, we approximate Eq. (\ref{eq:TM3_Wpm}) by
\eqna
W_{\pm} &=& [\omega^2 - \eta_{n,\nu}-|\gamma_{n}|^2\Sigma]    \\
 && \pm \sqrt{(2\zeta k_x n K_g)^2 +
\left(|\gamma_{n}|^2 \Sigma^2+|\gamma_{2n}|\cos{\varphi_\gamma}V\right)^2 +|\gamma_{2n}|^2\sin^2{\varphi_\gamma}V^2 }\nonumber
\label{eq:rootsTM}
\eqne
with
\begin{equation}
e^{i\varphi_\gamma}=\frac{\gamma_{2n}\gamma_n^{\star\,2}}{|\gamma_{2n}\gamma_n^2|}\,.
\end{equation}
The reflection amplitude is given by
\beq
r(\omega,k_x) = r_0(\omega,k_x) - \frac{i \epsilon_{I}|\gamma_{n}|^2(\mathcal{C}^{(+)})^2}{k_z}\frac{W_0}{W_+W_-}\;\;,\;\;
W_0=\omega^2 -  \eta_{n,\nu}+|\gamma_{2n}| \cos\varphi_\gamma V\,.
\eeq
The above expressions for $W_{\pm}$  and $r(\omega,k_x)$ have
 the same form as in   the TE case (cf.\,Eqs. (\ref{solev}),\,(\ref{sieps})).
We  again find that the interference of the contributions  of the two eigenvalues $W_\pm$ of the matrix $W$ gives rise to a zero in the resonance part of the reflection amplitude.  Also the  role played by the phase $\varphi_\gamma$ is identical to that of the TE phase $\varphi_\epsilon$.
 In particular for values
$$ \varphi=0,\pi $$
the width of one of the resonances (corresponding to $W_{+}$) goes to zero at $k_x=0$. This is illustrated in Fig. \ref{refTMvarphi02D}. In Figs. \ref{refTMvarphi0}  we show an example of resonance dominance compared to the exact results.
\begin{figure}[ht] \centering
\includegraphics[width=.48\linewidth]{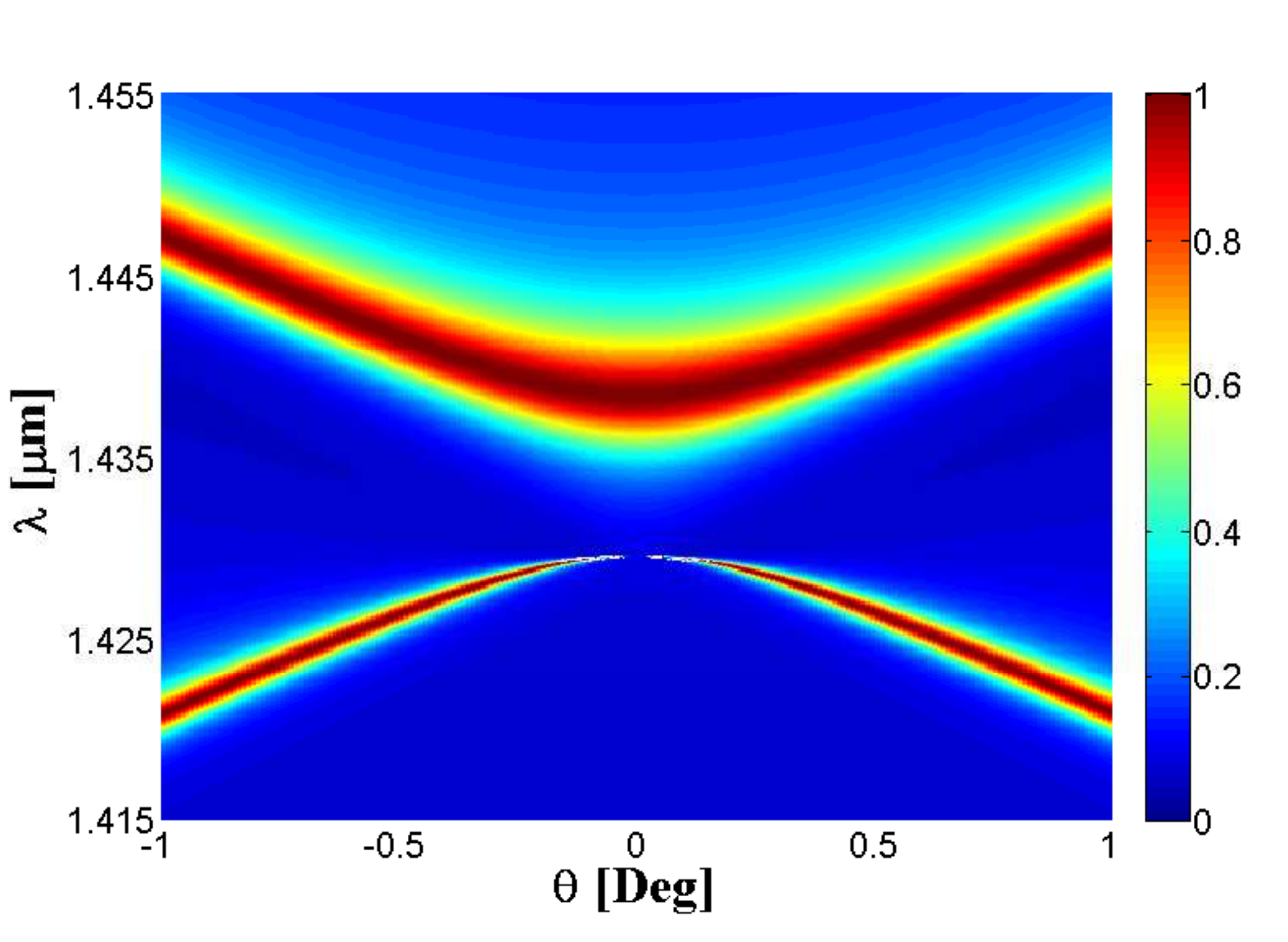}\hspace{.3cm}
\includegraphics[width=.48\linewidth]{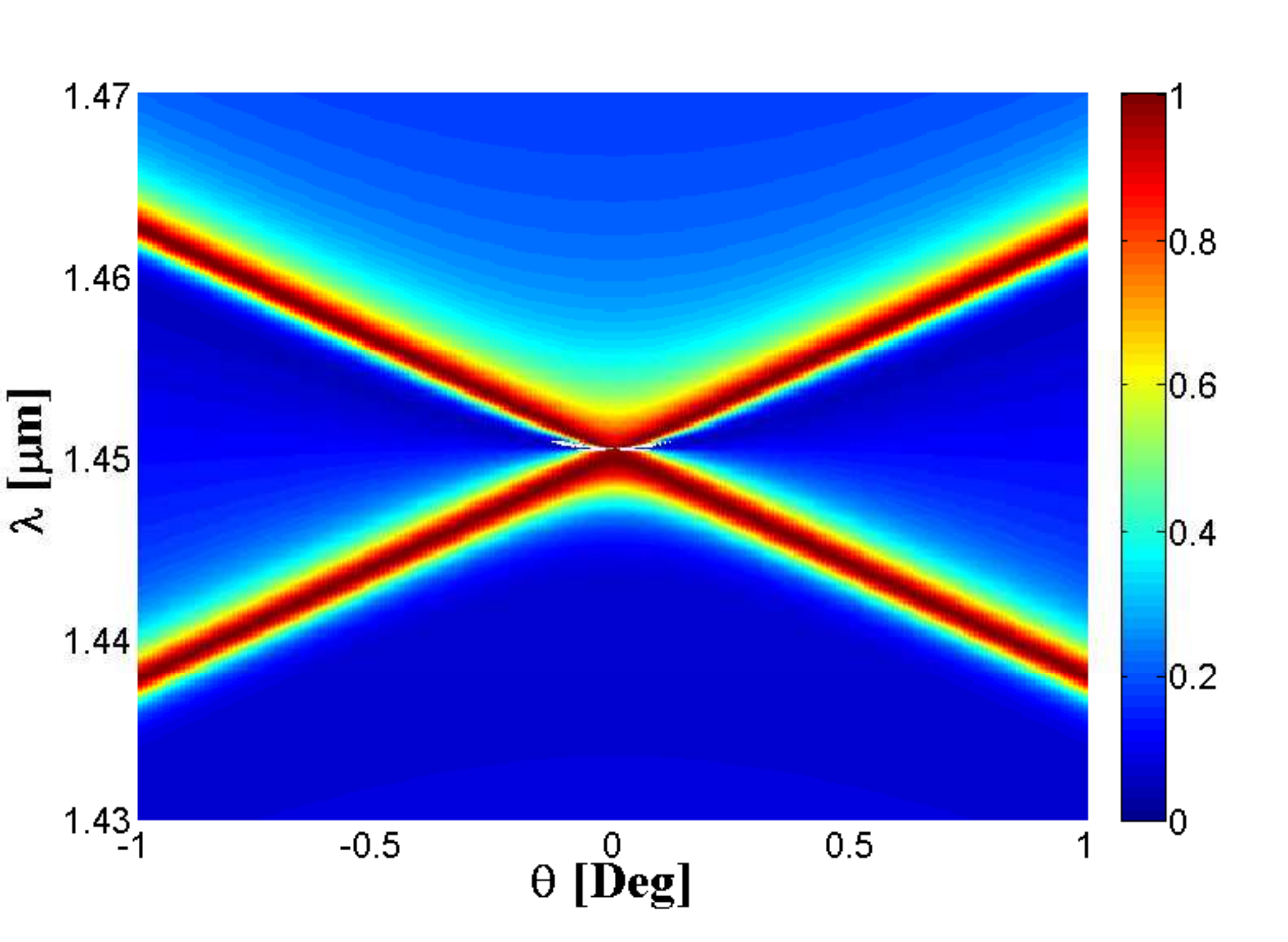}
\caption{Contour lines (exact results)  of the reflectivity  for TM polarization in the $\theta\,$-$\,\lambda$ plane.  Structure and  parameters as given in Eqs.\,(\ref{size}-\ref{gratin}) apart of the grating period $0.87\mu m$ and  $50\%$ duty cycle (left figure), $75\%$ duty cycle (right figure).}
\label{refTMvarphi02D}
\end{figure}
\begin{figure}[ht] \centering
\includegraphics[width=.45\linewidth]{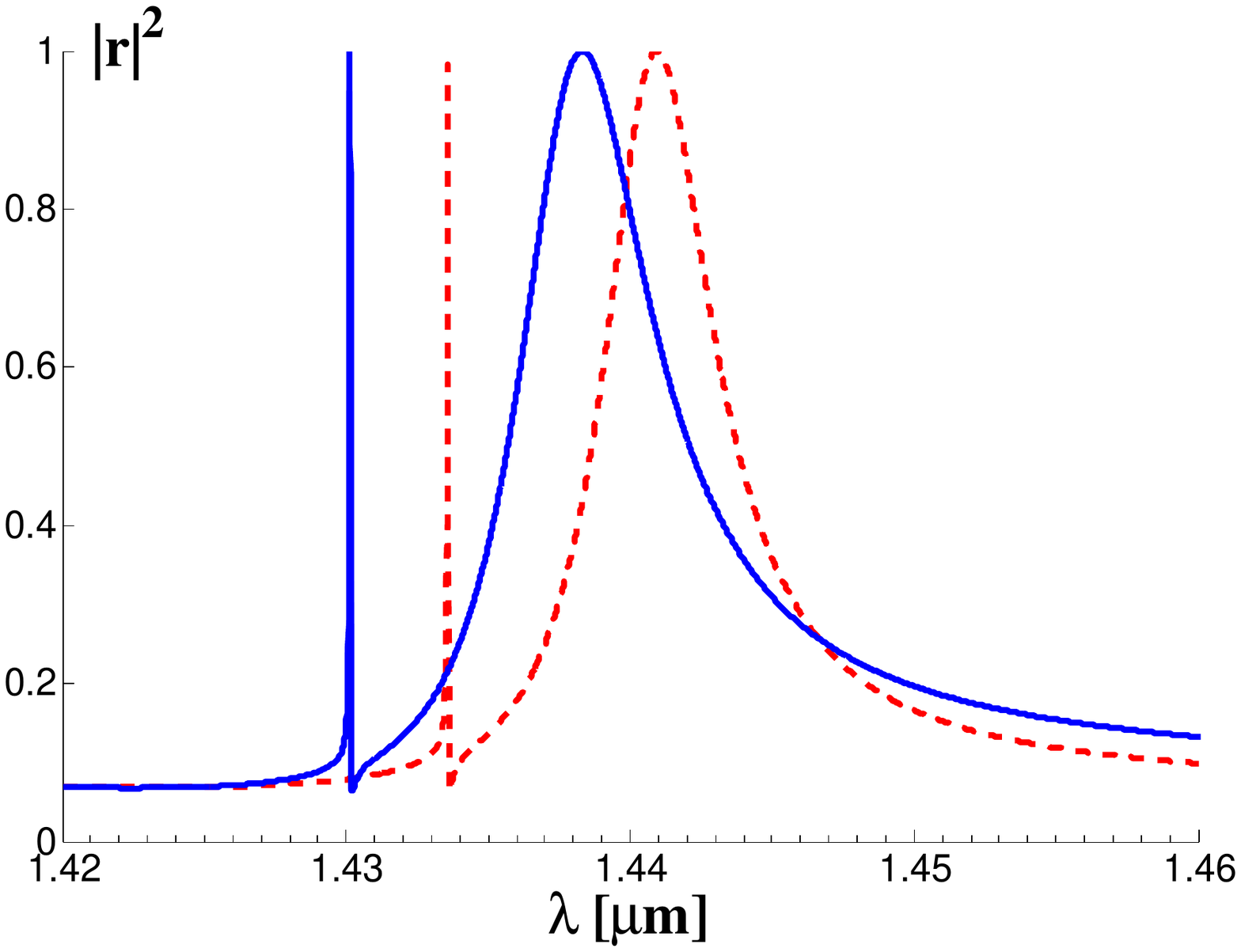}\hspace{.5cm}
\includegraphics[width=.45\linewidth]{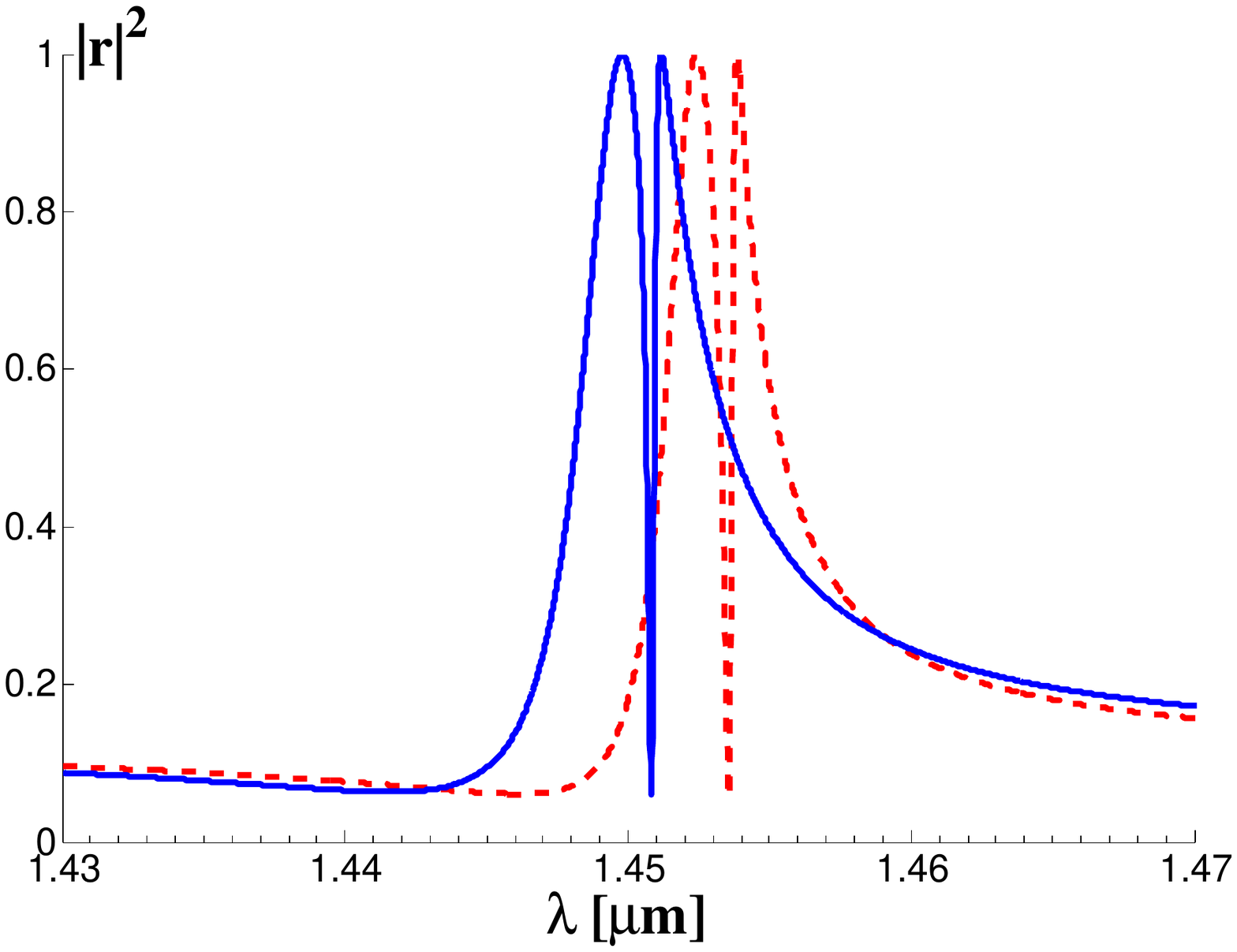}
\caption{Reflectivity for TM polarization as a function of the wavelength at $\theta=  0.05^\circ$  for the same structure and parameters as in Fig. \ref{refTMvarphi02D}  with $50\%$ duty cycle (left figure), $75\%$ duty cycle (right figure). Solid lines: exact values (vertical cut in Fig. \ref{refTMvarphi02D}),  dashed lines: resonance dominance approximation (cf., Eq.(\ref{eq:rootsTM}) with $\varphi_\gamma=0$).}
\label{refTMvarphi0}
\end{figure}

Note that the  dependence on the duty cycle of the TM results  is qualitatively very  different from the TE case, Figs. \ref{thresh2} and \ref{th3}.  The apparent band gap at $k_x=0$  closes at a very different value of the duty cycle. The source of this difference is clearly seen in our theory.  Let us define the band gap as the difference between the two resonant
frequencies at normal incidence.  This   corresponds to the difference between  the real parts of $W_{+}$ and $W_{-}$.
When the band gap size is significantly larger than the FWHM of the wider of the two resonances, the two resonances are well separated and thus
the band gap becomes clearly visibly as in the left part of Fig. \ref{refTMvarphi02D}. For the band gap  of the order or smaller than the wider resonance width  the resonances overlap and appear merged.

According to Eq. (\ref{solev}) and (\ref{eq:rootsTM})  the difference between $W_{+}$ and $W_{-}$ for $\varphi=0,\pi$ is given by
\beq \label{eq:bandgaps}
2\omega^2 \left(|\epsilon_{\nu}|^2 \omega^2\,\Sigma +|\epsilon_{2\nu}|   V\right)\;\;\;{\rm{and}} \;\;\; 2\left(|\gamma_{n}|^2 \Sigma+|\gamma_{2n}|V\right)
\eeq
for TE and TM respectively. When the duty cycle is close  to 50\% we have  vanishing  $|\epsilon_{2\nu}|$  so that the direct coupling term $V$ does not contribute in the TE case.  At the same time $\gamma_{2n}$ does not vanish in the inverse Fourier  transform matrix approach which we adopted as explained  in Section \ref{sec:TMresonances}. This difference between the TE and TM is however not as important as the difference in the expressions for  $\Sigma$.   They are respectively given by Eqs. (\ref{shga}) and (\ref{eq:TMexpressions})  with the  reference to expressions (\ref{gr1}) and (\ref{eq:identity}).  Comparing we notice a peculiar term present only in the TM case.  This is the last term in (\ref{eq:identity}).  It is proportional to the $\delta$-function and produces a "contact" term
\beq
-1/\gamma_0\int_{I_g} dz'  [\partial_{z'}\mathcal{H}_{n,\nu}(z')]^2
\label{contact}
\eeq
in the expression for $\Sigma$ in the TM case. We call this term "contact" since in it the mode functions
$\partial_{z}\mathcal{H}_{n,\nu}(z)$ "interact" at one point.  This is in contrast to expression (\ref{solev}) for the TE or  (\ref{eq:rootsTM})  with only the first term of (\ref{eq:identity}) for TM. In those cases the mode functions interact via a finite range interaction caused by the propagation in the extended mode.

The contact term is real and  therefore contributes
to the band gap. Numerical evidence shows that this is the dominant contribution for the  50\% duty cycle  and is significantly larger than the wider resonance width given by the imaginary part of $\Sigma$. This is the reason the apparent band gap is present at this duty cycle in the TM case.
 In order to complete the argument and to understand why the real part of $\Sigma$ does not give rise to the same effect, we also compare the corresponding quantities for TE and TM case. To estimate the above contribution to the TM Re$\Sigma$ we write in the thin grating case
$$ |\gamma_n|^2 \int_{I_g} dz'  [\partial_{z'}\mathcal{H}_{n,\nu}(z')]^2  \approx  |\gamma_n|^2\ell_g \mathcal{H}_{n,\nu}^{\prime\,\,2}(\ell_g/2).$$
We compare this estimate with the corresponding TE result
$$ |\epsilon_\nu|^2\omega^4 \text{Re} \Sigma \approx \frac{1}{6}|\epsilon_\nu|^2 \omega^4\ell_g ^3 \mathcal{E}_{\eta_0}^2 (\ell_g/2)$$ recalling also that we have to divide the TE result by  $d\eta_0/d\omega^2 \approx 3.5$ before we can interpret is as a shift.


As one changes to higher or lower
values of the duty cycle the second term, the direct interaction $V$ begins to play a role in  Eq. (\ref{eq:bandgaps}). Here again the explicit expressions of V, cf., (\ref{v2}) and (\ref{eq:TMexpressions}) show the clear difference between the TE and the TM waves. In the former $V$ is real, positive and therefore contributes to the increase of the band gap. For the TM case $V$ is a difference of two positive terms and can be either positive, i.e. increasing the gap or negative, i.e. leading to its decrease. In fact we find that the apparent TM  band gap closes at  75\% duty cycle, cf. the right part of Fig. \ref{refTMvarphi02D}  as well as at 17 \%, unlike in the TE case where the band gap closes only for a single value of the duty cycle.

 The overall agreement between the resonance dominance and the exact results in the TM case is not as satisfactory as for the TE case. As an example we show in Fig.~\ref{refTM2} the TM results for the same complex grating as in the TE case, cf., Fig.~\ref{epII} and Eq. (\ref{par2}) with $\lambda_g=0.87\mu m$.  It is seen that the  band gap is significantly smaller  and the relative widths of the two resonances are  different  in the resonance dominance approximation as compared to the exact results. In our understanding there is a number of effects in the TM case contributing to such a discrepancy vis-a-vis an excellent agreement which is obtained in the TE  case.  Firstly, it is known, cf., Ref. \cite{Li96} that  numerical convergence  as a function of number of truncation orders is much slower in the TM case than the TE case.  We note  that the grating parameters as in Eq. (\ref{par2}) lead to $|\gamma_2|\gg|\gamma_1|$ raising the relative importance of the higher order components.

 Secondly, we believe that the issue of the special boundary conditions (cf., the end of Section \ref{sec:TMresonances})   in the TM case is not accounted properly in our way of implementing the resonance dominance approximation.
 We plan to address this and related problems of the TM polarized scattering in the future  work.

\begin{figure}[ht] \centering
\includegraphics[width=.47\linewidth]{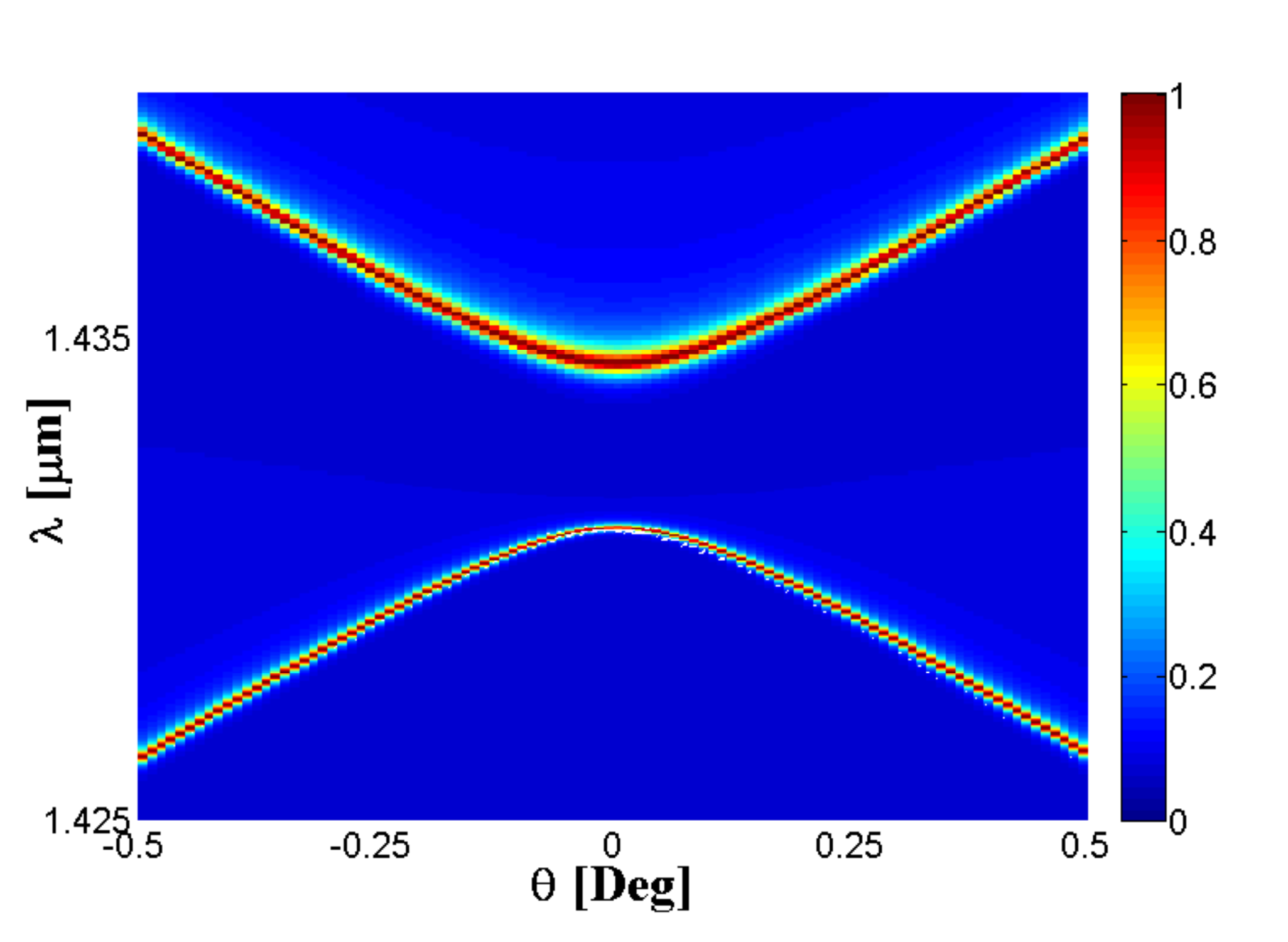}
\includegraphics[width=.45\linewidth]{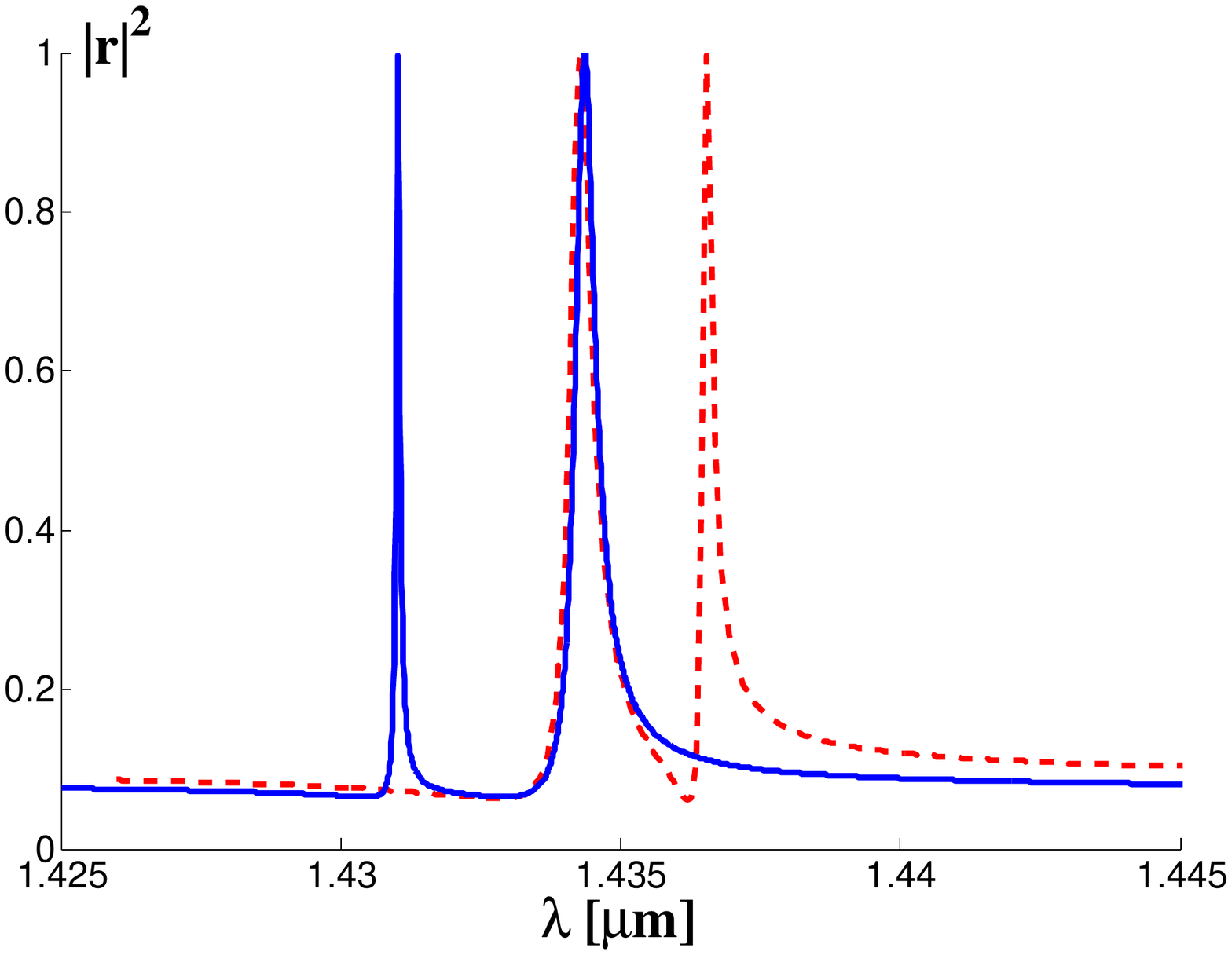}
\caption{Left: contour lines (exact results) of the reflectivity  for TM polarization in the $\theta\,$-$\,\lambda$. Right: reflectivity for TM polarization as a function of the wavelength at $\theta=  0.005^\circ$. Solid line: exact values (vertical cut in the left figure),  dashed line: resonance dominance approximation (cf., Eq.(\ref{eq:rootsTM}). Same structure and parameters as in Fig. \ref{refTMvarphi02D} apart the grating parameters which are given by Eq. (\ref{par2}) with $\lambda_g=0.87\mu m$.}
\label{refTM2}
\end{figure}

\subsection{Comparison with coupled resonances in quantum mechanics}
The physics of coupled guided mode resonances of light scattering off photonic crystal slabs discussed above is closely related to  the physics of interacting (overlapping) resonances or bound states in quantum mechanics. This connection is  obvious in the limit of vanishing coupling $\epsilon_\nu$ to the extended mode. Then the matrix $W$ (\ref{Msig})  is hermitian and the two (real) eigenvalues $W_\pm$ of  $W$  can be written as
\begin{equation}
W_\pm = \frac{1}{2}\big[W_1+W_2 \pm\sqrt{(W_1-W_2)^2+4\omega^4|\epsilon_{2\nu}|^2V^2}\big]\,, \quad W_{1,2}=\eta_0+(k_x\pm \nu K_g)^2\,.
\label{lvrp}
\end{equation}
It's counterpart is the quantum mechanical two level system defined by the Hamiltonian matrix
\begin{equation}
H=
    \left(\begin{array}{ccc}
    E_1 & v \\
v & E_2 \\ \end{array}\right)\,,
\label{2lev}
\end{equation}
with the energy eigenvalues $E_\pm$
\begin{equation}
E_{\pm} = \frac{1}{2}\Big(E_1+E_2\pm \sqrt{(E_1-E_2)^2+4v^2}\Big)\,.
\label{hs}
\end{equation}
  The level repulsion, i.e., increase of   the gap of the two eigenvalues with increasing $v^2$ or $V^2$ is the characteristic  property of both systems.  To establish the  correspondence in the general case is more involved since the  matrix $W$ (\ref{Msig}) is not hermitian. The  non-hermitian contribution  arises due to the non-vanishing imaginary part of $\Sigma$. The (identical) contributions of Im$(\Sigma)$ to the diagonal elements of $W$ is due to the decay into the extended mode. Similarly the imaginary  part  of the off diagonal elements, i.e., in the coupling between the $m=\pm\nu$ modes,  $\Sigma$ appears since this coupling is generated by intermediate excitation of the extended mode. Nevertheless $W$ is not a general complex matrix. The conservation of flux restricts the non-hermitian part. Separating hermitian and non-hermitian contributions
$$W=W_h - iW_{nh},$$
one easily verifies that the non-hermitian part possesses the following structure
$$W_{nh}=    w^\dagger w,$$
 with
$$w= \omega^2 \sqrt{\text{Im}\Sigma}\,\left(\begin{array}{ccc}
    0 & 0 \\
\epsilon_{-\nu} & \epsilon_\nu \\ \end{array}\right)\,, $$
In quantum mechanical scattering on many body systems such as nuclei or atoms,  the   effective Hamiltonian  whose eigenvalues  are given by the resonance positions and widths   exhibits exactly this structure of its non-hermitian part \cite{FESH92},  \cite{MAWE69}, \cite{BREN9}.


\section{Conclusion}
The focus of our studies has been on the nature and the properties of  resonances which are generated in  scattering of light off photonic crystal slabs. We have established that these resonances are Feshbach resonances,\,i.e.,\,they are the optical analog of a particular type of resonances formed in  quantum mechanical scattering  of particles off atoms or atomic nuclei.  In quantum mechanics,   bound states of   many body systems in the absence of the coupling to the projectile are the progenitors of the resonances. In classical optics the role of the progenitors is played by the guided modes of the  dielectric slabs in the absence of the coupling to the extended modes where the grating region is replaced by a homogeneous layer with an effective dielectric constant.  Turning on  the respective couplings, bound states and guided modes are converted into resonances. Close to the resonance, the interaction between projectile and target is dominated by resonance formation. We have shown in our investigations that  resonance dominance is not only an important concept but also a  quantitative tool in the analysis of  scattering of light  off photonic crystal slabs.

In the formal part of our studies we have employed  techniques developed in the context of nuclear reactions to establish the resonances observed in scattering of light off photonic crystal slabs as Feshbach resonances. Despite of the common framework of resonance dominance we have identified significant differences in the implementation for  TE and TM polarizations.  We have derived expressions for the observables, the reflection and transmission amplitudes and the strengths of the magnetic and electric fields. The various shapes of the reflectivity as function of the wave length  are properly reproduced. Also  strong distortions of the Breit-Wigner form are correctly described in resonance dominance  and shown to result from the interference between background scattering and resonance formation.  In comparison with results of exact numerical evaluations  the accuracy of resonance dominance has been established to  range,   depending on the particular application,  from a few to 15 \% .

In comparison with exact numerical methods,  the power of the resonance dominance approach   rests upon the analytical formulation of the observables. It offers the possibility to study in detail the dependence of reflection and transmission amplitudes and the strength of the electromagnetic fields  in terms of the properties of photonic crystal slabs and the kinematics of the incident light. We have demonstrated  this  potential of resonance dominance in the analysis of the rather involved phenomena appearing in  the excitation of interacting resonances. In particular, we have been able to identify the source for the significant differences in the interaction of TE and TM resonances and the relevant parameters of the grating layer which control the shape of the reflectivity of overlapping resonances. We have derived analytical expressions for the band gap and the related curvature as a function of the angle of incidence and have study analytically the weird behavior of the electromagnetic field enhanced by the overlapping resonances.
Needless to say that for design issues this analytical procedure can be reversed,\,i.e.\,the properties of the photonic crystal slabs can be identified  which optimize certain requirements on observables.

\section{Appendices}
\subsection{Feshbach projection operators for light scattering}  \label{app:feshbach}

Since  Feshbach's seminal work \cite{FESH58,FESH62,FESH92} on the theory of nuclear reactions, the main tool   for such investigations is the formulation in terms of projection operators acting in an appropriately defined  Hilbert space. We now will sketch a reformulation of this approach  in the context of  differential equations describing the classical scattering  of light, \cite{FAJO02,TTHK99,NEST98}.

We will first consider   the  TE case. It is convenient to begin with the  coupled equations  (\ref{syseq}).  We will rewrite them in the form
\beq \label{eq:coupledTE}
\sum_{m=-\infty}^{\infty}[-\partial_{z}^{2}+(k_{x}+nK_g)(k_{x}+mK_g)]\delta_{nm}+
\omega^{2}(\delta_{nm}-\epsilon_{m-n}(z))]E_m(z)=
\omega^{2}E_n(z)
\eeq
or in the matrix-operator notation
\beq
h\psi=\omega^2\psi
\eeq
where
\beq \label{eq:fullh}
h_{nm}=[-\partial_{z}^{2}+(k_{x}+nK_g)(k_{x}+mK_g)]\delta_{nm}+
\omega^{2}(\delta_{nm}-\epsilon_{m-n}(z))]
\eeq
and
\beq
 \psi(z)\equiv
\{... ,\; E_{-m}(z),\; ...,\; E_{-1}(z),\; E_0(z),\; E_1(z),\;  ...,\;  E_m(z),\; ...\}
\eeq

We define projection operators acting on the array $\psi(z)$
\eqna
P\psi \equiv\psi_P &=&{\rm array\;\;obtained\;\;from}\;\; \psi\;\; {\rm by\;\;setting\;\;to\;\;zero\;\;all\;\;components\;\;apart \;\;of}\;\; \psi_0\;\;, \nonumber \\
Q\psi\equiv\psi_Q &=& {\rm array\;\; with \;\;the \;\; same \;\; components \;\; as}\;\;
 \psi \;\;{\rm but \;\; with}\;\; \psi_0=0
\eqne
Clearly
\beq
P^2=P\;\;, \;\; Q^2=Q\;\;, P+Q=1
\eeq
In a way which is standard in the Feshbach formalism \cite{FESH58},\cite{FESH92} we can act with $P$ and with $Q$ on our basic equation $h\psi=\omega^2\psi$. Using the above properties of $P$ and $Q$  we obtain coupled equations  of the Feshbach formalism
\eqna \label{eq:PQeqs2}
h_{PP}\psi_P+h_{PQ}\psi_Q=\omega^2\psi_P  \\
h_{QQ}\psi_Q+h_{QP}\psi_P=\omega^2\psi_Q  \nonumber
\eqne
where we defined
$$ PHP=h_{PP}\;\;,\;\;QHQ=h_{QQ}\;\;,\;\;PhQ= h_{PQ}\;\;,\;\; QhP=h_{QP}.$$
Clearly $h_{PP}=h_{00}$, $h_{QQ}$ is the matrix $h_{nm}$ with $n=0$ row and $m=0$ column excluded. The operators $h_{PQ}$ and $h_{QP}$ are  matrices  consisting respectively of  just $n=0$ row and $m=0$ column both without the $m=n=0$ element.
We can formally solve the second equation
\beq \label{eq:solveQsect} \psi_Q=(\omega^2-h_{QQ})^{-1}h_{QP}\psi_P
\eeq and insert into the first equation
Converting the resulting equation into an integral form we obtain
\eqna \label{eq:solvePsect}
\psi_P(z)&=&\psi_0^{(+)}(z)+\int dz' g_0(z,z')\langle z'|h_{PQ}|\psi_Q\rangle = \\
&=& \psi_0^{(+)}(z)+
\int dz' g_0(z,z')\langle z'|h_{PQ}(\omega^2-h_{QQ})^{-1}h_{QP}|\psi_P \rangle \nonumber
\eqne
where $\psi_0^{(+)}$ is the properly chosen solution of $(\omega^2-h_PP)\psi_0^{(+)}=0$ and $g_0(z,z')$ is the (properly chosen)  Green's  function of this equation, i.e.  $(\omega^2-h_PP)g_0=1$.  The meaning of "properly chosen"  for the present problems is explained in Appendix \ref{sec:bagr}.  We have also assumed that there is no unperturbed solution in the $Q$ sector, i.e. $\psi_Q=0$ for $h_{PQ}=0$.

So far we made no approximations. Let us consider the eigenfunctions  of $h_{QQ}$
\beq \label{eq:eigeninproj}
h_{QQ}\phi_\nu= \tilde{\eta}_\nu\phi_\nu
\eeq
 and assume that the geometrical parameters of the optical system and the kinematics of the light scattering  are such that there exist a range of discrete values of $\tilde{\eta}_\nu$ (guiding modes).   We will further assume that $h_{PP}$  has   a range of continuum eigenvalues (radiating modes) which overlap with the discrete modes of $h_{QQ}$ and moreover that $\omega^2$ lies within this combined range. In these circumstances the "off diagonal" terms  $h_{PQ}$ and $h_{QP}$  couple the  guiding modes    to the radiating modes turning the former into Feshbach resonances.  Note that we use $\tilde{\eta}_\nu$ to distinguish from $\eta_\nu$ used in the TE section of the main text.

A generalized version  of the  resonance dominance approximation used in the present work amounts to using  one or few resonant terms in the full spectral decomposition Green's function in the $Q$ sector
\beq
\frac{1}{\omega^2-h_{QQ}}=\sum_{\nu}\frac{|\phi_\nu\rangle\langle\phi_\nu|}{\omega^2-\tilde{\eta}_\nu}+G_c
\eeq
where $G_c$ is the contribution due to the continuum eigenfunctions. Using as an example two resonating eigenstates  we obtain
\beq
\frac{1}{\omega^2-h_{QQ}}\approx\frac{|\phi_1\rangle\langle\phi_1|}{\omega^2-\tilde{\eta}_1}+
\frac{|\phi_2\rangle\langle\phi_2|}{\omega^2-\tilde{\eta}_2}
\eeq
which means (cf., (\ref{eq:solveQsect}) that
\beq
 \psi_Q=\sigma_1 \phi_1+\sigma_2\phi_2 \;\;,\;\; \sigma_i=(\omega^2-\tilde{\eta}_i)^{-1}\langle\phi_i|h_{QP}|\psi_p\rangle\;,\;i=1,2
 \eeq
Following the standard route one inserts this into (\ref{eq:solvePsect})
\beq
\psi_P(z)=\psi_0^{(+)}(z)+\sum_{i=1,2}\sigma_i\int dz' g_0(z,z')\langle z'|h_{PQ}|\phi_i\rangle,
\eeq
then multiplies and integrates  by $\int dz \langle\phi_j| h_{QP}|z\rangle...$  with $j=1$ and $j=2$ and obtains coupled  equations for $\sigma_i$
\begin{equation} \label{eq:coupledgeneralizedeqsforsigmas}
\left(\begin{array}{cc}
\omega^2 - \tilde{\eta}_{1}-\Sigma_{11} &
-\Sigma_{12} \\
- \Sigma_{21} &
\omega^2 - \tilde{\eta}_{2}-\Sigma_{22}\end{array} \right)\left(\begin{array}{c} \sigma_1 \\ \sigma_2 \end{array}\right)=
\left(\begin{array}{c} \mathcal{C}_1^{(+)} \\ \mathcal{C}_2^{(+)} \end{array}\right)
\end{equation}
Here
\beq
\Sigma_{ij}=\langle\phi_i| h_{QP}g_0h_{PQ}|\phi_j\rangle     \;\;\; , \;\;\; \mathcal{C}_i^{(+)}=\langle\phi_i| h_{QP}|\psi_0^{(+)}\rangle \;, \;\; i=1,2
\eeq
In the TM scattering one can follow exactly the same route by simply noting that the basic set of equations (\ref{eq:eq2forxin}) is already in the form $h\psi=\omega^2\psi$  with $h_{nm}\equiv\Theta_{nm}$ and
 \beq
 \psi(z)\equiv
\{... ,\; H_{-m}(z),\; ...,\; H_{-1}(z),\; H_0(z),\; H_1(z),\;  ...,\;  H_m(z),\; ...\}
\eeq

It is important to recognize that equations  (\ref{eq:coupledgeneralizedeqsforsigmas}) are based on assumed exact eigenfunctions $\phi_1$ and $\phi_2$ of $h_{QQ}$, Eq. (\ref{eq:eigeninproj}). In reality $h_{QQ}$ is almost as complicated as the full $h$ in Eq. (\ref{eq:fullh}) and further approximations are required. In our treatment  of overlapping resonances leading to  (\ref{sigsy}) and (\ref{eq:coupledTMeqsforsigmas}) we have drastically truncated  the problem and replaced the exact $\phi_1$ and $\phi_2$ by  the eigenfunctions  of  $h_{11}$ and $h_{22}$.
In this way the coupling between these components via $h_{12}$ and $h_{21}$ as well as the coupling to other components are neglected.
In fact we observe that equation  (\ref{eq:coupledgeneralizedeqsforsigmas}) is similar in form  to the equations (\ref{sigsy})  and (\ref{eq:coupledTMeqsforsigmas}) apart from  the absence of  the term $V$ in the off diagonal elements of the matrix W.  This term approximately accounts in Eqs. (\ref{sigsy})  and (\ref{eq:coupledTMeqsforsigmas}) for the direct coupling between the approximate guided modes $\phi_1$ and $\phi_2$.

\subsection{Background fields and Green's function}\label{sec:bagr}
In this section we will derive the necessary expressions for the background field and the associated Green's function. As in the main section we assume that the dielectric function $\epsilon_0(z)$ approaches 1 for $ z\to -\infty$.
\subsubsection{TE waves}
We start with Eq.(\ref{gzp}) for the Green's function $g_0(z,z^\prime)$ in the TE scattering.    Let us denote by $E_0^{(\pm)}(z)$
two independent solutions of  the  homogeneous equation (\ref{hom}).  Using these solutions the Green's function can be expressed as
\begin{equation}
g_0(z,z^\prime)= -\frac{1}{W}\big( \theta (z^\prime-z) E_0^{(+)}(z^\prime)\,E_0^{(-)}(z)+\theta (z-z^\prime) E_0^{(+)}(z)\,E_0^{(-)}(z^\prime)\big)\,.
\label{gr1}
\end{equation}
where $W$ is the ($z$-independent) Wronskian
\beq \label{Wrsk}
W= \frac{d E_0^{(+)}(z)}{dz} E_0^{(-)}(z)- E_0^{(+)}(z)\frac{d E_0^{(-)}(z)}{dz}.
\end{equation}
This  is easily verified by observing that by construction the Green's function satisfies Eq.(\ref{gzp}) for $z\neq z^\prime$.
Furthermore the prefactor $-1/W$ is found by integrating (\ref{gzp}) over $z$ in an infinitesimal interval around $z'$.

The choice of  the   two linearly independent solutions $E_0^{(\pm)}(z)$   is dictated by the type of the Green function
one needs.  It is convenient to require  that $g_0(z,z')$ as a function of z  for the values of
 $z'$  inside the grating behaves as a reflected (transmitted)   wave in the $z\rightarrow-\infty$  ($z\rightarrow \infty$) region far away from the grating.  This leads to the boundary conditions
\begin{equation}
  \mathop {\lim }\limits_{z \to \infty }  E_0^{(+)}(z) =t_0^+(\omega) e^{ ik^+_z z}\,,\quad \mathop {\lim }\limits_{z \to -\infty }E_0^{(-)}(z) =t^-_0(\omega) e^{- ik^-_z z}\,
\label{asyp}
\end{equation}
with  $k_z^\pm$ denoting  the z-component of the incident wave vector in the two asymptotic regions
\begin{equation}
k_z^{+}=\omega \sqrt{\epsilon_{\text{IV}}-\sin^2\theta}\,,\quad k_z^{-}=\omega \sqrt{\epsilon_{\text{I}}-\sin^2\theta}\,.
\label{asmo}
\end{equation}
These equations describe  electric fields incident from above ($E_0^+$) or below ($E_0^-$) (cf.\,Fig.\,\ref{dispn}).
Their normalization is fixed by the amplitude of the incident wave which we choose as
\begin{equation}
\mathop {\lim }\limits_{z \to -\infty } E_0^{(+)}(z)=  e^{ ik_z^- z} + r_0^+(\omega) e^{- ik_z^- z }\,,\quad \mathop {\lim }\limits_{z \to \infty } E_0^{(-)}(z) =  e^{- ik_z^+z} + r_0^-(\omega) e^{ ik_z ^+ z }\,.
\label{asym}
\end{equation}
Evaluating $W$ in region IV  yields
\beq
W=-2ik_z^+ t_0^+(\omega)
\eeq
 Equally well the  Wronskian can be calculated in  region I  yielding the relation
\begin{equation}
\big|r_0^+(\omega)\big|^2+\frac{k_z^+}{k_z^-}\big|t_0^+(\omega)\big|^2=1.
\label{unita}
\end{equation}
The two solutions  $ E_0^{(\pm)}(z)$ are related to each other
\begin{equation}
E_0^{(-)}(z)= \frac{1}{t_0^{+\,\star}(\omega)}\big( E_0^{(+)\,\star}(z) - r_0^{+\,\star} E_0^{(+)}(z)\big)\,.
\label{epm}
\end{equation}

The Green's function at  large negative  $z$ is needed in the computation of the reflection amplitude
\begin{equation}
\mathop {\lim }\limits_{z \to -\infty }  g_0(z,z^\prime) = \frac{1}{2i k_z^-}e^{-ik^-_z z}\,E_0^{(+)}(z^\prime)\,.
\label{asg0}
\end{equation}
We also need the spectral representation of the Greens-function
\begin{equation}
g_0(k_z,z,z^\prime)=\frac{1}{2\pi}\int dk^\prime_z \frac{E^{(+)}_0(k_z^\prime,z)E^{(+)\,\star}_0(k_z^\prime,z^\prime)}{k_z^{2}- k_z^{\prime\,2}+i\epsilon}\,.
\label{specreps}
\end{equation}
Here we have made explicit the dependence of the fields and the Green's function on the incident wave vector.
Again it is obvious that for $z\neq z^\prime$ the Green's function satisfies Eq.(\ref{gzp}).  With the help of the completeness relation
\begin{equation}
\int dk^\prime_z \,E^{(+)}_0(k_z^\prime,z)\,E^{(+)\,\star}_0(k_z^\prime,z^\prime) = 2\pi \delta(z-z^\prime)\,,
\label{complt}
\end{equation}
the correct form of the singularity in Eq.(\ref{gzp}) is confirmed. Using the decomposition of the denominator into principal value and $\delta$ function contributions
$$\frac{1}{k_z^{2} - k_z^{\prime\,2}+ i\epsilon} = \frac{P}{k_z^{2} -k_z^{\prime\,2}} -i\pi \delta\big(k_z^{2} -k_z^{\prime\,2}\big)\,,$$
the imaginary part of the Green's function is obtained
\begin{equation}
\text{Im}\, g_0(k_z,z,z^\prime) = -\frac{1}{4k^-_z}\Big[E^{(+)}_0(k_z,z)\,E^{(+)\,\star}_0(k_z,z^\prime)+E^{(+)\,\star}(k_z,z)
\,E^{(+)}_0(k_z,z^\prime)\Big].
\label{Imag}
\end{equation}

\subsubsection{TM waves}
The Green's function $g_0(z,z^\prime)$ for the TM case,  Eq. (\ref{eq:TM-green}), is found   in the same  way as for the TE scattering.
 We denote two independent solutions of the homogeneous equation (i.e. of Eq. (\ref{eq:TM-green}) with the right hand side set to zero) by $H_0^{(\pm)}(z)$. Then
\beq \label{eq:greenfunforTM}
g_0(z,z')=
\nu \left[H_0^{(+)}(z)H_0^{(-)}(z') \theta(z-z')+ H_0^{(+)}(z')H_0^{(-)}(z) \theta(z'-z) \right]
\eeq
with a z-independent constant $\nu$ which is found by integrating (\ref{eq:TM-green}) over $z$ in an infinitesimal interval around $z'$. The only subtlety relative to the TE case is the presence
of the $\partial_z\gamma(z)\partial_z$ in the defining equation (\ref{eq:TM-green}). This leads to two modifications relative to the TE case. On one hand the  expression  for $\nu$  is now
\beq \label{eq:expressionfornu}
\nu=-\frac{1}{\gamma_0(z)W(z)}.
\eeq
where  $W(z)= H_0^{(+)}\partial_z H_0^{(-)} - H_0^{(-)}\partial_z H_0^{(+)}$ is the Wronskian of the two solutions. On the other  hand the Wronskian of  any two solutions like $H_0^{(\pm)}(z)$ is not a constant  but obeys the so called Abel formula written in our case as  $\gamma_0(z)W(z)=constant$. Together we have that $\nu$ is indeed independent of $z$ and can be calculated in a convenient point.

We again choose the solutions $H_0^{(+)}(z)$  and $H_0^{(-)}(z)$  such that  $g_0(z,z')$ behaves as a reflected (transmitted)   wave when $z'$ is finite and $z\rightarrow-\infty$  ($z\rightarrow \infty$). This leads to the conditions
 \beq
  z \to  - \infty\,:\quad   H_0^{(-)}(z)\to t^-_0 e^{- ik^-_z z} \,; \quad z \to   \infty\,: \quad H_0^{(+)}(z) \to t^+_0 e^{ ik^+_z z}
 \eeq
with
\beq
  z \to   \infty\,: \quad H_0^{(-)}(z)\to e^{ -ik^+_z z}+r^+_0 e^{ ik^+_z z} \,; \quad z \to  - \infty\,:\quad H_0^{(+)}(z)\to e^{ ik^-_z z}+r^-_0 e^{- ik^-_z z}.
\eeq

At $z\to-\infty$ we have $\gamma(z\to-\infty) = 1/\epsilon_{I}$ and  $W(z\to-\infty)= - 2ik^-_zt^-_0$ so that
 \beq
 \nu=-\frac{i\epsilon_{I}}{2k^-_z t^-_0}
 \eeq
Using in Eq. (\ref{eq:expresforH0}) the asymptotic expressions for $H_0^{(\pm)}(z)$  and  the Green's function (\ref{eq:greenfunforTM})  we obtain the result (\ref{eq:asymtofH0forTM}).

\subsection{Dealing with singular  expressions in the description of TM resonances}\label{app:singular}
The derivative terms in the operators  $\Theta_{k0}$ and $\Theta_{0k}$ contain combinations
$\partial_z\epsilon_{k}(z)\partial_z$  which produce $\delta$-function like singularities when acting on  $H_0^{\pm}(z)$, $\mathcal{H}_{k,\kappa}$ and $g_0(z,z')$.  Here we will explain how we have regularized such expressions.
Let us consider as an  example the  expression (\ref{eq:TMexpforSigma}) for $\Sigma$.  Our strategy is to start with
narrow but continuous transition layers between the dielectric layers.  We then transform this expression in such a way that in the limit of  sharp boundaries i.e. zero transition layers width it will not contain ambiguous singularities.  Let us demonstrate this
for the term of (\ref{eq:TMexpforSigma})  which contains   the differential operator parts of both  $\Theta_{k0}$ and $\Theta_{k0}$
\eqna
&& \int dz dz' \mathcal{H}_{k,\kappa}(z)\partial_z\gamma_k(z)\partial_z g_0(z,z') \partial_{z'}\gamma_{-k}(z')\partial_{z'}\mathcal{H}_{k,\kappa}(z') =  \nonumber \\
&=&\int dz dz' [\partial_z\mathcal{H}_{k,\kappa}(z)]\gamma_k(z)[\partial_z \partial_{z'}g_0(z,z')]\gamma_{-k}(z')[\partial_{z'}\mathcal{H}_{k,\kappa}(z')]\rightarrow \\
&\rightarrow &  |\gamma_k|^2 \int_{I_g} dz dz' [\partial_z\mathcal{H}_{k,\kappa}(z)][\partial_z \partial_{z'}g_0(z,z')][\partial_{z'}\mathcal{H}_{k,\kappa}(z')]
\eqne
where we started with infinite  integration range assuming that $\gamma_{\pm k}(z)$ tend to zero fast but
 continuously outside the grating layer.
 In integrations by parts  the surface terms vanish while the resulting integral has only finite discontinuities  in the limit of sharp boundaries. Such discontinuities  do not cause ambiguities so that the limit can safely be taken.  The other two terms containing  differential operators in (\ref{eq:TMexpforSigma})  can be regularized in the same manner.

The above procedure can be conveniently summarized by stating that expressions like (\ref{eq:TMexpforSigma})  should be used with the operators $\Theta_{nm}$ replaced by
\beq
\Theta_{nm}\rightarrow \gamma_{n-m}[\overleftarrow{\partial_z}\;\overrightarrow{\partial_z}+(k_z+nK_g)(k_x+mK_g)]
\eeq
where the arrow above the derivative indicates that it acts on the function to the right or to the left of it depending on the direction of the arrow.

We also note a useful identity
\eqna \label{eq:identity}
\partial_{z}\partial_{z'} g_0(z,z') &=& \nu [\partial_{z}H_0^{(+)}(z)\partial_{z'}H_0^{(-)}(z') \theta(z-z')+\partial_{z'}H_0^{(+)}(z')\partial_{z}H_0^{(-)}(z) \theta(z'-z)] -  \nonumber \\
&& \;\;\;  - 1/\gamma_0(z)\delta(z-z')
 \eqne
which is obtained by differentiating Eq. (\ref{eq:greenfunforTM}) and using (\ref{eq:expressionfornu}).

\section*{Acknowledgments}

F.L. is grateful for the support  and the hospitality at  the  Department of Condensed Matter, Weizmann Institute.  This work is supported in part by the Albert Einstein Minerva Center  for Theoretical Physics and by a grant from Israeli Ministry of Science.

\end{document}